\documentclass[11pt]{article}
\usepackage{amsmath,amsfonts,array,amscd}
\newcommand{\balpha}{{\vec{\alpha}}}
\newcommand{\bs}{{\vec{s}}}

\newcommand{\adH}{{\delta_H}}

\newtheorem{theorem}{Theorem}[section]          
\newtheorem{corollary}[theorem]{Corollary}
\newtheorem{lemma}[theorem]{Lemma}
\newtheorem{proposition}[theorem]{Proposition}
\numberwithin{equation}{section}                

\newcommand{\RR}{\mathbb{R}}
\newcommand{\Rl}{\mathbb{R}}
\newcommand{\CC}{\mathbb{C}}
\renewcommand{\Re}{\,\mathrm{Re}\,}   

\renewcommand{\vec}[1]{\boldsymbol{#1}}

\newcommand{\Hess}{\mathop{\rm Hess\,}}
\def\idty{{\mathchoice {\mathrm{1\mskip-4mu l}} {\mathrm{1\mskip-4mu l}} %
{\mathrm{1\mskip-4.5mu l}} {\mathrm{1\mskip-5mu l}}}}

\newcommand{\eps}{\varepsilon}

\newcommand{\eqlabel}[1]{\label{#1}}
\newcommand{\thlabel}[1]{\label{#1}}

\newenvironment{proof}{\noindent {\bf Proof: }}{\QED\medskip}
\newenvironment{proofof}{\noindent {\bf Proof of }}{\QED\medskip}
\def\QED{{\hspace*{\fill}{\vrule height 1ex width 1ex }\quad}
    \vskip 0pt plus20pt}

\newcommand{\NN}{\mathbb N}
\newcommand{\ZZ}{\mathbb Z}

\newcommand{\be}{\begin{equation}}
\newcommand{\ee}{\end{equation}}
\newcommand{\bea}{\begin{eqnarray}}
\newcommand{\eea}{\end{eqnarray}}
\newcommand{\bean}{\begin{eqnarray*}}
\newcommand{\eean}{\end{eqnarray*}}
\newcommand{\beq}{\begin{equation}}
\newcommand{\eneq}{\end{equation}}
\newcommand{\bey}{\begin{eqnarray}}
\newcommand{\eey}{\end{eqnarray}}
\newcommand{\beyn}{\begin{eqnarray*}}
\newcommand{\eeyn}{\end{eqnarray*}}
\newcommand{\beann}{\begin{eqnarray*}}
\newcommand{\eeann}{\end{eqnarray*}}

\newcommand{\g}{\gamma}
\newcommand{\e}{\varepsilon}
\newcommand{\s}{\sigma}

\newcommand{\Tr}{\mbox{\rm Tr\, }}
\newcommand{\Ind}{\mbox{\rm Ind}}

\newcommand{\cF}{{\cal F}}

\newcommand{\cV}{{\cal V}}

\newcommand{\ran}{\rangle}
\newcommand{\lan}{\langle}

\newcommand{\rcite}{\cite}

\newcommand{\db}{{\mkern2mu\mathchar'26\mkern-2mu\mkern-9mud}}

\newcommand{\Av}{\mathop{\rm Av}\limits}

\newcommand{\bdot}{{  \bullet}}

\newcommand{\donothing}[1]{{}}

\newcommand{\fu}{\mathfrak{u}}

\newcommand{\fq}{\mathfrak{q}}
\newcommand{\fn}{\mathfrak{n}}

\newcommand{\fv}{{v}}

\newcommand{ \cI }{\mathcal{I}}

\newcommand{\bq}{{\bf q}}

\newcommand{\bu}{{\bf u}}

\newcommand{\bx}{{\bf x}}
\newcommand{\by}{{\bf y}}

\newcommand{\bw}{{\bf w}}
\newcommand{\bm}{{\bf m}}

\newcommand{\bA}{{\bf A}}
\newcommand{\bW}{{\bf W}}
\newcommand{\bJ}{{\bf J}}

\newcommand{\bua}{{\bf u}_{\alpha^+}}
\newcommand{\bwma}{{\bf w}_{M, \alpha^-}}

\newcommand{\ubw}{\underline{\bf w}}
\newcommand{\unabla}{\underline{\nabla}}

\newcommand{\ubA}{\underline{\bf A}}

\newcommand{\ubB}{\underline{\bf B}}

\begin{document}

\rightline{September 2002, revised June 2003}

\vspace{2.5cm}
\begin{center}{\Large\bf
Derivation of the Euler Equations\\[20pt]
from Quantum Dynamics}

\end{center}

\vspace{3cm}

\renewcommand{\thefootnote}{\fnsymbol{footnote}}

{\large \centerline{Bruno Nachtergaele \footnote[1]{Department of
Mathematics, University of California, Davis,
bxn@math.ucdavis.edu. Research partially
supported by NSF   \# DMS-0070774} and Horng-Tzer Yau\footnote[2]{Courant
Institute, New York University, yau@cims.nyu.edu. Research partially
supported by NSF  \# DMS-0072098, the Veblen fund from the
Institute for Advanced Study and a Fellowship from the MacArthur
Foundation.}
\renewcommand{\thefootnote}{\arabic{footnote}}}
} \vspace{1cm}
\begin{abstract}
We derive the Euler equations from quantum dynamics for a class of
fermionic many-body systems. We make two types of assumptions. The
first type are physical assumptions on the solution of the Euler
equations for the given initial data. The second type are a number
of reasonable conjectures on the statistical mechanics and
dynamics of the Fermion Hamiltonian.
\end{abstract}
\vspace{2cm}

\newpage

\tableofcontents

\section{Main notations}

\begin{tabular}{p{1.5cm}<{\quad\dotfill}p{12cm}}
$\mu$
  &Typical index labeling the five conserved quantities.\\
$j$
  &Typical index referring to time, $j=0$, and space, $j=1,2,3$, components.\\
$\bw$
  &$\bw=(\bw^\mu)=(w^\mu_{j,x})$,$\mu = 0, \ldots 4,  j = 1, 2, 3$,
   are the components of the current densities. $\mu=0$ is the particle current,
   $\mu=1,2,3$, indexes the three components of the momentum current,
   $\mu=4$ is the energy current. $j=1,2,3$ refers to the three spatial
   directions of the current.\\
$\bW$
  &All quantities denoted by $\bW_*$, are understood to be given by
  $l^3\bw_*$, for any subsript ${}_*$. \\
$\ubw$
  &Underlined vectors have an additional component,
   referring to time, or, in the case of the currents,
   the conserved quantities.\\
$\unabla$
  &$\unabla=(\partial_t, \nabla)$ is the four-component gradient, inclding
  the derivative with respect to time.\\
$\bu$
  &$\bu = (u^0, \cdots, u^4)= \underline{\bw}_0 := \bw_{0}$ are the five
  conserved
  quantities: particle number, three components of the momentum, and energy.\\
$u^0_x$
  &$u^0_x= n_x$ is the quantum observable for the particle number density at
  microscopic space point $x$.\\
$\fu_x$
  &$\fu_x=( u^1_x, u^2_x, u^3_x)$, the quantum observables for the three
  components of the momentum density at $x$.\\
$u^4_x$
  &$u^4_x= h_x$ is the quantum observables for the energy
   density at $x$.\\
$\bA$
  & $\bA=(A^\mu_j)$ are the classical currents appearing the RHS of the Euler
  equations
    ($\mu=0, \cdots, 4, j=1, 2, 3$). They are defined by
    \beann
    A_j^0 &=&q^j \\
    A^i_j &=&\delta_{ij} P + q_i q_j /q _0\\
    A^4_j &=& q^j (q_4 +P)/q_0 \ .
    \eeann\\
$\underline{\bA}$
  &$\underline{\bA} =(A^\mu_j)$ are the classical currents augmented with
  the conserved quantities ($j=0$, vanishing current in the time
  direction):
  $\underline{\bA}_0= \bq= {\bA}_0$.\\
$\bq$
  &$\bq=(q^0, \cdots, q^4)$: $q^0=\rho$ is the classical particle density,
  $q^1, q^2, q^3$ are
   the three components of the classical momentum density,
   and $q^4=e$ is
   the classical energy density. These quantities are a function of macroscopic
   space and time. This notation is also used for the expectation value of these
   quantities in a quantum Gibbs state.\\
$\fq$
   &$\fq=(q^1, q^2, q^3)$ are
   the three components of the classical momentum density.\\
$\fv$
   &$\fv=\fq/\rho$ is the classical, macroscopic mean velocity per particle.\\
$\tilde{e}$
   &$\tilde{e} = e /\rho$ is the classical, macroscopic mean energy per
   particle.\\
\end{tabular}

\newpage
\begin{tabular}{p{1.5cm}<{\quad\dotfill}p{12cm}}
$P(e,\rho)$
   &The thermodynamic pressure as a function of the energy and
   particle densities. Appears in the Euler equations and is defined
   as the quantum statistical pressure for the Fermion system under
   consideration.\\
$\e $
   &$\e $ is the scaling parameter relating the macroscopic coordinates
   $X,T$, with the microscopic coordinates $x,t$: $X=\e  x, T=\e  t$.
   The hydrodynamic limit is the limit $\e  \to 0$.\\
$\cI$
   &The embedding from a collection of independent subcubes of periodic
   boundary condition to the
   cube $\Lambda_{\e^{-1}}$.\\
$I$
   &The embedding from a subcube of periodic
   boundary condition to the
   cube $\Lambda_{\e^{-1}}$.\\
$\bu_{x, \ell}^+$ & local conservative quantities in a cube of
size $\ell$ centered at $x$.\\

$\Lambda_\ell$ & cube of width $\ell$ centered at the origin. \\

$\lambda_\e (t, x)$ & $\lambda_\e (t, x) = \lambda (\e t, \e x)$\\
\end{tabular}

\subsection{A convention}

In the course of our arguments, we will encounter a large (but
finite!) number of error terms. Therefore, we introduce a common
notation that conveys all relevant information about these error
terms. For $k\geq 1$, $y$ any list of symbols, let $\Omega_y(x)$
denote a real-valued function of $x\in\Rl^k$, with the property
that
$$
\lim_{x_k}\lim_{x_{k-1}}\cdots\lim_{x_1}\Omega_y(x)=0
$$
where the limits are determined by the names $x_1,\ldots,x_k$ of
the variables. E.g., any term denoted by
$\Omega_{\kappa,X}(\e , \ell)$ has the property
\begin{equation}\label{vancon}
\lim_{\ell\to \infty} \lim_{\e \to 0}
\Omega_{\kappa,X}(\e , \ell)=0
\end{equation}
The limits are to be taken in the specified order and the
quantities denoted by $y=(y_1,\ldots,y_l)$ are kept fixed in the
limits. The limit for $\e $ is $\e \to 0$, as this is
the limit we are considering and similarly the limit for $\ell$ is
unambiguously $\ell\to\infty$. The actual value of any
$\Omega_y(x)$ may vary from occurrence to occurrence.

For quantities used in error bounds about which no claim of convergence
to zero is made, we will usually use the notation $C_y$, where $y$ lists
the relevant parameters the constant may depend on.

\newpage

\section{Introduction}

The fundamental laws of non-relativistic microscopic physics are
{\em Newton's} and {\em Schr\"odinger} equations in the classical
and the quantum case respectively. These equations are impossible
to solve for large systems and macroscopic dynamics is therefore
modeled by phenomenological equations such as the {\em  Euler} or
the {\em Navier-Stokes} equations. Although they were derived
centuries ago from  continuum  considerations, they are in
principle consequences  of  the microscopic physical laws and
should be viewed as secondary equations. It was first observed by
C. Morrey \rcite{Mor} in the fifties that the Euler equations
become `exact' in the Euler limit, provided that the solutions to
the  Newton's equation  are `locally' in equilibrium. Morrey's
original work was far from rigorous and the meaning of `local
equilibrium' was not clear.  It is nevertheless a very original
idea and it contributed significantly to the later development of the
hydrodynamical limits of interacting particle systems, see \cite{Sp} for a
review. Instead of considering  general classical dynamical systems with
two body interactions, a different approach
is to  prove as much as possible for some simplified
models. Outstanding examples are the works by Boldrighini, Dobrushin,
and Suhov \cite{BDS}, and Sinai \cite{Sin} in the case of one space
dimension, and the more recent work by Eyink and Spohn \cite{ES} who
study a $d$-dimensional classical system of non-interacting particles.
In terms of a rigorous proof of Morrey's idea, however,
significant progress has only been made rather recently \cite{OVY}.
This long delay is mostly due to a serious lack of tools for
analyzing many-body dynamics, in the classical case and even more
so in the quantum case.

In this paper, we derive the Euler equations from microscopic
quantum dynamics, extending the relative entropy method
of \cite{Yau, OVY} to the quantum cases.
Our main result was announced in \cite{NY}.
As we want to consider the genuine quantum
dynamics for a system with short-range pair interactions, we
cannot take a semiclassical limit. Although one-particle quantum
dynamics converges to Newtonian dynamics in the semiclassical
limit, this limit does not commute with the
scaling  limit needed for the Euler equation.
This is most clearly seen in the pressure function, for which
quantum corrections survive at the macroscopic scale. In fact, one
of the conclusions of our work is that under rather general
conditions, the pressure function is the only place where the
quantum nature of the underlying system, in particular the
particle statistics, survive in the Euler limit.

The Euler equations have traditionally been derived from  the
Boltzmann equation both in the classical case and in the quantum
case, see Kadanoff and Baym \cite{KB} for the quantum case.  Since
the Boltzmann equation is valid only in  very low density regions,
these derivations  are not  satisfactory, especially in the
quantum case where the relationship  between  the quantum dynamics
and the Boltzmann equation is not entirely clear. There were,
however, two  approaches based directly on quantum dynamics. The
first was due to Born and Green \cite{BG}, who used an early
version of what was later called the BBGKY hierarchy, together
with mo\-ment methods and some truncation assumptions.  A bit
later, Irving and Zwanzig \cite{IZ} used the Wigner equation,
moment methods and truncations to accomplish a similar result.
These two approaches rely essentially on the moment method with
the Boltzmann equation replaced by the Schr\"odinger equation.
Unlike in the Boltzmann case, where one can do asymptotic analysis
to justify this approach, it seems unlikely that this can be done
for the Schr\"odinger dynamics.

One of the benefits of our approach is that we develop a
general strategy applicable to  all situations where a number of reasonable
assumptions are satisfied. We believe that our general assumptions, which are
discussed in detail in Section \ref{sec:assumptions}, hold for a large class
of physical models. We regard proving the properties that we assume as an
important, although rather challenging, research project in quantum statistical
mechanics.

The main novelty of our work lies in the fact that, for the first  time, the
relative entropy method is applied to a quantum mechanical system.  This
requires solving a number of technical problems which, not surprisingly, all
stem from the fact that the local observables corresponding to the globally
conserved quantities of the dynamics, are represented by non-commuting
operators. This is mainly discussed in Section \ref{sec:local_ergo}.

Although our goal is a derivation of the Euler equations, the relative entropy
method indeed constructs an approximate solution to the underlying many body
dynamics based on solution of Euler equations and the concept of local Gibbs
states. It thus establishes the key role played by the Euler equations: they
are not just a set of conservation laws but,  with the correct choice for the
pressure function, they actually dictate the leading approximation to the many
body classical or quantum dynamics. Thus, the Euler equations may also be used
to  obtain information about the solutions of the many-body Schr\"odinger
equation.

\subsection{Schr\"odinger and Euler dynamics}

We begin by considering $N$ particles on $\RR^3$, evolving
according to the Schr\"odinger equation
$$
i \partial_t \psi_{t}(x_1, \cdots, x_N)
=   H  \psi_{t}(x_1, \cdots, x_N)
$$
where the Hamiltonian is given by
\begin{equation}
H = \; \;
\sum_{j=1}^N  \frac { - \Delta_{j}}{ 2} +
\sum_{1\le i < j \le N} W (x_i-x_j) \; .
\label{Ham}\end{equation}
Here, $W$ is a two-body short-ranged stable isotropic pair
interaction and  $\psi_{t}(x_1, \cdots, x_N)$ is the wave function
of particles  at time $t$. We only consider Fermions (such as
electrons, but for simplicity we ignore spin) and thus the state
space ${\cal H}^N$ is the subspace of antisymmetric functions in
$L^2(\RR^{3N})$, i.e., $ \psi (x_{\s_1}, \cdots, x_{\s_N})=
(-1)^{\s} \psi (x_1, \cdots, x_N), $ for any permutation $\s$ of
$\{1, \cdots, N \}$. It is more suitable not to fix the total
number of particles and to use the second quantization
terminology. In fact, it would be extremely cumbersome to work
through all arguments without the second quantization formalism.
The state space of the particles, called the Fermion Fock space,
is the direct sum of ${\cal H}^N$:
${\cal H} : =   \oplus_{N=0}^\infty \; {\cal H}^{N}$.

Define the annihilation  and creation operators $a_x$ and $a_x^+$
by
\bea
(a_x \Psi)^{N} ( x_1, \cdots, x_N) & = & \sqrt {N+1}
\Psi^{N+1}
(x, x_1, \cdots, x_N) \label{a} \\
(a_x^+ \Psi)^{N} ( x_1, \cdots, x_N) & = & \frac {1 }{\sqrt {N}}
\sum_{j=1}^N  (-1)^{j-1} \delta(x-x_j) \Psi^{N-1} (x_1, \cdots,
\widehat {x_j}, \cdots, x_N) \; , \label{a+}
\eea
where, as usual,
$\widehat{\; \;}$ means ``omit''. $a_x$ and $a_x^+$ are to be
interpreted as operator-valued distributions \cite{BLT}. The
annihilation  operator  $a_x$ is simply the adjoint of $a_x^+$
with respect to the standard inner product of the Fock space with
Lebesgue measure $dx$. These operators satisfy the canonical
anticommutation relations
\begin{equation}\label{car}
[ a_x, a^+_y ]_+ : = a_x a^+_y  + a_x a^+_y = \delta (x-y) , \;\;
[ a^+_x, a^+_y ]_+ = [ a_x, a_y ]_+=0\; ,
\end{equation}
where $\delta$ is the delta distribution. The derivatives of these
distributions with respect to the parameter $x$ are denoted by
$\nabla a_x$ and $\nabla a^+_x$. With this notation, we can
express the Hamiltonian as $H= H_0 + V$ where the kinetic energy
is given by
$$
    H_0 = \frac 1 2 \int \nabla a_x^+  \nabla a_x \, \db x
$$
and the potential energy
$$
V = \frac{1}{2} \int \int dx dy W(x-y)
a_{x}^+ a_{y}^+ a_{y} a_{x}   \; .
$$
It is more convenient to put the Schr\"odinger equation into the
operator form, which is sometimes called the
Schr\"odinger-Liouville equation. Denote the density matrix of the
state at time $t$ by $\g_t$. Only normal states, which can be
represented by density matrices, will be considered in the time
evolution. Then the Schr\"odinger equation is  equivalent to

\be\label{sch}
i\partial_t \g_t = \delta_H  \gamma_t\; \qquad \mbox{with }
\delta_H \gamma_t : = [H , \g_t ] .
\ee

The  conserved quantities of the dynamics are the  number of
particles ,   the three components of the  momentum and the
energy. The local densities of these quantities are denoted by
$\bu =(u^\mu), \mu=0, \cdots, 4$, and are given by the following
expressions:
\bea
u^0_x = n_x &=& a_x^+ a_x  \nonumber\\
u^j_x= p_x^j &=& \frac i 2   [ \nabla_j a^+_x  a_x- a^+_x
\nabla_j a_x  ], \qquad j=1, 2, 3,  \label{u-def}\\
u^4_x= h_x &=& \frac 1 2 \nabla a^+_x \nabla a_x + \frac 1 2 \int
dy W(x-y) a_{x}^+ a_{y}^+ a_{y} a_{x} \nonumber
\eea
We also introduce the notation $\fu= (u^1, u^2, u^3)$. This
convention will be  followed for the rest of the paper. We use the bold face
for the vector of the conservative quantities and use the frac
for the vector consisting only the components $1, 2, 3$.

Let $\Lambda_\ell$ denote a cube of width $\ell$ centered at the origin.
The subscript $\ell$ may be omitted if it plays no active role.
We shall adopt the convention that unbounded observables on
$\Lambda$ will be defined with periodic boundary conditions. E.g.,
the number of particles in $\Lambda$, the total momentum, and the
total energy of the particles in $\Lambda$, respectively, are
defined by \bea
N_\Lambda &=& \int_\Lambda dx\, n_x \nonumber\\
P^j_\Lambda &=& \int_\Lambda dx\, p^j_x,\quad j=1, 2, 3 \nonumber\\
H_\Lambda &=& \int_\Lambda dx\, h_x \nonumber \eea
In other words, we shall always
view  $\Lambda$ as a three-dimensional
torus.

We slightly generalize the definition of the grand canonical Gibbs
states to include a parameter for the total momentum of the
system: the Lagrange multiplier $\alpha$. We will work under the
assumption that the temperature and chemical potential are in the
one-phase region of the phase diagram of the system under
consideration such that the thermodynamic limit is unique. The
finite volume Gibbs states are then given by the following
formula:
\be \label{finitebiggs}
\omega_{\beta, \alpha, \mu}^\Lambda(X)=
\frac{\Tr Xe^{ -\beta( H_{0,\Lambda}+V_\Lambda-\alpha \cdot
P_\Lambda- \mu N_\Lambda)}}{\Tr
e^{-\beta(H_{0,\Lambda}+V_\Lambda-\alpha P_\Lambda- \mu
N_\Lambda)}}
\end{equation}
The infinite volume Gibbs states $\omega_{\beta, \alpha, \mu}$
are the limiting points
of the finite volume ones.
It is convenient to denote the parameters $(\beta, \alpha, \mu)$
by $\vec \lambda= (\lambda^\mu), \mu=0, \cdots, 4$ with
$\lambda^0= \beta \mu, \lambda^j = \beta \alpha^j, \lambda^4 =
\beta $. Define (notice the sign convention)
\be
 {\vec \lambda}\cdot  \bu
= \sum_{\mu=0}^3  { \lambda}^\mu \,   u  ^\mu\, - \, { \lambda}^4
\,   u  ^4
\label{dot-convention}\ee
and
$$
\lan  {\vec \lambda},  \bu   \ran_\Lambda = |\Lambda|^{-1}
\int_\Lambda  d x   {\vec \lambda}(x) \cdot  \bu   (x)
$$
These notations allow us to give a compact formula for the unique,
translation invariant Gibbs state (defined with constant $\vec
\lambda$), as well as for the states describing local equilibrium
(defined with $x$-dependent $\vec \lambda$):
\begin{equation}\label{gibbs}
\omega_{\vec \lambda}^\Lambda =  e^{ |\Lambda| \lan
{\vec \lambda}, {\bu   } \ran_\Lambda }/Z_{\Lambda} ({\vec
\lambda} )
\end{equation}
where $Z_{\Lambda}({\vec \lambda} ) $ is the partition function
\begin{equation}\label{part}
Z_{\Lambda} ({\vec \lambda} ) = \Tr e^{|\Lambda|
\lan {\vec \lambda}, {\bu   } \ran_\Lambda }
\end{equation}
The pressure as a function of the constant vector $\vec \lambda$,
is defined by
$$
\psi ({\vec \lambda} ) =  \lim_{\Lambda \to \infty} |\Lambda|^{-1} \log
Z_{\Lambda} ({\vec \lambda} )
$$
Denote the expectation value of the conservative
quantities in an infinite-volume equilibrium state, $\omega_{\vec \lambda}$
introduced following \eqref{finitebiggs},
by $\bq = (q^0, \cdots, q^4)$. The we have
\be
\frac {\partial \psi } {\partial \lambda^\mu} =
\omega_{\vec \lambda}(  u  ^\mu ) \label{dual}
\ee
Explicitly,
\beann
\rho
&=&\omega_{\vec \lambda}(n_x)=\lim_{\Lambda\to\RR^3}\frac{1}{\vert\Lambda\vert}
\omega_{\vec \lambda}^\Lambda(N_\Lambda)\\
\fq
&=&\omega_{\vec \lambda}(p_x)=\lim_{\Lambda\to\RR^3}\frac{1}{\vert\Lambda\vert}
\omega_{\vec \lambda}^\Lambda(P_\Lambda)\\
e&=&\omega_{\vec \lambda}(h_x)=\lim_{\Lambda\to\RR^3}\frac{1}{\vert\Lambda\vert}
\omega_{\vec \lambda}^\Lambda(H_\Lambda)
\eeann
Notice that $\fq$ and $e$ are momentum and energy per volume.

Again, we will work under the assumption that these parameters
stay in the one-phase region, the limiting Gibbs state is unique
and these definitions are unambiguous. Although momentum is
preferable as a quantum observable, we also introduce the velocity
in order to be able to compare with the classical case. The
velocity field $\fv(x)$ has to be defined as a mean velocity of
the particles in a neighborhood of $x$. Therefore we have
$\fv(x)=\fq(x)/\rho(x)$. We also introduce the energy per particle
defined by $\tilde e  = e /\rho$. The usual Euler equations are
written in terms of $\rho, \fv$, and $\tilde e $.

In order to derive the Euler equations, we need to perform a
rescaling. So we shall put all particles in a torus $\Lambda_{\e^{-1}}$
of size $\e^{-1}$ and use $(X,T) = (\e x, \e t)$ to denote the
macroscopic coordinates. For all equations in this paper periodic
boundary conditions are implicitly understood.

The Euler equations for the five conserved quantities, which arise
in the limit $\epsilon \to 0$, are given by
\bea
\frac{\partial
\rho}{\partial T} +\sum_{j=1}^3 \frac{\partial}{\partial
X_j} (\rho \fv_j)  &=& 0 \nonumber \\
\frac{\partial \, (\, \rho \fv_k \, ) }{\partial T} +\sum_{j=1}^3
\frac{\partial}{\partial X_j} \left [ \; \rho \fv_j  \fv_k  \;
\right ] +\frac{\partial}{\partial X_k}P(e , \rho)
&=& 0 \label{euler}\\
\frac{\partial \, (\, \rho \tilde e  \, ) }{\partial T} +
\sum_{j=1}^3 \frac{\partial}{\partial X_j}\left [ \; \rho \tilde e
\fv_j  + \fv_j P(e , \rho) \; \right ]
&=& 0  \nonumber
\eea
These equations are in form identical to the classical ones
but all physical quantities are computed quantum mechanically. In
particular, $P(e , \rho)$ is the thermodynamic pressure computed
from quantum statistical mechanics for the microscopic system. It
is a function of $X$ and $T$ only through its dependence on $e$
and $\rho$. If no velocity dependent forces act between the
molecules of the fluid under consideration (we consider only a
pair potential), the pressure is independent of the velocity.

The conservative quantities  $\bq = (q^0, \cdots, q^4)$, related to density,
momenta and energy as follows:
\be
q^0 = \rho \ , \quad q^i   = \rho \, \fv^i
\ , \quad q^4 =e=
 \rho \, \tilde e  \ ,
\label{2.17} \ee
In other words $q^1, q^2, q^3,$ and $q^4$ are
momenta and energy per {\it volume} instead of {\it per particle}
as in the usual Euler equation \eqref{euler}. We rewrite the Euler
equations in the following  form
\be \frac {\partial q^\mu }{
\partial T} +\sum_{i=1}^3 \nabla_i^X \big[ A^\mu_i (q)\big] = 0 \ ,
\quad  \mu =0,1,2,3,4\ . \eqlabel{2.15}
\end{equation}
The matrix $A$ is determined by comparison with the Euler
equations: \bea
A_j^0 &=&q^j \nonumber\\
A^i_j &=&\delta_{ij} P + q_i q_j /q _0\label{A-def}\\
A^4_j &=& q^j (q_4 +P)/q_0 \ . \nonumber
\eea

\subsection{Local equilibrium}

To proceed we need a microscopic description of local equilibrium
and a microscopic prescription to compute the pressure from
quantum statistical mechanics. Suppose we are given macroscopic
functions $\bq  (X)$. We wish to find a local Gibbs state with
the conserved quantities given by $\bq  (X)$. The local Gibbs
states are states locally in equilibrium. In other words,  in a
microscopic neighborhood of any point $x\in T^3$ the state is
given by a Gibbs state. More precisely, we wish to find a local
Gibbs state with the expected values of the energy, momentum, and
particle number per unit volume at $X$ given by $\bq  (X)$. To
achieve this, we only have to adjust the parameter $\vec \lambda$
at every point $X$. More precisely, we choose ${\vec \lambda} (X)$
such that the equation \eqref{dual} holds at every point, i.e.,
$$
\frac {\partial \psi ({\vec \lambda} (X))} {\partial {
\lambda}^\mu (X)} = q^\mu (X).
$$
If we denote the solution to the Euler equation by $q(X, T)$, then
we can choose in a similar way a local Gibbs state with given
conserved quantities at the time $T$. Define
the local Gibbs state \be \omega_t^\e = \frac {1}{c_\e (t)}\exp
\left [\,\e^{-3}  \lan
 {\vec \lambda}(\e t,\e \cdot), \, \bu
 \ran_{\Lambda_{\e^{-1}} } \right ]
\label{2.27}
\end{equation}
where $c_\e (t)$ is the normalization constant. Clearly, we have
that $\omega_t^\e ( u  _x^\mu ) = q^\mu (\e x, \e t)$ to leading
order in $\e$. Our construction of local equilibrium states is consistent
with the abstract framework discussed in \cite{SS}.

Later we will need the following relation for the
normalization constant $c_\e (t)$: \be \frac d {dt} \log c_\e (t)
= \e \Tr \omega_t^\e  \big[\, \e^{-3} \lan  \partial_t {\vec \lambda}
(\e t,\e \cdot), \, \bu    \ran_{\Lambda_{\e^{-1}} }  \big]
\label{2.28}
\end{equation}
Since the inner product almost exclusively taken on $\Lambda_{\e^{-1}}$,
we shall drop this subscription or replaced it by
$\lan \; , \;  \ran_{\e^{-1}}$ for the rest of this paper.

In summary, the goal is to show that, in the limit $\e \to0$,
the following diagram commutes:
$$\begin{CD}
\bq(X,0)@>\mbox{Euler}>>\bq(X,T)\\
@V\mbox{local equilibrium}VV @AA\parbox{4cm}{limit
$\e \downarrow 0$
of expectation of locally averaged observables}A\\
\gamma_0@>\mbox{Schr\"odinger}>>\gamma_{\e ^{-1}T}
\end{CD}$$

As smooth solutions of the Euler equations are guaranteed to exist
only up to a finite time \cite{KM}, say $T_0$, we will formulate
our assumptions on the dynamics of the microscopic system for a
finite time interval as well, say $t\in [0, T_0/\e]$. Note the
cutoff assumptions below would hold automatically for lattice
models.

\subsection{Assumptions and the main theorem}\label{sec:assumptions}

Our main result is stated in Theorem \ref{thm-1} below. First, we
state the assumptions of the theorem with some brief comments.
There are three kinds of assumptions.

The first category of assumptions could be called {physical\/}
assumptions on the solution of the Euler equations that we would
like to obtain as a scaling limit of the underlying dynamics, and
on the pair interaction potential of this system.

\bigskip \noindent {\bf I. One-phase regime:} We assume that the pair
potential, $W$, is $C^1$ radial  and supported in a ball of
radius $R$.
Furthermore, we assume that $W$ is stable in the sense
that
\begin{equation}\label{superstable}
W(x)=W_0(x)+W_1(x)\hbox{ where } W_0\ge 0\ ,
W_0(0)>0\hbox{ and } W_1 \hbox{ is positive definite.}
\end{equation}
Here, the positive-definiteness of $W_1$ refers too the sesquilinear
form, not the function itself. I.e.,
$$
\sum_{1\leq i,j\leq n}W_1(x_i-x_j)\overline{z}_iz_j,
\mbox{ for all}, n\geq 1, x_i, x_j\in\Rl^3, z_i,z_j\in \CC.
$$
Such potentials automatically satisfy
the usual super-stability property \cite{Rue}. In particular,
they are stable, i.e., there is a constant
$B\geq 0$ such that,  for all $N\geq2$, $x_1,\ldots,x_N\in\Rl^3$,
$$ \sum_{1\leq1<j\leq N}W(x_i-x_j)\geq -BN $$

Of the Fermion system with potential $W$ we assume that there is
an open region $D\, \subset \, \Rl^2$, which we will call the {\em
one-phase region\/}, such that the system has a unique limiting
Gibbs state and a regular pressure function for all values of
particle density and energy density $(\rho,e)\in D$.

The solution of the Euler equations we consider, $q(X,T)$, will be
assumed to $C^1$ in $X$ for $T\in [0,T_0]$, and have local
particle and energy density in the one-phase region for all times
$T\in [0,T_0]$. I.e., $(\rho(X,T), e(X,T))\in D$, for all $X\in
\Lambda_1$ and $T\in [0,T_0]$.

The next category of assumptions is on the local equilibrium
states for the Fermion system that we construct and on their
time-evolution under the Schr\"{o}dinger equation.

\bigskip
\noindent {\bf II. Cutoff assumptions:} Suppose that $\gamma_t$ is
the solution to the Schr\"odinger equation \eqref{sch} with a local
equilibrium state as initial condition, constructed with the
parameters derived from a solution of the Euler equations (with
the appropriate pressure function) for times $t$ in a finite
interval $[0,T_0]$, that does not leave the
one-phase region.  We make the following two assumptions.

\noindent {\sl 1. High-momentum cutoff assumption:} Let $N_p(t)=
\Tr \gamma_t  a_p^+  a_p$, where $a^\#_p$ is the Fourier transform
of $a^\#_x$. Then there is a constant $c > 0$ such that for all $t
\le T_0/\e$,
\be
\e^d  \int dp   e^{ c p^2} N_p(t) \le C_{T_0}
\label{maxwellian}\ee
where $C_{T_0}$, is constant only depending on $T_0$.

\bigskip
\noindent {\sl 2. Non-implosion assumption:} There is a constant
$C_{T_0}$ (not necessarily the same as $C_{T_0}$ in the previous paragraph)
such that for all $t \le T_0/\e$ \be
\Tr \left\{\gamma_t  \e^{d} \int_{\Lambda_\epsilon} dx n_x
\left [  \int_{|x-y| \le 2 R}
n_y  d y \right ]^2 \right\}\le C_{T_0} \eqlabel{2bound}
\end{equation}
where $R$ is the range of the interaction $W$.

Finally, we have an assumption on the set of the time-invariant
ergodic states of the Fermion system. To state this assumption we
need the notion of {\em relative entropy\/}, of a normal state
$\gamma$ with respect to another normal state $\omega$. Let
$\gamma$ and $\omega$ denote the density matrices of these states.
The relative entropy, $S(\gamma\mid\omega)$, is defined by
$$
S(\gamma|\omega)= \begin{cases} \Tr \left\{\gamma(\log \gamma -\log
\omega)\right\} & \mbox{if } \ker \omega \, \subset \,  \ker \gamma\\
+\infty &\mbox{otherwise}
\end{cases}
$$
For a pair of translation invariant locally normal states, one can
show existence of the relative entropy density \cite{OP}, defined
by the limit
$$
s(\gamma|\omega)= \lim_{\e\downarrow 0}\e^3
S(\gamma_{\Lambda_{\e^{-1}}}\mid\omega_{\Lambda_{\e^{-1}}})\; ,
$$
where $\gamma_{\Lambda_{\e^{-1}}}$ and $\omega_{\Lambda_{\e^{-1}}}$ denote the
density matrices of the normal states obtained by restricting
$\gamma$ and $\omega$ to the observables localized in
$\Lambda_{\e^{-1}}=\e^{-1}\Lambda_1$. The existence of the limit can be
proved under more general conditions on the finite volumes, but
this is unimportant for us.

\bigskip
\noindent {\bf III. Ergodicity assumption (``Boltzmann
Hypothesis''):} All translation invariant,  ergodic with respect to space
translations, stationary (i.e., time invariant) states to the
Schr\"odinger equation with the Hamiltonian
$H$ are Gibbs states with the same Hamiltonian
provided they satisfy the following assumptions: {\em 1) the
density and energy is in one  phase region}. {\em 2) The  relative
entropy density with respect to some Gibbs state is finite}.

We expect that the cutoff assumptions  hold for the solutions $\gamma_t$
of the Schr\"odinger equation that we employ, but for now
there is no complete proof that it holds for
Gibbs states other than the free Fermi gas.
For Gibbs states in the high temperature region we expect
these assumptions can  be proved  by using some
type of cluster expansion methods. A partial result in this direction
has been obtained  recently, in the case of Bosons, by Gallavotti,
Lebowitz, and Mastropietro in \cite{GLM}.

For the rest of this paper, we shall assume this cutoff
assumptions for the solution to the Schrodinger equations as well
as the Gibbs states in the one phase regions considered in this
paper.

We wish to point out that in the treatment of  the classical case  in
\cite{OVY} the cut-off assumption 2 was not needed. There is however no proof
for the cut-off assumption 1 even in the classical case. (In \cite{OVY}, the
usual quadratic kinetic energy was replaced by one with bounded derivatives
with respect to momentum. So the cut-off assumption 1 is not needed too.)

The cutoff assumptions are technical in nature. For Fermion models on a lattice
instead of in the continuum, no cut-off assumptions are required. The Boltzmann
hypothesis on the other hand is a fundamental problem in statistical physics. A
version of it was proved to hold for a classical ideal gas by Eyink and Spohn
in \cite{ES}. Gurevich and Suhov \cite{GS} proved that a stationary Gibbs state
to a classical dynamics with a Hamiltonian $H$ has to be a Gibbs state with the
same Hamiltonian. Under the assumption that the stationary measures velocity
distribution has no correlation (a weaker assumption than in \cite {GS}), the
Boltzmann hypothesis was proved for classical gas with two-body interaction
\cite{OVY}.

Our main result is the following Theorem. We also expect it to
hold for Bosons with a super-stable interaction.

\begin{theorem} \thlabel{thm-1}
Suppose that $\bq (X, T)$ is a smooth solution to the Euler
equation in one phase region up to time $T\le T_0$. Let
$\omega^\e_t$ be the local Gibbs state with conserved quantities
given by $\bq (X, T)$. Suppose that the  cutoff assumptions and
the ergodicity assumption hold. Let $\gamma_t$ be the solution to
the Schr\"odinger equation \eqref{sch} and $\gamma_0 =\omega^\e_0$
(Note that $\gamma_t$ depends on $\e$). Then 
we have
$$
\lim_{\e \to 0} \sup_{0\leq t\leq \e^{-1}T_0}s(\gamma_t|\omega^\e_t) = 0
$$
In other words, $\omega^\e_t$ is a solution to the Schr\"odinger
equation \eqref{sch} in entropy sense. In particular, for any smooth
function $f$ on $\Lambda$, we have, for all $0\leq T\leq T_0$,
$$
\lim_{\e \to 0} \e^3\int_{\Lambda_{\e^{-1}}} d x f(\e x)\left[ \gamma_{\e^{-1}T}
(\bu   _x) - \bq (T, \e x) \right ] = 0
$$
\end{theorem}

The proof of this theorem will be given in Section \ref{sec:conclusion}.

Illustrated with a diagram, the main theorem says
$$\begin{CD}
{\omega}^\eps_{{\bq_0}} @> \mbox{Euler equation}>> {\omega}^\eps_{{\bq_T}}\\
@VVV @VV
\lim_{\e\downarrow 0} s(\gamma_{\e ^{-1}T} | {\omega}^\eps_{{\bq_T}})=0V\\
\gamma_0@>\mbox{Schr\"odinger equation}>>\gamma_{\e ^{-1}T}
\end{CD}$$
Notice that we have proved more than just convergence to the Euler equation.
We have shown that the local-equilibrium Gibbs state constructed from the
evolution of the Euler equations solves the many-body Schr\"odinger equation,
approximately in entropy sense.

\subsection{Outline of the proof}

The basic structure of our proof follows the relative entropy
approach  of \cite{OVY, Yau}. The aim is to derive a differential
inequality for the relative entropy between  the solution to the
Schr\"odinger equation and a time-dependent local Gibbs state
constructed to reproduce the solution of the  Euler equations. The
time derivative of the relative entropy can be expressed as an
expectation of the local currents with respect to the solution to
the Schr\"odinger equation. Since we do not know the solution
well-enough, this expectation can not be computed.

{\sl Step 1: Replace the local microscopic currents by macroscopic
currents.} The basic idea in hydrodynamical limit is first to show
that the local space time average of the solution is time
invariant. From the Boltzmann hypothesis, ergodic time invariant
states are Gibbs. For Gibbs states, we can replace the local
microscopic currents by macroscopic currents.  This is the first
step. In the quantum setting, there are several crucial issues we need
to address.

{\sl 1a: Construct a commuting version of the local conserved
quantities.} Recall macroscopic currents are functions of the
local conserved quantities, i.e., density, momentum and
energy. For the microscopic quantum system, the local conservative quantities
are operators which commute only up to boundary terms. In order to
express the macroscopic currents as functions of the local
conserved quantities, we need either to prove that the
non-commutativity does not affect the macroscopic
currents or we need to construct some commuting version of the
local conserved quantities. As the first approach seems very
difficult to carry out, we follow the second one and construct a
commuting version of local conservative quantities in section 4.

{\sl 1b: Restriction to the one phase region.} Since the Boltzmann
hypothesis holds only in the one phase region, we have to exclude
the region outside the one phase region. To perform this restriction
to the one-phase region , we would normally multiply the observables by
some cutoff function. In our case however, the cutoff function does not
commute with the local currents. This seemingly trivial multiplication by a
cutoff function illustrates the kind of technical problems
we have to address in this work. Our approach to this is presented in
Section \ref{sec:local_ergo}

{\sl 1c: Virial Theorem. } Even assuming the local ergodic states
are Gibbs in the one phase region,  we still have to compute the
macroscopic currents from the microscopic currents. This requires
a virial Theorem, which we provide in Section \ref{sec:virial}.

{\sl Step 2: Estimate all errors by local conservative quantities.
}  As will become clear, the errors associated with
the cutoff of the one phase region are difficult to control
directly. We shall  bound them by the local conserved
quantities. This will be carried out in sections 5 and 6.

{\sl Step 3: Derive a differential inequality of the  entropy with
error term given by a large deviation formula. } After Step 2, we
have an expression of the derivative of the entropy in terms of
local (commuting) conservative quantities. Since these quantities
commute, by an entropy inequality, we can bound it by a
large deviation expression. Notice that it is crucial that we
control everything by commuting objects. There is no large
deviation theory for non-commuting observables.

After this step, the standard relative entropy method provides the
rest of the argument.
Technically speaking, the main difficulty to study a quantum
mechanical system, in comparison with a classical one,  can be
traced back to the non-commutativity of the algebra of
observables. E.g., suppose $A$ and $B$ are two self-adjoint
operators representing observables of the system. A simple
inequality, such as $|A+B| \le |A|+|B|$, which is used numerous
times in estimates for classical systems, is false, and so is
$\vert AB\vert\le \vert A\vert\,\vert B\vert$. Therefore, there
are essentially no absolute values taken in our proof and we
estimate all quantities by commuting versions of the locally
conserved quantities. Of course, these inequalities hold with the
absolute value replaced by the norm. However, we will frequently deal
with error terms that are expectations of unbounded observables, such
as, e.g., the high-momentum contributions to the energy. Clearly,
norm estimates are useless in this situation.

\section{Relative entropy identity and high momentum cutoff}

\subsection{Entropy identity}

The first step in the derivation of a diffential inequality for
the relative entropy is the compute the derivative.
Suppose $\gamma_t$ is a solution to the Schr\"odinger equation.
Recall that one has
\begin{equation}
\frac{d}{dt}\Tr  A(t) B(t)= \Tr    A^\prime(t) B(t) +  A(t)
B^\prime(t),
 \qquad \frac{d}{dt} \Tr e^{A(t)}=\Tr e^{A(t)} A^\prime(t)
\label{derivative_trace_2}
\end{equation}
and
\begin{equation}
\frac d {dt} S(\gamma_t) = 0
\end{equation}
Thus we have  for any time-dependent density matrix $\rho_t$
the identity
\be \frac d {dt} S(\gamma_t|\rho_t) = \Tr \gamma_t \left \{
 - i \adH  - \partial_t  \right \} \log \rho_t \eqlabel{re}\ .
\end{equation}

This identity replaces the relative entropy inequality in
\cite{OVY,Yau}. Thus, by \eqref{2.27},
\be \frac d {dt} s(\gamma_t|\omega^\e_t) =
\Tr \gamma_t \left \{ - i \adH  - \partial_t   \right \} \left \{
 \lan  {\vec \lambda_\e } (t, \cdot ) ,   \bu    \ran_{\e^{-1} }
 - \e^3  \log c_\e (t) \right \} ,
\eqlabel{re2}
\end{equation}
where
$ \lan \;  ,   \;     \ran_{\e^{-1} }
 = \lan \;  ,  \;  \ran_{\Lambda_{\e^{-1}} }$.
 
Let $\bw$ denote the current tensor with components $w^\mu_{k,x}$
defined by
\bea
{ w }^0_{k,x} &=&
p^k_x=\frac{i}{2}
[\nabla_k a^+_x  a_x-a^+_x\nabla_k a_x]\label{theta0}\\
{ w }^j_{k,x} &=&  \frac{1}{2}\left[\nabla_j  a_x^+  \nabla_k   a_x
+ \nabla_k  a_x^+  \nabla_j   a_x\right]\nonumber\\
&& -\frac 1 2 \int\!dy\, \Big [ W'(x-y) \frac{ (x-y)_j  (x-y)_k} {|x-y|}\Big ]
  a_x^+  a_{y}^+ a_{y} a_{x},\quad k=1,2,3\label{thetaj}\\
{ w }^4_{k,x} &=& -\frac i 4   \Big [  \nabla_k a_x^+  \Delta a_x
 - \Delta a_x ^+     \nabla_k a_x  \Big ]
 + \frac i 4 \int\! dy\,  W(x-y)  \Big [ \nabla_k  a_x^+  a_{y}^+ a_{y} a_{x}
- a_{x}^+ a_{y}^+ a_{y}  \nabla_k  a_x  \Big ]\\
&&
  - \frac i 4  \int\!dy\,\Big [ W'(x-y) \frac{ (x-y)_k  (x-y)_j} {|x-y|}\Big ]
  \Big [a_x^+    \nabla_j   a_{y}^+  a_{y} a_{x}
   -  a_{x}^+ a_{y}^+   \nabla_j a_y  a_{x} \Big ]\label{theta4}
\eea where we have used the rotation invariance of the potential
to write
$$
W'(x) = \frac {d W (r)}{d r}|_{r = |x|}\ .
$$
We have the following proposition.
See Section \ref{sec:calculations} for the derivation of these
expressions for the current.

\begin{proposition}\label{prop:theta}
Let $\Omega_{\vec \lambda}(\e)$ be defined by the equation
\be \eqlabel{re3}
i \adH  \lan
{\vec \lambda}_\e  ( t, \cdot),   \bu    \ran_{\e^{-1} }
=  \e \sum_{j=1}^3   \lan  \nabla_j {\vec \lambda_\e } (t, \cdot
),  \; { \bw }_{j}(t) \ran_{\e^{-1} }  \, + \, \Omega_{\vec
\lambda}(\e)\,
\end{equation}
Then $\Omega_{\vec \lambda}(\e)$ is an error term which satisfies
the condition
$$
\lim_{\e \to 0} \Tr \gamma \,  \Omega_{\vec \lambda}(\e) = 0
$$
\end{proposition}

These expressions of the microscopic currents seemingly bear no
relationship to the macroscopic currents in the Euler equations,
even when one assumes that $\gamma_t$ is locally Gibbs. This
difficulty already appears in the classical case. But by
reasonably straightforward computation and application of a
quantum version of the virial theorem  \ref{thm:virial} one can
show that indeed these currents correspond to the standard Euler
equations given in \eqref{euler}.

Define $\nabla_0 = \partial_t$ and ${ w}^\mu_{0, x}=  u  ^\mu_x$.
We have
\begin{equation} \label{re4}
 \left \{  - i  \adH - \partial_t   \right \} \e^3  \lan
 {\vec \lambda_\e } (t, \cdot ),   \bu    \ran_{\e^{-1} }
=  - \e \sum_{j=0}^3   \lan  \nabla_j {\vec \lambda_\e } (t, \cdot
)\, , \; { \bw }_{j}(t) \ran_{\e^{-1} } := \e G({\vec \lambda_\e},
a^+, a)
\end{equation}
where $\lambda_\e (t, x) = \lambda (t, x)$.
Introduce the notations $\unabla = (\nabla_0, \nabla)$,
$\ubw = (\bw_0, \bw)$ and
\begin{equation}
\ubA\, \bdot \,  \ubB = \sum_{j=0}^3 \sum_{\mu=0}^3  { A}^\mu_j \,
B^\mu_j\, - \, \sum_{j=0}^3 {A}^4_j \,   B^4_j \; .
\end{equation}
Then we can rewrite the last expression as
\begin{equation} \label{re44}
 \left \{ - i  \adH - \partial_t   \right \} \e^3  \lan
 {\vec \lambda_\e } (t, \cdot ),   \bu    \ran_{\e^{-1} }
=  - \lan  \unabla {\vec \lambda_\e } (t, \cdot
)\, \bullet  \; { \ubw } (t) \ran_{\e^{-1} }
\end{equation}
If we wish to emphasize the dependence on the operator, we shall
write ${ w}^\mu_{j,x} = { w}^\mu_{j,x}(a^+, a)$. From now on, we
shall drop the subscript $\e^{-1}$ in $\lan \, , \, \ran_{\e^{-1}}$.

\subsection{High-momentum cutoff}\label{sec:momentum-cutoff}

Most of the estimates we need are obtained using bounded versions
of the creation and annihilation operators, i.e., suitable
so-called smeared operators. Physically, this corresponds to
introducing a high-momentum cutoff. The precise form of the cutoff
will be important for us, as we will have strict requirements on
the behavior of the error terms for the proof to go through.

Let $\hat  \phi _M$ be a smooth function such that

1. $ |\hat   \phi _M (p) -1|\le e^{-M^2} \hbox{ for $|p| \le M$
and } |\hat  \phi _M (p)| \le e^{-M^2} $  for $|p| \ge 2 M$.

2. The support of $ \phi _M$  is bounded in a ball of radius
$e^{M^2}$.

To construct such a function, let $g$ be a smooth function
supported in $|x| \le 2$ such that
$$
\hat g(p) \le C [ 1+ p^2]^{-3}
$$
Define $g_\lambda(x) = g(x/\lambda) \lambda^{-3/2}$. Let $h_M$ be
a smooth function such that
$$
\hat  h_M (p) =1 \hbox{ for $|p|
\le M$ and } \hat h_M (p)  = 0,\; \mbox{ for } |p| \ge 2 M\; .
$$
Let
$$
 \phi _M= (g_\lambda \ast g_\lambda)h_M
$$
Notice that $\int   \phi _M = 1$. Let $\lambda= e^{M^2}$. Then we
can check easily the properties 1 and 2. Although  $ \phi _M$  is
supported in a ball of radius $e^{M^2}$, its mass is concentrated
in a ball of radius $M^{-1}$. More precisely, there is a constant
$c$ such that
$$
\int_{|x| \ge r}  h_M(x) dx \le e^{ -c r M}
$$

Define
$$
a_{x, M}^+  = \int  \phi _M (x-y) a^+_y = a^+_{ \phi _{x,M}} ,
\quad a_{x, M} = \int \overline {  \phi (x-y)} a_y =a_{\overline {
\phi _{x,M}}}
$$
where $ \phi _{x,M} =  \phi _M (x-y)$. In our setting  $\overline
{ \phi _{x,M}} =  \phi _{x,M}$.  Define
$$
\nabla  a_{x, M}^+ =a^+_{\nabla  \phi _{x,M}} , \qquad \nabla
a_{x, M} =a_{\nabla  \phi _{x,M}}
$$
Notice that $\nabla  a_{x, M}^{\pm}$ and $a_{x, M}^{\pm}$ are
bounded operators {\it localized in a ball of radius  $e^{M^2}$}.

We now perform the preliminary truncation. Denote the cutoff
version of the current by
\begin{equation}\label{wm}
{ \bw }_{j, x, M}={ \bw }_{j,x}(a^+_M, a_M)
\end{equation}
Notice that ${ \bw }_{j, x, M}$ is bounded.
The difference between the kinetic energy term in the energy current
with and without cut-off can be calculated using
$$
\nabla a_x^+  \Delta a_x - \nabla  a_{x, M}^+  \Delta a_{x, M} =
\nabla  b_{x, M}^+  \Delta a_x + \nabla  a_{x}^+  \Delta b_{x, M}
$$
where
$$
b_{x, M}^+ =  a_x^+ - a_{x, M}^+
$$

\begin{lemma}\thlabel{cutoff-vel}
For any state $\gamma$ that satisfies the cutoff assumptions and
for $G$ defined by \eqref{re4}, we have
$$
\e^d  \Tr \gamma \int dx \left [ \nabla  b_{x, M}^+  \Delta a_x +
\nabla  a_{x}^+  \Delta b_{x, M} \right ] \le   e^{ - c M^2}
$$
\end{lemma}

\begin{proof}
By using the Fourier transform, we have
$$
\int dx \nabla  b_{x, M}^+  \Delta a_x = \int dp (1-\hat  \phi
(p)) p^2 a^+_p a_p
$$
Let $N_p(t)= \Tr \gamma  a_p^+  a_p$. Then
$$
\Tr \gamma \int dp (1-\hat  \phi _M (p)) p^2 a^+_p a_p = \int dp
(1-\hat  \phi _M (p)) p^2 N_p
$$
The lemma is thus a simple consequence of the Chebeshev inequality
and the definition of $ \phi _M$.
\end{proof}

From the Schwarz inequality \bean
&&\Tr \gamma  \e^{d}
\int dx \nabla a_x^+  \int dy W(x-y) a_y^+ a_y a_x 
\\
&\le & \e^{d} \left \{ \Tr \gamma   \int dx \nabla a_x^+ \nabla
a_x \right \}^{1/2} \left \{ \Tr \gamma  a_x^+ \left [  \int dy
W(x-y) n_y \right ]^2 a_x \right \}^{1/2} \le C \eean
Thus for any state $\gamma$ satisfy the cutoff assumptions, we have
\be
\Tr \gamma \e^{d}   \int dx dy W(x-y)
 \left [ \nabla a_x^+  a_y^+ a_y a_x -
\nabla  a_{x,M}^+  a_{y, M}^+ a_{y, M} a_{x, M} \right ] \le e^{ -
c M^2} \eqlabel{4Wbound}
\end{equation}

\begin{lemma}\label{lcutoff}
For any state $\gamma$ satisfy the cutoff assumptions, we have \be
\Tr \gamma \e^{d}   [ G({\vec \lambda}, a^+, a) -
 G({\vec \lambda}, a^+_M, a_M)] \le  e^{ - c M^2}
\eqlabel{4bound}
\end{equation}
\end{lemma}

Thus  we have \be \frac d {dt} s(\gamma_t|\omega^\e_t) =  \e \Tr
\gamma_t   G({\vec \lambda_\e}, a^+_M, a_M) -\e^3  \frac d {dt }
\log c_\e(t) + {\cal E}_M^1  \eqlabel{re5}
\end{equation}
with  ${\cal E}_M^1  \le C e^{-c M^2} $. The precise form of the last
estimate is crucial as we shall see later on.

We recall the crucial relative entropy
inequality \cite{OP}: for all
self-adjoint observables $h$, and for any $\delta > 0$, :
\begin{equation}
\gamma(h) \le \delta^{-1} \log\Tr \; e^{ \delta h+ \log \omega }
+\delta^{-1} S(\gamma|\omega) \eqlabel{entineq}
\end{equation}
A proof  will be given in the  appendix.

\section{Construction of local commuting observables}\label{sec:commuting}

The conserved quantities commute as global observables on
$\Lambda$ with periodic boundary conditions. In fact, this is
essential for the classical equations of motion to make sense.
For any bounded quasi-local observable $X$ on the Fock space define
\be
\hat X (\bq ) = \Tr \omega_{\vec \lambda}  X
\label{Xhat}\ee
where the chemical potential ${\vec \lambda}$ is
the dual of $\bq $ in the sense of \eqref{dual}. We define $\hat X$
only for arguments in the one phase region.

Local averages of the densities of the conserved quantities, however,
do not commute due to boundary effects. Therefore, we cannot extend the
functions $\hat X$ to functions of the operator-valued local densities
of the conserved quantities, which is what we would like to do. To
circumvent this difficulty, in this section we construct
commuting versions of the local conserved quantities.

We have for any smooth function $J$
$$
\Tr \gamma_t \lan  J(\e t, \e \cdots),
X \ran = \Tr \gamma_t \Av_{x \in \Lambda_{\e^{-1}}}  J (\e t, \e x)
\Av_{|z-x| \le \ell/2} \tau_z X + \Omega_\ell (\e)
$$
where
$\Av_{\cdot}(\cdot)$, stands for the {\em average} of its argument over the
domain indicated in the subscript, and $\lim_{\e \to 0} \Omega_\ell (\e)  = 0$
for any $\ell$ fixed.

Denote $ \bu_{\ell} := \bu _{\Lambda_\ell} := \int_{\Lambda_\ell}
\; \tau_x  \,  \bu    \, dx $ the local conservative quantities
and we would like to replace the microscopic current $ \Av_{|z-x|
\le \ell/2} \tau_z X$  by certain  function of the local
conservative quantities $\tau_z \bu_\ell$. Unfortunately  the
components of $\bu_\ell$ do not commute and functions of $
\bu_\ell$ are not well-defined. In fact, even the definition of
$\bu_\ell$ is ambiguous since we did not specify the boundary
condition. Intuitively, the components of $ \bu_\ell$  actually
commute up to boundary terms, and the ambiguity should be
negligible in the limit $\ell \to 0$. Since it is rather difficult
to control these boundary terms in a simple way,  we construct in
the following  a commuting version of the local conservative
quantities.

\subsection{Construction of an isometric embedding}

Let $f$ be a smooth function with
$$
f(s) = 1/\sqrt 2   \quad \text{  if } \;  s \le 0 \qquad = 0 \quad
\text { if }  \; s \ge 1
$$
and
$$
f(1)=f'(1)=f^{''}(1)= 0 \; .
$$
For any given $\eta, 0 < \eta < 1/2$, let
$$
g(t)= \left [ 1 - f^2 \left ( \frac {\frac 1 2 - t} \eta \right )
\right ]^{1/2}, \qquad 0 \le t \le 1/2
$$
$$
g(t)=  f\left ( \frac {t- \frac 1 2  } \eta \right ) ,  \qquad  t
\ge 1/2,
$$
$$
g(t) = g(-t), \quad t \in \mathbb  R \; .
$$
Then $g$ is smooth, supported in $|t|\le 1/2 + \eta$ and
$$
\sum_{j \in \mathbb  Z}  g^2(t+j) = 1
$$
Let $\chi (x) = g(x^1)g(x^2) g(x^3)$. Then
\begin{equation}\label{=11}
\sum_{j \in \mathbb  Z^3}  \chi^2(t+j) = 1
\end{equation}

Let  $\alpha^{\pm} := \tau_\alpha \Lambda_\ell^{\pm}$
be a cube of size $\ell \pm 4 \ell \eta $
centered at $\alpha$.
Let
$$
\chi_\alpha (x) = \chi( (x-\alpha)/\ell )
$$
be a smooth function supported in  $ \tau_{\alpha} \Lambda_{\ell+
2\ell \eta} \, \subset \,  \alpha^+$ and $\chi_\alpha (x) = 1 $
in $\tau_{\alpha} \Lambda_{\ell- 2\ell \eta} \supset \alpha^-$. We
collect these relations in the following:
\begin{align}\label{relation}
&   \alpha^- \, = \, \tau_{\alpha} \Lambda_{\ell- 4\ell \eta}
 \, \subset \,    \tau_{\alpha} \Lambda_{\ell- 2\ell \eta}
\, \subset \,  \{x:  \chi_\alpha (x) = 1  \} \nonumber \\
 & \, \subset \,  \{x:  \chi_\alpha (x) \not= 0  \} \, \subset \,
\tau_{\alpha} \Lambda_{\ell+ 2\ell \eta} \, \subset \,
\tau_{\alpha} \Lambda_{\ell+ 4\ell \eta} \,= \,  \alpha^+
\end{align}

There is a wide range of choices for $\eta$. The main restrictions
we needed are
$$
\eta \ell \to \infty, \qquad \eta \to 0\; .
$$
We shall choose, for simplicity of notation
$$
\eta = \ell^{-1/2}
$$
for the rest of this paper.

Recall the configuration space ${\cal S} (\Lambda)$ is the space
$$
{\cal S} (\Lambda) = \{ \vec x : = (x_1, \cdots, x_n): n \in \{0\} \cup \NN,
x_j \in \Lambda \text{ for all } j\}
$$
Denote by $\Gamma (\Lambda)$ the space of antisymmetric
functions from the
configuration space ${\cal S} (\Lambda)$ to the complex numbers.
With the standard $L^2$ inner product,
$\Gamma (\Lambda)$ is a Hilbert space.

Define ${
I^{\Lambda}_\alpha}$ from $\Gamma (\alpha^+)$ to $\Gamma
(\Lambda)$ by
$$
(  I^{\Lambda}_\alpha \psi ) ( x_1, \cdots,
x_n) = \bigg [  \;  \prod_j \chi_{\alpha} (x_j) \; \bigg ]   \psi
(x_1, \cdots, x_n)
$$
Usually we shall take $\Lambda= \Lambda_{\e^{-1}}$
and  omit the labels $\Lambda$ and $\alpha$ whenever they are
obvious or unimportant.
It is crucial that  $\bigg [  \,  \prod_j    \chi_{\alpha_j} (x_j) \,
\bigg ]$ is symmetric w.r.t. permutations of ${\vec x}$ so that
$ I^{\Lambda}_\alpha \psi$ is antisymmetric as a function of
${\vec x} $.
Define $I^\ast$ to be the adjoint of $I$, i.e., we have
$ ({ I }^\ast f, \; g) = (f,  \; { I } g)$.

Let $X$ be an observable $X$ on $\alpha$ defined by
$$
X=\int_{\alpha^+}dx_1\cdots dx_k dy_1\cdots dy_k\,
f(x_1,\ldots,x_k;y_1,\ldots,y_k)a^+_{x_1,\alpha}\cdots
a^+_{x_k,\alpha}a_{y_k,\alpha}\cdots a_{y_1,\alpha}\quad.
$$
where  $f$ is a distribution (kernel) with support in
$(\alpha^+)^{\times 2k}$. Here we have labelled the operators
by $\alpha$ to emphasized the cube $\alpha$.   We can check
the identity:
\bey\label{Idef}
I ^*X I  &= &
\int dx_1\cdots dx_k dy_1\cdots dy_k\,
\chi_\alpha(x_1)\cdots\chi_\alpha(x_k)\chi_\alpha(y_1)
\cdots \chi_\alpha(y_k) \nonumber \\
&&\quad\times f(x_1,\ldots,x_k;y_1,\ldots,y_k) a^+_{x_1}\cdots
a^+_{x_k}a_{y_k}\cdots a_{y_1}
\eey
as an operator on the torus $\Lambda_{\e^{-1}}$.

{From} this definition the pull-backs of all
observables we need, i.e., $I ^*X I$ for a conserved quantity or current given
by $X$, can easily be computed. By using the
appropriate distribution kernels $f$, observables involving
derivatives are included. E.g.,
\bey\label{4.1}
 I ^* \left[\int_{\alpha^+} dx f(x) \nabla a^+_x a_x\right] I
&=& I ^*\left[-\int dy \delta^\prime(y-x)\int_{\alpha^+} dx f(x)
a^+_y a_x\right] I \nonumber \\
&=& -\int dy \delta^\prime(y-x)\int dx f(x)
\chi_\alpha(x)\chi_\alpha(y)
a^+_y a_x\\
&=&\int dx f(x) \left[\chi_\alpha(x)^2 \nabla a^+_x a_x +
\chi_\alpha(x)\nabla\chi_\alpha(x) a^+_x a_x\right] \nonumber \eey
%
%
For the kinetic energy we have \bey\label{4.2}
 I ^*\left[\int_{\alpha^+} dx f(x) \nabla_j a^+_x \nabla_j a_x\right] I
&=& \int dx f(x)\bigl[\chi_\alpha(x)^2  \nabla_j a^+_x \nabla_j
a_x
+ (\nabla_j \chi_\alpha(x))^2 a^+_x a_x\\
&&\quad + \chi_\alpha(x)\nabla_j\chi_\alpha(x)(\nabla_j a^+_x a_x
+ a^+_x \nabla_j a_x) \bigr] \nonumber \eey If we take $f =
1_{\alpha^-}(x)$, we have $f(x)\chi_\alpha(x)= 0$. Together with
$\chi_\alpha (x) = 1 $  in $\tau_{\alpha} \Lambda_{\ell- 2\ell
\eta} \, \supset \,  \alpha^-$, we have \bey\label{4.4}
 I ^* \left[\int_{\alpha^-} dx  \nabla a^+_x a_x\right] I
&=& \int_{\alpha^-} dx\left[ \nabla a^+_x a_x\right] \\
 I ^*\left[\int_{\alpha^-} dx  \nabla_j a^+_x \nabla_j a_x\right] I
&=& \int_{\alpha^-} dx  \nabla_j a^+_x \nabla_j a_x  \nonumber
\eey


\subsection{Commuting local conserved quantities}

Let $H_{\alpha^+}, P_{\alpha^+}$ be
the total energy and momentum operators on $\alpha^{+}$ with {\it
periodic boundary
condition}. 
Then $H_{\alpha^{ + }}$ and $P_{\alpha^{ + }}$ commute with each
other and also with the number operator $N_{\alpha^{ + }}$. Denote
$$
\bua
= \ell^{-3} ( P_{\alpha^{ + }}, N_{\alpha^{ + }}, H_{\alpha^{ + }})
$$
Since the image of ${ I }$ is in the domain of $H_{\alpha^{+ }} $
and $P_{\alpha^{ + }}$, the operator ${I }^* \bua I $
is well-defined.
Since the components of $\bua$ commute, the function
$$
{ I }^*  \hat X (   \bua )  { I }
$$
is now well-defined. We shall use the notation
\begin{equation}\label{4.8}
\bu_{x, \ell}^+ := \bua, \quad
n_{x, \ell}^+:=
\ell^{-3}  N_{\alpha^{+}}, \; h_{x, \ell}^+:=
\ell^{-3}  H_{\alpha^{ +}}
\end{equation}
when  $\alpha$ is centered at $x$. When $x= 0$, we shall omit
the subscript $x$.

\subsection{Local average of currents}

Let
$$
{ \bw }_{\alpha^\pm}= \ell^{-3} \int_{\alpha^\pm} dx { \bw
}_{x},\quad \bW_{\alpha^\pm}=  \int_{\alpha^\pm} dx { \bw }_{x}
$$
be the average over the cube $\alpha^{\pm}$ of the currents $\bw_x$,
where we have divided the integration by $\ell^{3}$ which is
approximately the volume to the cube $\alpha^\pm$. by definition ${
\bw }_{x}$ is an operator on the torus $\Lambda_{\e^{-1}}$. Since ${ a
}^\sharp_x $ can be viewed as an operator on $\alpha^+$ with
periodic boundary condition for $x \in \alpha^+$, ${ \bw
}_{\alpha^\pm}$ can be understood  as an operator on the cube
$\alpha^{+}$ as well. We shall use the same symbol in both
contexts.

Recall the cutoff version of the current ${ \bw }_{ x, M}= { \bw
}_{x}(a^+_M, a_M)$ \eqref{wm}. Thus we can define the cutoff
version of the current
\begin{equation}\label{Wm}
{\bw}_{ M, \alpha^\pm}= \ell^{-3} \int_{\alpha^\pm} dx { \bw
}_{M,x}
\end{equation}
Here ${\bw}_{ M, \alpha^\pm}$ can be viewed as an
operator either on the torus $\Lambda_{\e^{-1}}$ or $\alpha^+$ with
periodic boundary condition.

Recall that $a_{x, M}^{\sharp}$ are bounded operators {\it
localized in a ball of radius  $e^{M^2}$ centered at $x$}. Thus
the support of $a_{x, M}^{\sharp}$ is contained in $\tau_\alpha
\Lambda_{\ell- 2\sqrt \ell }$ for $ x \in \alpha^- = \tau_\alpha
\Lambda_{\ell- 4\sqrt \ell } $, as long as $\sqrt \ell >
e^{M^2}$, which we shall assume from now on. Since
$\chi_\alpha(x)=1$ for $x\in \tau_\alpha \Lambda_{\ell- 2\sqrt
\ell }$, following the proof of \eqref{4.4} we have the identity

\begin{equation}\label{wid}
I^*  \bwma I =  \bwma ;
\end{equation}
here  $\bwma$ is understood as an operator on the torus
$\Lambda_{\e^{-1}}$ on the right side and as an operator on $\alpha^+$
on the left side.
Define the notation
$$
{ \bw }_{M, x, \ell}^- = \ell^{-3} \int_{\Lambda_{x, \ell}^- } dy\,
 { \bw
}_{M, y}
$$
where $\Lambda_{x, \ell}^- = \tau_x \Lambda_{\ell- 4 \sqrt \ell}
= \tau_x \Lambda_{\ell}^-$.


From \eqref{wid}, the boundedness of $a_{M, x}^\sharp$ and
simple  counting of the number of terms, we have the following
lemma.

\begin{lemma}\label{Ilemma}
For any state $\gamma$%
, and any smooth function $J$, we have \be
 \Tr \gamma \langle
J(\eps\cdot),\bw_M\rangle
=
 \Tr \gamma   \Av_x  J_\e ( x)\,
I^* { \bw }_{M, x, \ell}^- I +\Omega_M(\eps,\ell)
\label{Iestimate} \ee where the
error term $\Omega_M(\eps,\ell)$ vanishes in the sense given by
\eqref{vancon} and
$$
\Av_x= \e^3 \int_{\Lambda_{\e^{-1}}}  dx
$$
\end{lemma}

Applying this Lemma to a smooth function $J(\e t, \e x)$ and
average over $t$, we have
\be  \Av_{t \le T/\e}\Tr \gamma_t  \langle J(\eps\cdot, \e
t),\bw_M\rangle =\Av_{t \le T/\e} \gamma_t   \lan \,
J_\e(t, \cdot)\,, \,  I^* { \bw }_{M, \cdot, \ell}^- \  I  \,
\ran +\Omega_M(\eps,\ell)
\eqlabel{Iestimate-t} .
\ee

\section{Bounds on the currents}

The aim of this section is to show that the currents with momentum cutoff, i.e., the quantities  ${\vec W}_{M,\alpha^\pm}$, can be bounded by a multiple of the Hamiltonian plus particle number. This is the content of the following lemma.

\begin{lemma} \thlabel{keycutlemma}
The following operator inequalities hold:
\begin{equation} \eqlabel{keycutest}
{\vec W}_{M,\alpha^+} \le C  M\left [ \; H_{\alpha^+} +
N_{\alpha^+} \; \right ]
\end{equation}
Note that the dependence on $M$ in the right hand side is linear.
Similar inequality holds if ${\vec W}_{M,\alpha^+}$ on the left
side is replaced by ${\vec W}_{M,\alpha^-}$.
\end{lemma}

As it is essential for the proof of this lemma, we first recall a standard stability result based on the superstability conditions we have assumed on the potential $W$. Stated in words, the result says that a large class of two-body quantities can be bounded in terms of the two-body interaction and the particle number. We state this
result as a lemma for functions but it obviously extends to the corresponding second quantized observables.

\begin{lemma}\label{lem:stability}
{Suppose} $U$ {is a positive bounded function
with compact support on} $\RR^3$
and $W$ is a superstable potential stated in the
sense of \eqref{superstable} .   {Then there is a} $\delta > 0$
{such that}
$$
\delta \sum_{\alpha\ne \beta} ^N U(x_\alpha - x_\beta)\le
\sum_{\alpha\ne \beta}^N W(x_\alpha - x_\beta) + N \ .
$$
\end{lemma}

The proof of Lemma \ref{lem:stability}  is contained in Ruelle's book \cite{Rue} or see \cite{OVY}.

\begin{proofof}{\bf Lemma \ref{keycutlemma}:} First, we treat the one-particle (i.e., quadratic in the $a^\#_x$'s) terms and show that they can be bounded by the kinetic
energy term and chemical potential term of the Hamiltonian.

We will use the following inequalities several times without
further reference: for any pair of bounded  operators $A$ and $B$,
and $c>0$, one has
$$
A^*B+B^*A \leq c A^*A + c^{-1}B^*B
$$
and
$$
A+B \leq \vert A\vert +\vert B\vert\quad.
$$
From the last inequality it follows that for a bounded family of
self-adjoint operators $A_x$, and a real valued $L^1$-function $f$,
one has
$$
\int f(x) A_x dx\leq \int \vert f(x)\vert   \vert A_x\vert
dx\quad.
$$
Note that one cannot replace the LHS of these inequalities by
their absolute values unless all terms commute.

We start with bounding the momentum components of $\bW_M$: \bey
W^0_{k,M,\alpha^+}&=&\frac{i}{2}\int_{\alpha^+}dx\,
\left[\nabla_k a^+_{x,M} a_{x,M}- a^+_{x,M} \nabla_k a_{x,M}\right]\nonumber\\
&\leq&\int_{\alpha^+}dx\, \nabla_k a^+_{x,M} \nabla_k a_{x,M}
+a^+_{x,M} a_{x,M} \label{startboundW0}\eey We would like to
obtain bounds by multiples the Hamiltonian and the number operator
{\em without} momentum cut-off $M$. For this we use the following
inequalities:
\bey
a^+_{x,M} a_{x,M} &\leq& C\vert\phi_M\vert * a_x^+ a_x\label{removeM1}\\
\nabla  a_{x,M}^+   \nabla a_{x, M}&\le& C M  |\nabla  \phi _M|*
a_x^+ a_x
\label{removeM2}\\
\nabla  a_{x,M}^+  \nabla  a_{x, M}&\le& | \phi _M|* \nabla a_x^+
\nabla a_x
\label{removeM3}\\
\Delta   a_{x,M}^+  \Delta  a_{x, M}&\le& C M |\nabla  \phi _M|*
\nabla a_x^+  \nabla a_x\label{removeM4}
\eey
In the above
expressions the convolutations are with respect to the variable
$x$ in the RHS. The four inequalities are proved in almost
identical fashion. E.g., the first inequality is obtained as
follows:
\beyn
a_{x,M}^+   a_{x, M}&=&  \int dz dw   \phi_M(x-z) a_z^+ \phi_M(x-w) a_w\\
&\le& \frac 12 \int dz \int dw  | \phi_M(x-z)|
| \phi_M(x-w)|\, [ a_z^+ a_z + a_w^+ a_w]\\
&\le& | \phi_M|* a_x^+ a_x \eeyn where we have used that $\int
\vert\phi_M\vert =1$. For \eqref{removeM2} and \eqref{removeM4} one
also has to use $\Vert \nabla \phi_M\Vert_1\leq C M$, for a
suitable constant $C$, but otherwise the proofs are the same. Now,
we can finish the bound of $W^0_{k,M\alpha}$, by using
\eqref{removeM1} and \eqref{removeM3} in \eqref{startboundW0}. We obtain
\beyn W^0_{k,M,\alpha^+} &\leq&\int_{\alpha^+}dx\, \vert
\phi_M\vert(x-y) \left[
\nabla_k a^+_y \nabla_k a_y + a^+_y a_y\right]\\
&\leq& C (H_{0,\alpha}+N_\alpha) \eeyn By stability of the
potential the kinetic energy term in this bound can be replaced by
the full Hamiltonian, up to an adjustment to the constant $C$.
This completes the proof of the lemma for $W^0_{k,M\alpha}$.

For the other one-particle terms of $\bW_{M,\alpha}$ one proceeds
in the same way. E.g., for the last inequality one starts from
$$
- i\left[\nabla_k a^+_{x,M} \Delta a_{x,M}- \Delta a^+_{x,M}
\nabla_k a_{x,M} \right]\leq M\nabla_k a^+_{x,M} \nabla_k a_{x,M}
+M^{-1} \Delta a^+_{x,M} \Delta a_{x,M}
$$
The rest of the argument is the same. In summary, the results are
\beyn \frac{i}{2}\int_{\alpha^+}dx\, \left[\nabla_k a^+_{x,M}
a_{x,M}- a^+_{x,M} \nabla_k a_{x,M}\right]
&\leq& C(H_{\alpha^+}+N_{\alpha^+})\\
\int_{\alpha^+}dx\,\nabla_j a^+_{x,M}\nabla_k a_{x,M}
&\leq& C(H_{\alpha^+}+N_{\alpha^+})\\
- \frac{i}{4} \int_{\alpha^+}dx\, \left[\nabla_k a^+_{x,M} \Delta
a_{x,M}- \Delta a^+_{x,M} \nabla_k a_{x,M} \right] &\leq& C M
(H_{\alpha^+}+N_{\alpha^+}) \eeyn

The two-particle terms (quartic in the $a_x^\#$'s) appearing in
\eqref{thetaj} and \eqref{theta4}, we can follow the same procedure.
E.g., to bound the middle term of \eqref{theta4}, we start from \bey
&&\frac{i}{4}\left[ \nabla_k  a_{x,M}^+  a_{y, M}^+ a_{y, M} a_{x,
M}
- a_{x,M}^+  a_{y, M}^+ a_{y, M} \nabla_k a_{x, M} \right]\nonumber\\
&&\leq  M a_{x,M}^+  a_{y, M}^+ a_{y, M} a_{x, M} + M^{-1} \nabla
a_{x,M}^+  a_{y, M}^+ a_{y, M} \nabla a_{x, M}
\label{quarticSchwarz}\eey The first term can further be bounded
in a way similar to \eqref{removeM1}: \beyn a_{x,M}^+  a_{y, M}^+
a_{y, M} a_{x, M}
&\leq&\int dz \vert\phi_M(z-y)\vert a_{x,M}^+  a_y^+ a_y a_{x, M}\\
&=& \int du \phi_M(u-y)a_y^+ a_{x,M}^+ a_{x, M}  a_y \\
&\leq&\int du\, dv \vert\phi_m\vert(u-y) \vert\phi_m\vert(v-x)
a^+_x a^+_y a_y a_x \quad . \eeyn The same quantity appears in
\eqref{thetaj}. In both cases, after integration, we get something
of the form:
$$
M\int_{\alpha^+}dx\int_{\alpha^+}dy [\int du\, dv
\vert\phi_M\vert(u-x)G(u-v)\vert\phi_M\vert(v-y)]a^+_x a^+_y a_y
a_x  \quad,
$$
where $G$ is a non-negative function of compact support.
To estimate this term, we use Lemma \ref{lem:stability} to obtain
a bound of the last expression of the form $CM$ times
the potential energy.

The second term in the RHS of \eqref{quarticSchwarz} gives rise to
$$
M^{-1}\int_{\alpha^+}dx\int_{\alpha^+}dy M[\int du\, dv
\vert\nabla\phi_M\vert(u-x)G(u-v)\vert\phi_M\vert(v-y)] a^+_x
a^+_y a_y a_x
$$
and something of the same form for the third term in \eqref{theta4},
which, with another application of Lemma \ref{lem:stability},
can also be bounded by the $CM$ times the
potential energy. This completes the proof of the lemma.
\end{proofof}

\section{Local ergodicity}\label{sec:local_ergo}

Recall that by assumption the solution up to time $t \le T_0/\e$
of the Euler equations has density and energy taking values in a
compact set strictly contained in the one phase region of the
phase diagram of the fermion systems. Let  $\sigma^{\kappa}$ be a
smooth function supported in the one phase region such that
$\sigma^{\kappa}=1 $ on this compact set. Furthermore, we require
that as $\kappa \to 0$, $\sigma^{\kappa} $ becomes the
characteristic function of a compact neighborhood of this set
contained in the one phase region. Since the phase transition
region depends only on the density and energy, ${\sigma^\kappa}$
needs to depend only on the density and energy. We will take
$\sigma^\kappa(e,n)$ of the form $\sigma_1^\kappa(e)
\sigma_2^\kappa(\rho)$ where  $\sigma_1^\kappa$ and
$\sigma_2^\kappa$ are some smoothed versions of the characteristic
functions on a set of sufficiently high $e$ and sufficiently low
$\rho$, respectively.

The aim of this section is to prove the following theorem. Recall
$\hat X$ is defined in \eqref{Xhat}.

\begin{theorem}\label{local-ergodicity}
For all smooth functions $J$, and $X$ any one of the components of
$\ubw_M$ we have
\bea &&
\Av_{t \le T/\e} \gamma_t   \Av_x
J(\e t, \e  x)\, I^* { X }_{x, \ell}^- I
\nonumber\\
&\le& \Av_{t \le T/\e} \Tr \;  \gamma_t
\Av_x J(\e t, \e  x) \; \big \{
I^* \big (\, \tilde \sigma^\kappa \hat X \tilde
\sigma^\kappa) (\bu_{x, \ell}^+) I \; +\; I^*
(1-\sigma^\kappa ( \bu_{x, \ell}^+))   { X }_{x, \ell}^-
(1-\sigma^\kappa (\bu_{x, \ell}^+))
I \, \big \}
 \nonumber\\
&&+ \Omega_{J,\kappa,X}(\e, \ell, a)
\label{corXtohatX}\eea where
$\tilde \sigma^\kappa=\sqrt{\sigma^\kappa(2-\sigma^\kappa)}$.
\end{theorem}

The function $\tilde
\sigma^\kappa$ behaves  essentially the same way as
$\sigma^\kappa$, i.e., it is a smooth version of a characteristic
function  supported in the one-phase region.

\bigskip

As a first step towards the proof of Theorem \ref{local-ergodicity}, we
partition $\Lambda_{\eps^{-1}}$ into cubes of
size $a\eps^{-1}$, where $a$ is a sufficiently small
positive constant. 
For any $z\in\Lambda_{\eps^{-1}}$, let $Q= \Lambda_{z, a \e^{-1}}$ denote
the cube of size $ a \e^{-1}$ centered at $z$.
For any bounded
quasi-local observable $Z$, define the average of $Z$ in the cube
$Q$ by
$$
Z_Q=  \Av_{y \in Q } \tau_y Z
$$
We also divide the time interval $[0,\eps^{-1}T]$, into disjoint
intervals of size $ 2 a \e^{-1}$ and label the centers by $t_1,
\cdots t_n$, $n=T/(2a)$ (the $n$-th interval is $[t_n-a,t_n+a]\cap
[0,\eps^{-1}T]$).

Since $J $ is a smooth function,
\begin{equation}\label{6.1}
 \lan {J_\e}(t, \cdot),Z\ran = n^{-1}
\sum_{j=1}^n \Av_z \left [ \,  {J }(\e t_j, \e z ) \left \{ \,
\Av_{ |t - t_j|\le a/\e} \Tr \; \gamma_t  \;
 Z_{\Lambda_{z, a \e^{-1}}} \, \right \} \right ]  + \Omega_{Z,J}(a, \e)
\end{equation}
where $\lim_{a \to 0}\lim_{\e \to 0} \Omega_{Z,J}(a, \e)  = 0$
and the average is over $z \in a\e^{-1}\ZZ^3 \cap \Lambda_{\eps^{-1}} $.

For $Q, a, j$ fixed, define a family of states labelled by $\e$
consisting of the states defined by
$$
\gamma_\e^{Q, j}(Z) = \Av_{ |t - t_j|\le a/\e} \Tr \, \gamma_t  \,
Z_Q
$$
Then $\{\gamma_\e^{Q, j}\mid \eps >0\}$  is w$^*$-precompact
and, hence, has at least one limit point.

\begin{lemma}\label{lem:bounded-entropy}
Let $\overline{  \omega}$ be the Gibbs state on $\Lambda_{\e^{-1}}$
defined in \eqref{gibbs} with $\Lambda=\Lambda_{\e^{-1}}$
and the chemical potential $\vec \lambda : = \bar {\vec \lambda}$
is chosen to be
$$
 \bar {\vec \lambda} = Av_x {\vec  \lambda} (0, \e x)
$$
where ${\vec \lambda}(0,\cdot)$ are the parameters for the initial
condition  defined in \eqref{2.27}.
Then for any $t\ge 0$ the relative entropy
$$
s(\gamma_t \mid\bar{\omega}) \le C
$$
for some constant $C$ depending only on the initial value
${\vec  \lambda} (0, \cdot)$.
\end{lemma}

\begin{proof}
Recall the initial state is \be
 \omega^\e_0  = \frac 1 { c_\e (0)}\exp  \big[\,
\e^{-3} \lan {\vec \lambda}_\e (0,  \cdot ) \,  , \,  \bu \ran \big]
\eqlabel{2.27.1}
\end{equation}
Then we have
\be
s(\omega^\e_0| \bar \omega )  =  \int d x
\omega^\e_0 \bigg ( \lan {\vec \lambda}_\e (0, \cdot ) \,
, \,  \bu \ran
- \lan \bar {\vec \lambda}\,  , \,  \bu \ran \bigg )
+ \e^3 \log c_\e (0) -  \e^3 \log Z_{\Lambda_{\e^{-1}}}
\eqlabel{2-1}
\end{equation}
Since each term on the right side is bounded, we have
$s(\omega^\e_0| \bar \omega ) \le C$.
From a simple direct calculation, we know that
$s(\gamma^\e_t| \bar \omega )$
is a constant of motion. This proves the Lemma.
\end{proof}

%
%
%

\begin{lemma} \label{lem:entropy-eta}
Fix the parameter  $a$ and let $\eta$ be any limit point of
$\{\gamma_\e^{Q, j}\mid \eps
>0\}$. Then $\eta$ is a translation invariant, time invariant
state of the dynamics. Furthermore, 
the specific relative
entropy of $\eta$ with respect to the translation invariant state
$\omega_{\overline{\vec \lambda}}$, satisfies the bound
$$
s(\eta | \omega_{\overline{\vec \lambda}}) \le C_{\vec \lambda}
a^{-3}
$$
\end{lemma}

\begin{proof}
The invariance under space and time translations is an immediate
consequence of the scaling by $\eps^{-1}$. Since the proof
for quantum case is parallel to that of the
classical case, we refer the reader to \cite{OVY} for a
proof of the classical
case.
To show that the
specific relative entropy with respect to $\omega_{\overline{\vec
\lambda}}$ is finite, we start form Lemma
\ref{lem:bounded-entropy} stating that
the relative entropy
$
\eps^3 S(\gamma^\e _t\mid \omega^\e _{\overline{\vec
\lambda}})\leq C $
for a suitable constant $C$.

The operations of averaging over
translations in a cube $Q$ and over times in an interval
$[t_i-a/\eps,t_i+a/\eps]$, are completely positive, therefore, by
the monotonicity (or convexity) of the relative entropy (see,
e.g., \cite{OP}), we have
$$
\eps^3 S(\Av_{\vert t-t_j\leq a/\eps}\Av_{y\in
Q}\gamma^\e _t\circ\tau_y \mid
\omega^\e _{\overline{\vec \lambda}})\leq C
$$
The relative entropy is also monotone with respect to restriction
to the algebra of observables of a subvolume. Therefore we have
$$
\eps^d S(\Av_{\vert t-t_j\leq a/\eps}\Av_{y\in
Q}\gamma^\e _t\circ\tau_y \big \vert_{Q} \;  \bigg \vert \;
\omega^\e _{\overline{\vec \lambda}} \big \vert_{Q})\leq C
$$
Now, $\eta$ is a limiting point of $\{\gamma^{Q,j}_\eps\mid \eps
>0\}$, where
$$
\gamma^{Q,j}_\eps=\Av_{\vert t-t_j|\leq a/\eps}\Av_{y\in Q}
\gamma^\e _t\circ\tau_y\Big\vert_{Q}\quad .
$$
Therefore, by the lower semicontinuity of the specific relative
entropy, we can conclude \beyn s(\eta \mid \omega_{\overline{\vec
\lambda}}) &=&\lim_{\eps\to 0} \frac{1}{(2a \e^{-1})^3}
S(\gamma^{Q,j}_\eps\mid \omega^\e _{\overline{\vec \lambda}}
\Big\vert_{Q})\\
&\le& (2a)^{-3}\limsup_\eps \eps^3 S(\gamma^{Q,j}_\eps\mid
\omega^\e _{\overline{\vec \lambda}}) \leq C(2a)^{-3} \eeyn
\end{proof}

Consider any limiting point $\eta$ of $\{\gamma^{Q,j}_\eps\mid
\eps >0\}$. Since $\eta$ is translation invariant, we can
decompose it into ergodic components (with respect to space translations)
and there is a probability
measure $\mu$ supported on ergodic states $\omega$  such that
$$
\eta = \int  \omega \;  \mu(d {\omega})\quad.
$$
The key property of $\eta$ is the following lemma.

\begin{lemma}\label{prop:X2Xhat}
Let $\eta$ be as above, and $X\in\mathcal{A}_{\Lambda_0}$. Then
there is $\Omega_M(\ell,\kappa)$ such that
\be \left\vert\eta( I^\ast
X_\ell^- I)
-\eta\bigg [ \,
I^* \big (\, \tilde \sigma^\kappa \hat X \tilde
\sigma^\kappa) (\bu_{\ell}^+) I \; +\; I^*
(1-\sigma^\kappa ( \bu_{\ell}^+))   { X }_{\ell}^-
(1-\sigma^\kappa (\bu_{\ell}^+))
I \, \big \} \, \bigg ]
\right\vert \leq
\Omega_M(\ell,\kappa) \label{X2Xhat}\ee
where
$\tilde \sigma^\kappa=\sqrt{\sigma^\kappa(2-\sigma^\kappa)}$.
\end{lemma}

\noindent
{\bf Proof of Theorem \ref{local-ergodicity} assuming Lemma \ref{prop:X2Xhat}.}
\newline
Let
$$
Z = I^\ast
X_\ell^- I
- \bigg [ \,
I^* \big (\, \tilde \sigma^\kappa \hat X \tilde
\sigma^\kappa) (\bu_{\ell}^+) I \; +\; I^*
(1-\sigma^\kappa ( \bu_{\ell}^+))   { X }_{\ell}^-
(1-\sigma^\kappa (\bu_{\ell}^+))
I \, \big \} \, \bigg ]
$$
It is crucial that {\it $Z$ is a bounded and local
observable}.

Theorem \ref{local-ergodicity} now follows immediately
from  \eqref{6.1} and Lemma \ref{prop:X2Xhat}.

The rest of this section is devoted to prove
Lemma \ref{prop:X2Xhat}.
We shall drop the labels $\pm$ on $X$ and $\bu$ etc for the rest of
this section.

\subsection{General Properties of limiting states}

We now prove a number of results for the ergodic components of the
limit points $\eta$. At this point $\eta$ depends on a macroscopic
space point $z$, and a macroscopic time $t_j$, and in principle
also on the subsequence, but we will eventually see that $\eta$ is
in fact independent of the subsequence.

\begin{lemma}\label{lem:gamma_bound}
Let $\gamma_n, \gamma$ be normal states on a von Neuman algebra
$\mathcal{A}$, and $\gamma_n\to\gamma$ weakly. Suppose that $A$ is
a non-negative self-adjoint operator affiliated with
$\mathcal{A}$, such that $\gamma_n(A)$ is bounded by a constant
$M$, uniformly in $n$. Then, $\lim_{n\to\infty} \gamma_n(A)$
exists and satisfies
$$
\gamma(A)\leq \lim_{n\to\infty}\gamma_n(A)
$$
\end{lemma}
\begin{proof}
Let $A=\int\lambda dE_\lambda$ be the spectral resolution of $A$.
As $A$ is affiliated with $\mathcal{A}$, the projections
$P_k=\int_0^k dE_\lambda$ belong to $\mathcal{A}$, and
$\gamma_n(A)=\sup_k\gamma(A_k)$, where $A_k= AP_k$. The supremum
is finite by the assumptions. Therefore,
$$
\lim_n\gamma_n(A)=\lim_n\sup_k\gamma_n(A_k) \geq
\sup_k\lim_n\gamma_n(A_k) =\gamma(A)
$$
\end{proof}

Recall that $h_\ell$ and $n_\ell$ (we omit the superscripts $+$)
are the average of the local conservative quantities
$H_\ell$ and $N_\ell$ \eqref{4.8}.
Let $e$ and $\rho$ be determined by \be
e=\lim_{\ell\to\infty}\omega(I^*h_\ell I), \quad \text{and}\quad
\rho=\lim_{\ell\to\infty}\omega(I^*n_\ell I) \label{mean_values}
\ee Where necessary, we will indicate the dependence on $\omega$
by $e(\omega)$, and $\rho(\omega)$. In the following lemma we
prove the existence and finiteness of these limits when the parameter
$a$ is fixed.

\begin{lemma}\label{lem:e_and_rho_finite}
For $\mu-$almost all states $\omega$, the limits $e$ and $\rho$
of \eqref{mean_values} are finite. For the state $\eta = \int
\omega \; \mu(d {\omega})$, we have the following bounds \be
\limsup_{\ell\to\infty} \eta(I^* h_\ell I) \leq C a^{-3}e_0, \quad
\limsup_{\ell\to\infty} \eta(I^* n_\ell I) \leq C a^{-3}\rho_0,
\label{e-n_bounds}\ee
\end{lemma}

\begin{proof}
We have $\Av_{\text{all
boxes}}\Av_Q\overline{\gamma^{Q,j}_\eps}(h_\ell) =e_0$, where
$e_0$ is the initial total energy. This is a direct consequence of
the fact that the energy is conserved by the dynamics. Therefore,
for each box $Q$, we have
$$
\Av_Q\overline{\gamma^{Q,j}_\eps}(h_\ell) \leq C a^{-3}e_0\quad.
$$
By Lemma \ref{lem:gamma_bound} with $A= I^*h_\ell I$, it follows
that
$$
\eta( I ^* h_\ell  I )\leq \lim_{\e \to 0}
\Av_Q\overline{\gamma^{Q,j}_\eps}(h_\ell)\leq \sum_Q
\Av_Q\overline{\gamma^{Q,j}_\eps}(h_\ell)\leq Ca^{-3}e_0    \quad
,
$$
which implies the bound for the energy \eqref{e-n_bounds}. The proof
for the particle density is the same. As $\eta = \int  \omega\,
\mu(d {\omega})$, it then also follows that $e(\omega)$ and
$\rho(\omega)$ are finite for $\mu-$ almost all $\omega$.
\end{proof}

Let $A$ be a bounded observable in the local algebra
$\mathcal{A}_{\Lambda_0}\, \subset \, \mathcal{A}_{\Rl^3}$. E.g.,
$A=\int dx\,dy f(x,y) a^+_x a_y$, where $f(x,y)=0$ unless
$x,y\in\Lambda_0$. For concreteness, we assume that $\Lambda_0$
contains the origin.
We will also use the notation
$$
\Av_\Lambda(A)=\frac{1}{\vert\Lambda\vert}\int_{\Lambda}dx\,
\tau_x(A)
$$

\begin{lemma}\label{lem:ergo_I}
Suppose $\lim_\ell\eta_\ell=0$. For every translation invariant
ergodic state $\omega$ on $\mathcal{A}_{\Rl^3}$, any bounded local
observable $A$ and any continuous function $f$, we have the limit
\be \lim_l \omega(I^*_\ell f(\Av_{\Lambda_\ell}A) I
_\ell)=f(\omega(A)) \label{ergo_I}\ee
\end{lemma}

\begin{proof}
The proof rests on the following property of $ I $: for
$A\in\mathcal{A}_{\Lambda_0}$, we have \be
 I ^*\tau_x(A)I=\tau_x(A),\quad\mbox{if } \tau_x(\Lambda_0)
\, \subset \, \Lambda_\ell^- \label{locality_I}\ee Denote by
$\Lambda_\ell^{\rm int}=\{x\in\Lambda_\ell\mid \tau_x(\Lambda_0)\,
\subset \, \Lambda_\ell^- \}$. Note that \be
\frac{\vert\Lambda_\ell\setminus\Lambda_\ell^{\rm int}\vert}{
\vert\Lambda_\ell\vert} \leq 2\eta_\ell+\frac{{\rm diam}\,
\Lambda_0}{l}=:\delta_\ell \label{delta_l}\ee and that
$\lim_\ell\delta_\ell=0$.

First, consider the function $f(x)=x$. Then, using the property
\eqref{locality_I}, we have
$$
\omega(I_\ell^*\Av_{\Lambda_\ell} (A) I)
=\omega(\frac{1}{\vert\Lambda_\ell\vert}\int_{\Lambda_\ell^{\rm
int}}\tau_x(A)) +\omega(I^*\left[\frac{1}{\vert\Lambda_\ell\vert}
\int_{\Lambda_\ell\setminus\Lambda_\ell^{\rm int}}\tau_x(A)\right]
I)
$$
Without loss of generality we may assume $\omega(A)=0$. Using the
definition of $\delta_\ell$ \eqref{delta_l}, and the isometry
property of $I$, we find
$$
\left\vert\omega(I^*\Av_{\Lambda_\ell} (A) I)\right\vert \leq
(1-\delta_\ell)\left\vert\frac{1}{\vert\Lambda_\ell^{\rm
int}\vert} \int_{\Lambda_\ell^{\rm
int}}dx\,\omega(\tau_x(A))\right\vert +\delta_\ell\Vert A\Vert
$$
The two terms in the RHS tend zero, the first due to the
ergodicity of $\omega$, the second because $\delta_\ell\to 0$.

Next, we prove by induction the result for $f(x)=x^n$, for all
$n\geq 1$. Suppose we have the result for $f(x)=x^{n-1}$, i.e.,
$$
\lim_\ell \omega(I^* (\Av_{\Lambda_\ell} A)^{n-1} I)=0
$$
Then, by the same arguments as above, we have the estimate
$$
\lim_\ell \left\vert\omega(I^*(\Av_{\Lambda_\ell} (A))^n
I)\right\vert \leq
(1-\delta_\ell)^n\left\vert\omega((\Av_{\Lambda_\ell^{\rm int}}
(A))^n) \right\vert +C(1-(1-\delta_\ell)^n)
$$
and the result follows by the ergodicity of $\omega$.
For arbitrary continuous functions $f$, \eqref{ergo_I} can now be
obtained by approximating $f$ by polynomials, uniformly on
$[-\Vert A\Vert,\Vert A\Vert]$. This proves the Lemma.
\end{proof}

Lemma \ref{lem:ergo_I} can trivially be extended as follows:

\begin{corollary}\label{cor:ergo_I}
For any bounded local observables
$X,Y,A\in\mathcal{A}_{\Lambda_0}$, and continuous functions $f$
and $g$, we have that
$$
\lim_\ell\left[\omega(I^* Xf(\Av_{\Lambda_\ell} A
)Yg(\Av_{\Lambda_\ell} A )I) -f(\omega(A))g(\omega(A))\omega(I^*
XY I)\right]=0
$$
\end{corollary}

\subsection{Extension to unbounded conserved quantities}

Lemmas \ref{lem:ergo_I} and Corollary \ref{cor:ergo_I} are general
properties of ergodic states applied to bounded observables. We
now show how the one-phase region cut-off functions, which depend
on unbounded but conserved quantities, can be included. This is a
difficult step and we will have to use the special forms of the
conserved quantities. The key technical estimate is contained in
Lemma \ref{ergodic3}. We remark that a naive application of Schwarz'
inequality to prove Lemma \ref{ergodic2}, would produce
expressions with six or more creation or annihilation operators about
which we have no control.

\begin{lemma} \label{ergodic2}
Let $\eta$ be any limiting point of $\{\gamma^{Q,j}_\eps \mid \eps
>0\}$, let $X$ be one of the components of $\ubw$, and let
$X_\ell$ the averaged version of $X$. Then the following limits
vanish: \bea &&\lim_{\ell \to \infty} \eta\left(I^* B_\ell X_\ell
[ \sigma^\kappa_1(h_\ell )\sigma_2^\kappa( n_\ell
)-\sigma_1^\kappa(e) \sigma_2^\kappa(\rho) ] I \right) = 0
\label{for_ergo_Ia}\\
&&\lim_{\ell \to \infty} \eta\left(I^* B_\ell X_\ell [
\sigma_2^\kappa(n_\ell ) -\sigma_2^\kappa(\rho) ] I \right) = 0
\label{for_ergo_Ib} \eea for  $B_\ell = 1$ or
$$
B_\ell =  \sigma^\kappa_1(h_\ell )\sigma_2^\kappa(n_\ell)
$$
Here $e=e(\eta)=\int e(\omega)\,d\mu(\omega)$, and similarly for
$\rho$. In particular, we have \be \lim_{\ell \to \infty} \eta
\left( I^* \Big [ \, \sigma^\kappa (h_\ell, n_\ell) X_\ell
\sigma^\kappa (h_\ell, n_\ell) \, \Big ] -  \Big [ \,
\sigma^\kappa (e, \rho) X_\ell \sigma^\kappa (e,\rho) \, \Big ]I
\right) = 0 \label{for_ergo_I} \ee and the same result holds if
$\sigma^\kappa (h_\ell,n_\ell) X_\ell \sigma^\kappa (h_\ell,
n_\ell)$ is replaced by $\sigma^\kappa (h_\ell, n_\ell) X_\ell $
or by $X_\ell \sigma^\kappa (h_\ell, n_\ell)$.
\end{lemma}

\begin{proof}
We start with the case $B_\ell = 1$.

Recall, $\sigma^\kappa(e,\rho)=\sigma_1(e)\sigma_2\rho)$. There
exist bounded functions $\tilde\sigma_1^\kappa$ and
$\tilde\sigma_2^\kappa$ such that
$\sigma_i^\kappa(x)-\sigma_i^\kappa(y)=(x-y)\tilde\sigma_i^\kappa(x,
y)$, for $i=1,2$. Using these functions we can write
\beann
\sigma^\kappa_1(h_\ell )\sigma_2^\kappa(n_\ell
)-\sigma_1^\kappa(e) \sigma_2^\kappa(\rho) &=&
(\sigma_1^\kappa(h_\ell
)-\sigma_1^\kappa(e))\sigma_2^\kappa(n_\ell )
+\sigma_1^\kappa(e)(\sigma_2^\kappa(n_\ell )-\sigma_2^\kappa(\rho))\\
&=&\sigma^\kappa_2(n_\ell )\tilde\sigma_1^\kappa(h_\ell ,
e)(h_\ell -e)
+\sigma_1^\kappa(e)\tilde\sigma^\kappa_2(n_\ell,\rho)(n_\ell
-\rho)
\eeann
Therefore, for a suitable bounded function $f$, for
any ergodic state $\omega$, we can write \bey
\omega(I^*\sigma^\kappa (h_\ell, n_\ell) X_\ell \sigma^\kappa
(h_\ell,  n_\ell))
&=&\omega(I^* X\sigma^\kappa_2(N_\ell)f(h_\ell, e)(h_\ell-e)I)\nonumber\\
&=&\omega \Big (I X \sigma^\kappa_2(n_\ell)
f(h_\ell) [h^B_\ell - e_B ]I\Big )\nonumber\\
&&\quad+ \omega \Big (I^*X \sigma^\kappa_2(n_\ell) f (h_\ell)
(h_\ell-h_\ell^B) I\Big ) \label{subMsuperM}\eey where
$$
h^B_\ell=\frac{1}{\vert\Lambda_\ell\vert} H^B_{\Lambda_\ell}
$$
with $H^B_{\Lambda_\ell}=H_{\Lambda_\ell}\Ind(H_{\Lambda_\ell}\leq
B \vert\Lambda_\ell\vert)$, so that $\Vert
H^B_{\Lambda_\ell}\Vert\leq B\vert\Lambda_\ell\vert$, and
$H^B_{\Lambda_\ell}\uparrow H_{\Lambda_\ell} $ as $B\to\infty$.
Introduce
$$
e_B(\omega)= \lim_{\ell\to\infty}\frac{1}{\vert\Lambda_\ell\vert}
\omega(I^* H^B_{\Lambda_\ell}I)
$$
and use Schwarz' inequality to obtain \beyn &&\omega \Big (I^* X
\sigma^\kappa_2(n_\ell) f(h_\ell)
\left[h_\ell^B - e\right]I\Big )\\
&&\quad\leq\left\vert\omega \Big (I^* X \sigma^\kappa_2(n_\ell)
f(h_\ell) \left[h_\ell^B
 - e_B\right]I\Big )\right\vert+(e_B(\omega)-e(\omega)) \;
\omega(I^*\sigma^\kappa_2(n_\ell) f(h_\ell)I)\\
&&\quad\leq \delta \omega \Big (I^* X \,
(\sigma^\kappa_2(n_\ell))^2 \, f (n_\ell)^2 X^* I\Big
)+\delta^{-1} \omega \Big(I^*
\left[h_\ell^B - e_B \right]^2 I \Big )\\
&&\quad +(e_B(\omega)-e(\omega)) \;
\omega(I^*\sigma^\kappa_2(n_\ell) f(h_\ell)I) \eeyn

Now, we integrate over $\omega$ with respect to the measure $\mu$,
and take absolute values. The
first term is uniformly bounded in $\ell$. As $h^B_\ell$ is
bounded, the integrand of the second term vanishes for each
$\omega$, in the limit $\ell\to\infty$, by Lemma \ref{lem:ergo_I}.
As the integrand is bounded uniformly in $\omega$, the integral
vanishes as well. The third term we use the argument of
\eqref{lem:e_and_rho_finite} to show that it vanishes in the limit
$B\to\infty$:
$$
\limsup_B\left\vert \eta(h^B_\ell-h_\ell)\right\vert \leq
\limsup_B\lim_\eps\Av_Q\overline{\gamma^{Q,j}_\eps}(h^B_\ell-h_\ell)
\leq Ca^{-3} \lim_\eps\eps^{d}\vert\gamma_{\eps,t}(H^B-H)\vert
$$
The RHS is independent of $t$ and $\eps$, and vanishes as
$B\to\infty$.

For the second term of \eqref{subMsuperM}, we first apply Schwarz'
inequaltity:
\beyn
&&\left\vert\omega \Big (I^* X
\sigma^\kappa_2(n_\ell) f(h_\ell)
\left[h_\ell-h^B_\ell \right]I\Big )\right\vert\\
&&\quad\leq \delta\omega \Big (I^*X \sigma^\kappa_2(n_\ell) f
(h_\ell) \left[h^B_\ell-h_\ell\right] f(h_\ell)
\sigma^\kappa_2(n_\ell) X I\Big ) + \delta^{-1} \omega \Big
(I^*\left[h^B_\ell-h_\ell\right]I\Big)
\eeyn

As before, the last term vanishes in the limit $B\to\infty$. Since
$(h_\ell-h^B_\ell)\leq h_\ell$, the first term is bounded by
$$
\omega \Big (I^*X \sigma^\kappa_2(n_\ell) f (h_\ell) h_\ell
f(h_\ell) \sigma^\kappa_2(n_\ell) X I\Big )
$$
As $f$ is bounded, we have that $f(h_\ell) h_\ell f(h_\ell) \le C
h_\ell$. Therefore, after integration over $\omega$, and with the
use of Lemma \ref{ergodic3}, we obtain the bound
$$
\delta(C\eta(I^*X (h_l + n_\ell)I) + C )
$$
which can be shown to be bounded in terms of the corresponding
expectation in $\gamma_{t,\eps}$, as before.

In conclusion, as $B$ and $\delta$ are arbitrary, we have proved
\eqref{for_ergo_Ia} for $B_\ell=1$. It is straightforward to adapt
the argument to prove also \eqref{for_ergo_Ib} and the case
$B_\ell=\sigma^\kappa_1(h_\ell) \sigma^\kappa_2(n_\ell)$.
\end{proof}

\subsection{Basic Estimate}

In the previous proof the following lemma was used. It provides a
bound on the Hamiltonian sandwiched by bounded operators.

\begin{lemma} \label{ergodic3} For $\mu$-almost all translation invariant
ergodic states  $\omega$, and $X_\ell$ the averaged version of one
of the components of $\ubw_M$ (which are all self-adjoint), we
have \be \omega\left( I^* X_\ell \sigma^\kappa_2(n_\ell)h_\ell
\sigma^\kappa_2(n_\ell) X_\ell I\right) \le C
\omega\left(I^*[h_{\ell} +  n_\ell]I\right) + C \label{energyb}
\ee where the constant is independent of $\e, \ell$ but may depend
on $a, M$.
\end{lemma}

\begin{proof}
Since $X_\ell$ is particle number preserving, $X_\ell$ commutes
with $n_\ell$. Therefore, we can rewrite rewrite the quantity we
need to estimate as
$$
\omega\left(I^*X_\ell \sigma^\kappa_2(n_\ell)h_\ell
\sigma^\kappa_2(n_\ell) X_\ell I\right) ) = \omega\left( I^*
\sigma^\kappa_2(n_\ell) X_\ell h_\ell X_\ell
\sigma^\kappa_2(n_\ell) I\right)
$$
$h_\ell$ is the sum of two terms, a kinetic energy and a potential
energy term, which we wil treat separately.

First, we consider the kinetic energy term: $\int_{\Lambda_\ell}
\nabla a^+_x \nabla a_x$, defined with periodic boundary
conditions. We start from the identity
\be
X_\ell\nabla a^+_x
\nabla a_x X_\ell =\nabla a^+_x X_\ell  X_\ell \nabla a_x + \nabla
a^+_x X_\ell [\nabla a_x,  X_\ell] + [X_\ell,\nabla a^+_x] \nabla
a_x X_\ell
\label{ergo3.1}\ee
Note that $X_\ell$ is a linear
combination of linear and quadratic terms in $a_{u, M}^+a_{v, M}$
(see (\ref{u-def},\ref{theta0}-\ref{theta4})). Therefore,
commutators of the form $[a_{x, M}^+a_{y, M}, a^+_x]$, $[a_{x,
M}^+a_{y, M}, \nabla a^+_x]$, etc., are bounded operators. More
precisely, there is a constant $C_M$. such that
\be
\Vert [X_\ell,
a^+_x]\Vert\le C_M \ell^{-3},\quad\mbox{and } \Vert[X\ell, \nabla
a^+_x]\Vert\le C_M \ell^{-3}\quad .
\label{ergo3.2}\ee
These
bounds will be used repeatedly in the following estimates. E.g.,
applied to the first term of \eqref{ergo3.1}, they yield
$$
\nabla a^+_x X  X \nabla a_x \le C_M  \nabla a^+_x  \nabla a_x \;
.
$$
To bound the second and third term we first apply Schwarz'
inequality: \bean &&\omega\Big (I^*\sigma^\kappa_2(\fn_\ell)
[X_\ell, \nabla a^+_x] \nabla a_x X_\ell  \sigma^\kappa_2(n_\ell) I \Big )\\
&&\quad\le \delta \omega\Big ( I^*\sigma^\kappa_2(n_\ell)
 [X_\ell,   \nabla a^+_x][X_\ell,   \nabla a^+_x]^* \sigma^\kappa_2(n_\ell)I
\, \Big )+\delta^{-1} \omega\Big (I^* \sigma^\kappa_2(n_\ell)
X_\ell \nabla a^+_x  \nabla a_x X_\ell  \sigma^\kappa_2(n_\ell)
I\Big ) \eean The first term of the RHS is bounded and the last
term can be re-absorbed into the quantity we started out the
estimate. Thus, for the kinetic energy term and any of the
$X_\ell$, we have an estimate of the form
$$
\omega \Big (I^* \sigma^\kappa_2(n_\ell) X_\ell h_{0,\ell} X_\ell
\sigma^\kappa_2(n_\ell) I \Big ) \le C \omega \Big ( I^*
\sigma^\kappa_2(n_\ell) h_{0,\ell} \sigma^\kappa_2(n_\ell) I \Big
)+ C \quad .
$$

Similarly, for the potential energy we start from the identity
\bean
&&X_\ell a^+_x a^+_y a_y a_x X_\ell\\
&&\quad =a^+_x a^+_y X_\ell  X_\ell a_y a_x + a^+_x a^+_y X_\ell [a_y a_x,
X_\ell] + [X_\ell,a^+_x a^+_y] a_y a_x X_\ell
\eean
and the bound
$$
a^+_x a^+_y X_\ell  X_\ell a_y a_x \le C  a^+_x a^+_y  a_y a_x\; .
$$
For the commutator terms we have
$$
a^+_x a^+_y X_\ell [a_y a_x,  X_\ell] =a^+_x a^+_y X a_y[a_x,
X_\ell]- a^+_x a^+_y X [a_y,  X_\ell]a_x
$$
which can be estimated using Schwarz' inequality:
\bean
&&2\Re
\omega\Big (I^*\sigma^\kappa_2(n_\ell) a^+_x a^+_y X_\ell
a_y[a_x,X_\ell]
\sigma^\kappa_2(n_\ell)I\Big )\\
&&\quad\le \omega \Big (I^*\sigma^\kappa_2(n_\ell) a^+_x a^+_y
X_\ell^2 a_y a_x\sigma^\kappa_2(n_\ell)I\Big ) + \omega \Big
(I^*\sigma^\kappa_2(n_\ell) [a_x, X_\ell] a_y^+ a_y[a_x, X_\ell]
\sigma^\kappa_2(n_\ell)I\Big )
\eean
We use $ a^+_x a^+_y X_\ell^2 a_y
a_x \le C_M a^+_x a^+_y a_y a_x$ for  the first term. The second
term we use the identity
$$
[a_x,  X]^* a_y^+ a_y[a_x,  X] = a_y^+[a_x,  X]^*[a_x,  X]a_y
+a_y^+ [a_x,  X]^* [a_y,[a_x,  X]] + [[a_x, X]^*,  a_y^+ ]a_y
[a_x,  X]
$$
The first term of the RHS is bounded by $C_M a_y^+ a_y$. The other
two terms can be bounded by $C_M a_y^+ a_y +C_M$ by repeating the
same procedure once more (first apply Schwarz' inequality, then
use \eqref{ergo3.2}). We conclude that
\bean
&&\int_{\Lambda_\ell}
dx \int_{\Lambda_\ell}  dy W(x-y) \omega \Big
(I^*\sigma^\kappa_2(n_\ell)  X_\ell
 a^+_x a^+_y a_y a_x X_\ell\sigma^\kappa_2(n_\ell) I\Big )\\
&&\quad\le C \int_{\Lambda_\ell}  dx \int_{\Lambda_\ell} dy
|W|(x-y) \left\{\omega \Big (I^*\sigma^\kappa_2(n_\ell)
a^+_x a^+_y a_y a_x \sigma^\kappa_2(n_\ell)I \Big )\right.\\
&&\quad + \left.\omega \Big (I^*\sigma^\kappa_2(n_\ell) a^+_y
a_y\sigma^\kappa_2(n_\ell)I \Big ) + C\right\}
\eean
Now from the super-stability estimate, we have
$$
C \int_{\Lambda_\ell}  dx \int_{\Lambda_\ell} dy |W|(x-y) a^+_x
a^+_y a_y a_x \le C \int_{\Lambda_\ell}  dx \int_{\Lambda_\ell} dy
W(x-y) a^+_x a^+_y a_y a_x + N_\ell
$$
Thus, \bean &&\ell^{-3}\int_{\Lambda_\ell}  dx \int_{\Lambda_\ell}
dy W(x-y) \int dy \omega \Big (I^* \sigma^\kappa_2(n_\ell)
X_\ell a^+_x a^+_y a_y a_x X_\ell\sigma^\kappa_2(n_\ell)I \Big )\\
&&\quad C \ell^{-3} \int_{\Lambda_\ell}  dx \int_{\Lambda_\ell}
dy W(x-y) \int dy \omega \Big ( I^*\sigma^\kappa_2(n_\ell)
a^+_x a^+_y a_y a_x \sigma^\kappa_2(n_\ell) I \Big )\\
&&\quad + C \int_{\Lambda_\ell}  dx \int_{\Lambda_\ell}  dy W(x-y)
\int dy \omega \Big ( \, I^* \sigma^\kappa_2(n_\ell) \fn_\ell
  \sigma^\kappa_2(n_\ell)  I  \, \Big )
\eean
The last term is bounded. Combining these estimates, we have
$$
\omega \Big (I^* \sigma^\kappa_2(n_\ell)  X_\ell h_{\ell} X_\ell
\sigma^\kappa_2(n_\ell) I\Big ) \le C \omega \Big (I^*
\sigma^\kappa_2(n_\ell)h_{\ell} \sigma^\kappa_2(n_\ell)I\Big )+ C
$$
Since $h_{\ell} \le h_{\ell} + C n_\ell$, $h_{\ell} + C n_\ell \ge
0$ and $[ h_{\ell}, n_\ell] = 0$, we have \bean &&\omega \Big (I^*
\sigma^\kappa_2(n_\ell) h_{\ell} \sigma^\kappa_2(n_\ell)I \Big )
\le \omega \Big (I^*[h_{\ell} + C n_\ell]^{1/2}
\sigma^\kappa_2(n_\ell) \sigma^\kappa_2(n_\ell) [h_{\ell} + C
n_\ell]^{1/2}
 I  \, \Big )\\
&&\quad \le C \omega \Big (I^* [h_{\ell} + n_\ell]I\Big )+ C \eean
We can prove that $\omega \Big ( I^*[h_{\ell} + n_\ell] I\Big )$
is bounded  by using Lemma \ref{lem:gamma_bound}.
\end{proof}

\subsection{Proof of main ergodic lemma}
We can now prove Lemma \ref{prop:X2Xhat}.

\begin{proof}
Recall the decomposition of $\eta$ into its spatially ergodic
components:
$$
\eta=\int\mu(d\omega)\omega
$$
Since $X$ is bounded, by Lemma \ref{ergodic2} there is
$\Omega_{\kappa,X}(\ell)$ such that \beann
&&\omega(I^*(1-\sigma^\kappa)X_\ell(1-\sigma^\kappa) I )
+\omega(I^*\sqrt{\sigma^\kappa(2-\sigma^\kappa)} \hat X
\sqrt{\sigma^\kappa(2-\sigma^\kappa)} I )\\
&&\quad = (1-\sigma^\kappa(\omega))^2\omega(I^* X_\ell  I )
+\sigma^\kappa(\omega)(2-\sigma^\kappa(\omega))\omega(I^*
\hat X  I ) +\Omega_{\kappa,X}(\ell)
\eeann
where
$\sigma^\kappa(\omega)
=\sigma^\kappa(\lim_\ell\omega(h_\ell),\lim_\ell\omega(n_\ell))$.
Therefore,
\beann
&&\eta(I^* X_\ell
 I )-\eta(I^*(1-\sigma^\kappa)X_\ell(1-\sigma^\kappa))
-\eta(I^*\sqrt{\sigma^\kappa(2-\sigma^\kappa)}
\hat{X}\sqrt{\sigma^\kappa(2-\sigma^\kappa)} I)\\
&&\quad =\int\mu(d\omega)[1-(1-\sigma^\kappa(\omega))^2-
\sigma^\kappa(\omega)(2-\sigma^\kappa(\omega))]\omega(I^* X_\ell  I )\\
&&\quad\quad
-\int\mu(d\omega)\sigma^\kappa(\omega)(2-\sigma^\kappa(\omega)) \omega\{
I^* (\hat{X}-X_\ell)  I  \}+\Omega_{\kappa, \ell}(X)
\eeann
As
$1-(1-x)^2-x(2-x)=0$, the first term vanishes identically. The
middle term vanishes by the hypothesis that the only ergodic
states of finite specific relative entropy in the one-phase region
are the Gibbs states. The support of the function
$\sigma^\kappa(\omega) (2-\sigma^\kappa)(\omega))$ is such that
only these Gibbs states contribute to the integral. The integrand
vanishes by the definition of $\hat{X}$ \eqref{Xhat}, since we have
$\omega(X)=\hat{X}(\lim_\ell\omega(\bu_\ell))$. This concludes
Lemma \ref{prop:X2Xhat}.
\end{proof}

\section{Relative entropy estimate}

We now summarize the estimates on the relative entropy
we have so far. For any $0\leq T\leq
T_0$, we  write
$$
s(\gamma_t\mid\omega^\e _t)\big\vert_{t=\e^{-1}T}
=\e^{-1}T\Av_{0\leq t\leq
\e^{-1}T}\frac{d}{dt}s(\gamma_t\mid\omega^\e_t)
$$
We compute the rate of change of entropy by
\eqref{re2}, \eqref{re4} and \eqref{re5} to have
$$
s(\gamma_t\mid\omega^\e _t)\big\vert_{t=\e^{-1}T}
= T\Av_{0\leq t\leq
\e^{-1}T} \left \{ \Tr \, \gamma_t \, G(\vec \lambda_\e, a^+_M, a_M)
- \e^2  \partial_t \log c_\e (t)\right \} + {\cal E}^1_M
$$
where $G$ is defined in
\eqref{re4} and ${\cal E}_M^1  \le C e^{-c M^2}$ \eqref{re5}.

Recall the meaning of the various length scales and cut-off
parameters: $\e$ is the ratio of the macroscopic to microscopic
length scale, $M$ is the high-momentum cut-off, $\ell$ is the
length scale in the isometry $ I$ employed to define
commuting local versions of the conserved quantities, $a$ is a
length scale for averaging needed to make use of local ergodicity,
$\kappa$ is the length scale used to smooth the characteristic
function of the one-phase region, and $\delta$ is a small
parameter used in applications of the entropy inequality.

\donothing{
\begin{theorem}\label{thm:diff-ineq}
Let $\kappa$ be smaller than the minimum distance between the
Euler solutions and the one-phase region. For any $M$, there is
$\e_0>0$, such that for any $0<\e<\e_0$, one can take $\ell$ large
enough, then $a$ small enough such that, for all $0\le t\le
\e^{-1}T_0$, we have the bound
$$
s(\gamma_t\mid\omega^\e_t)\leq \delta^{-1}M\e\int_0^t
s(\gamma_{t'}\mid\omega^\e_{t'})\,dt' +Ce^{-cM^2}
$$
\end{theorem}
\begin{proof}
We begin by applying Theorem \ref{ergodicity}
to the main term in \eqref{ent1}.
}

Recall the convention
$$
\ubA\, \bdot \,  \ubB = \sum_{j=0}^3 \sum_{\mu=0}^3  { A}^\mu_j \,
B^\mu_j\, - \, \sum_{j=0}^3 {A}^4_j \,   B^4_j \; .
$$
We  now apply Theorem \ref{local-ergodicity}
to estimate
$ \Tr \, \gamma_t \, G(\vec \lambda_\e, a^+_M, a_M) $ by
$$
\Av_{t \le T/\e} \Tr \, \gamma_t \, G(\vec \lambda_\e, a^+_M, a_M)
\le T_1+ T_2
$$
where $T_1$ and $T_2$ are  defined as follows:
\beann
T_1&=&-\Av_{0\le t\le\e^{-1}T} \gamma_t \Av_x \Big (   \,
\unabla\vec \lambda(\e t, \e x) \, \bdot \;
 I ^*  \big ( \,
\tilde \sigma^\kappa\hat  \ubw_{M, x}\tilde \sigma^\kappa \, \big )
 I
\,   \, \Big )\\
T_2&=& -\Av_{0\le t\le\e^{-1}T}   \gamma_t \Av_x \Big (   \,
\unabla\vec \lambda(\e t, \e x)\, \bdot  \;
 I ^* \, \big \{ \, (1-\sigma^\kappa)(\bu_{x, \ell})
 \ubw_{M, x}(1-\sigma^\kappa)(\bu_{x, \ell}) \, \big \} \,
 I \,  \, \Big )
\eeann
We need to compute $\hat \bw$ which we state as the following lemma.

\begin{lemma}\label{comcurrent}
We have the following identities
$$
\hat w_j^\mu = A_j^\mu, \quad j=0, \cdots 3, \; \; \mu=0, \cdots, 4.
$$
where the functions $A_j^\mu$ are given in \eqref{A-def}.
\end{lemma}

These relations follow directly from the definition of the Gibbs states, the
expressions for $A_j^\mu$ in \eqref{A-def}, the calculation of the currents
$w_j^\mu$ in Section \ref{sec:calculations} and the virial theorem proved in
Section \ref{sec:virial}.

By construction of the $\omega^\e_t$, the time derivative of $\log
c_\e(t)$ can be expressed as
$$
\frac{d}{dt} \e^2\log c_\e(t) = \e^3 \int dx (\partial_t {\vec
\lambda\cdot \bq})(\e t,\e x)
$$
where ${\bq }$ is the solution of the Euler equations that we
are considering.  Recall  the following identity  about the Euler
equations:
$$
\int  \, \sum_{j=1}^3  A _j ( \bq(X)\, )\, \cdot \nabla_j {\vec
\lambda} ( \bq(X)\, )\, dX = 0 \ .
$$
Recall also $A_0(\bq) = \bq$. We can rewrite
$$
\frac{d}{dt} \e^2 \log c_\e(t) = \e^3  \int dx (\partial_t {\vec
\lambda\cdot \bq})(\e t,\e x) =  \e^3 \int dx (\unabla {\vec
\lambda\cdot \ubA})(\e t,\e x)
$$
Together with Lemma \ref{comcurrent}, we have
$$
T_1-\frac{d}{dt}\e^2 \log c_\e(t)  =-\Av_{0\le t\le\e^{-1}T}
\gamma_t \Av_x \Big [   \,  \unabla\vec \lambda(\e t, \e x) \,
\bdot \; \big \{ \,
 I ^* \big ( \,
\tilde \sigma^\kappa\hat  \ubw_{M, x} \tilde \sigma^\kappa \, \big )
(\bu_{x, \ell}) I \; -\; \hat \ubw (\bq(\e t, \e x)) \,
\big \} \, \Big ]
$$

Denote
$$
\| \unabla  {\vec \lambda}  \|_\infty = \| \nabla  {\vec \lambda}
\|_\infty + \| \partial_t  {\vec \lambda}  \|_\infty
$$
and introduce the functions
$$
\Gamma_M^1 (\vec{\lambda},\bu) = \unabla {\vec \lambda}
\bdot \big[ \,  (\sigma^\kappa \hat {\ubw}_{ M} \sigma^\kappa) (
\bu)- \hat {\ubw}(\bq  ) \, \big ]\; ,
$$
$$
\Gamma^2 (\vec{\lambda},\bu) =  \| \unabla  {\vec
\lambda}  \|_\infty  \big[ \, (1-\sigma^\kappa) ( \bu) \big
( h+ n \big ) (1-\sigma^\kappa) ( \bu) \,
\big ] \; ,
$$
where $h, n$ are the energy and density components to $\bu$
and $\bq$ is the dual variable of $\vec \lambda$ defined
in \eqref{dual}.

We can bound  $\ubw_M$ in $T_2$ by the cutoff Lemma
\ref{keycutlemma}. Thus we  have
$$
T_1+T_2 -\frac{d}{dt}\e^2 \log c_\e(t)\le \Av_{0\le t\le\e^{-1}T}
\gamma_t \Av_x \Big (  I ^*
\, \{ -\Gamma_M^1 + M \Gamma^2\} (\vec{\lambda}(\e t, \e x ),
\bu_{x, \ell} ) I  \,
\Big )
$$
Therefore, we have
\begin{equation}\label{ent2}
s(\gamma_t\mid\omega^\e _t)\big\vert_{t=\e^{-1}T}
\le  T\Av_{0\le t\le\e^{-1}T}
\gamma_t \Av_x \Big (  I ^*   \, \{ -\Gamma_M^1
+M \Gamma^2\} (\vec{\lambda}(\e t, \e x ),  \bu_{x, \ell} ) I  \,
\Big ) + {\cal E}^1_M
\end{equation}
where ${\cal E}_M^1  \le C e^{-c M^2}$ \eqref{re5}.

\subsection{Reduction to large deviation}

Recall the standard thermodynamics pressure is defined by
$$
\psi(\vec \lambda )= \lim_{\ell \to \infty} \ell^{-3} \log
Z_{\ell, \vec \lambda}
$$
Define the entropy
$$
s(\bq^\prime )= \sup_{\vec \lambda} [ {\vec \lambda}{\bq^\prime }
- \psi({\vec \lambda})]
$$
and the rate function (notice we also use $I$ for the embedding
into the standard torus $\Lambda_{\e^{-1}}$)
$$
I({\bq^\prime },{\vec \lambda}) =  s( {\bq^\prime }) + \psi({\vec
\lambda}) -{\vec \lambda}\cdot {\bq^\prime }
$$
The rate function has the following property
$$
I(\bq^\prime  ,{\vec \lambda}\, ) \ge 0 , \qquad I(\bq  ,{\vec
\lambda}\, ) = 0
$$
where $\bq  =  \partial \psi (\vec \lambda)/\partial \vec
\lambda$. Furthermore, if the Gibbs state with chemical potential
$\vec \lambda$ is in  the one phase region, we have
$$
\Hess  I(\bq  ,{\vec \lambda}\, ) \ge c \idty
$$
for some $c > 0$.

The main large deviation estimate we shall use is given in the
following lemma. This lemma will be  proved
in Section \ref{sec:ldlg}.

\begin{lemma}\label{ldlg}
Suppose  $\vec \lambda$ is a bounded smooth function so the Gibbs
state with chemical potential $\vec \lambda (x)$ is in  the one
phase region for all $x$. For any bound smooth function $G$ that
satisfies the condition
\be\label{Gb}
|G ( {\vec \lambda}, \,  \bq )| \le C
(e+ \rho)
\ee
where $e$ is the energy and $\rho$ is the density. Then
there is a $\delta_0> 0$ depending only on $C$ and a convex
functional $\tilde I$ such that for all $0<  \delta \le \delta_0$
$$
\tilde I (\bq^\prime, \lambda) = I (\bq^\prime, \lambda)
$$
in a small neighborhood of $\bq  =  \partial \psi (\vec
\lambda)/\partial \vec \lambda$ and
$$
  \lim_{\ell \to \infty}\lim_{\e \to 0}  \gamma
\Av_x \Big (  I ^*   \, G(\vec{\lambda}(\e x),
 \bu_{x, \ell}^+ ) I  \, \Big )
$$
$$
\le \int d X  \sup_{{\bq^\prime(X) }}\big[   G ({\vec \lambda}(X))
, {\bq^\prime }(X)\, ) \, - \delta^{-1}\tilde I(\bq^\prime(X)
,{\vec \lambda}(X)\, ) \, \big] +\, \delta^{-1} \, \lim_{\e \to 0}
s(\gamma \mid \omega^{\e}_{\vec \lambda} )
$$
Here the sup is over all functions  $\bq^\prime(X)$.

\end{lemma}

\subsection{Conclusion of the relative entropy estimate and proof of the main
theorem}\label{sec:conclusion}

We now apply Lemma \ref{ldlg} to estimate  \eqref{entineq}. Since we
need the bound \eqref{Gb}, we set $G= M^{-1}
\{ -\Gamma_M^1 + M \Gamma^2\}  $. Thus we
have  for any $\delta \le \delta_0$
$$
-\Av_{0\le t\le\e^{-1}T} \gamma_t \Av_x \Big (  I ^* \,
 \{ -\Gamma_M^1 + M \Gamma^2\} (\vec{\lambda}(\e t, \e x),  \bu_{x, \ell}^+ ) I
 \, \Big )\le R_6 \, + \, \delta^{-1} M \, s(\gamma_t \mid
\omega^{\e}_t )
$$
where \be R_6 = \int d x \; \sup_{{\bq^\prime }} \; \big ( \,
\{ -\Gamma_M^1 + M \Gamma^2\} ({\vec \lambda}(x)) ,
{\bq^\prime }(x)\, ) \, - \delta^{-1}
M \tilde I(\bq^\prime(x)  ,{\vec \lambda}(x)\, ) \, \big)
\label{R6} \ee
where $\tilde I$ is related to the rate function
defined in Lemma \ref{ldlg}.
We now estimate the dependence of $\Gamma_M^1$ on $M$.

\begin{lemma} There is a constant $c > 0$ such that
$$
\Gamma_M^1 ({\vec \lambda},  \, {\bq^\prime }\, ) = \Gamma^1({\vec
\lambda} , {\bq^\prime }\, )+ e^{- c M^2}
$$
where
$$
\Gamma^1({\vec \lambda},  \, {\bq^\prime }\, ) = \unabla {\vec
\lambda} \bdot \big[ \,  (\sigma^\kappa \hat {\ubw} \sigma^\kappa)
( \bq^\prime)- \hat {\ubw}(\bq  ) \, \big ]
$$
\end{lemma}

This  lemma can be proved following the idea of the proofs of
Lemmas \ref{cutoff-vel} and \ref{lcutoff}. It is part of our
assumptions that the Gibbs states satisfy the cutoff assumptions.

Now, we can conclude the relative entropy estimate and the proof of
Theorem \ref{thm-1}.

\noindent
{\bf Proof of Theorem \ref{thm-1}.}

Recall $\bq  =  \partial \psi (\vec \lambda)/\partial \vec
\lambda$. Clearly,  $ \Gamma^1 (\bq (\e t, \e x) ,{\vec
\lambda}(\e t , \e x)) = 0$. The first derivative
$$
\frac { \partial \Gamma^1 ({\vec \lambda}(\e t , \e x) ,
{\bq }(\e t , \e x)\, )} {\partial {\bq } (\e t , \e x)}  = 0
$$
is equivalent to the Euler equation as checked in \cite{OVY}.
Recall $\Gamma^2({\vec \lambda}(X)\, , \,  {\bq^\prime }(X)\, )$
is nonzero only when $\bq^\prime(X)$ is away from $\bq (X)$. Thus
we have for $|\bq^\prime| \le C $
$$
  \{ -\Gamma^1 + M \Gamma^2\} ({\vec \lambda}(X)\, , \,
{\bq^\prime }(X)\, ) \le C M (\bq^\prime(X)  - \bq  (X))^2
$$
Furthermore, from the definition of $\Gamma_j$ we have for all
$\bq^\prime$
$$
 \{ -\Gamma^1 + M \Gamma^2\}({\vec \lambda}(X) ,
{\bq^\prime }(X)\, ) \le C  M \big ( \, |\bq^\prime|(X) \, + \, 1
\, \big )
$$
Since $\tilde I(\bq^\prime(X) \, , \, {\vec \lambda}(X)\, ) \ge 0$
and $\tilde I(\bq^\prime(X)  ,{\vec \lambda}(X)\, ) = 0$ only when
$\bq^\prime(X)= \bq(X)$, for $\delta$ small enough we  have
$$
 \sup_{{\bq^\prime }}\big[  \{ -\Gamma^1 + M \Gamma^2\}
 (\bq^\prime(X) \, , \, {\vec \lambda}(X)\, ) \, -
 \delta^{-1} M \tilde I(\bq^\prime(X) \, , \, {\vec \lambda}(X)\, )
\; \big ] \le e^{- c M^2}
$$

We thus have
$$
 s(\gamma_t |\omega^\e_t )
\le  \delta^{-1}  M \e \int_0^t  \left [  s(\gamma_{t'} | \omega^\e_{t'}) +
Ce^{-cM^2}+ \Omega_M(\ell, \kappa) +\Omega(\epsilon, a) + C_Me^{-cl^3}
\right ] dt' \nonumber
$$
By intergrating this inequality is (i.e, using Gronwall's inequality), and
using the fact that $\epsilon t\leq T_0$, we arrive at the bound
$$
s(\gamma_t |\omega^\e_t )
\leq \delta^{-1} M T_0 e^{\delta^{-1} M T_0}
\left[Ce^{-cM^2}+ \Omega_M(\ell,\kappa) +\Omega(\epsilon, a) + C_Me^{-cl^3}
\right].
$$
Taking the limits $ \lim_{\kappa \to 0}\lim_{a \to 0}\lim_{\ell \to \infty}
\lim_{\e \to 0}$, we get the inequality
$$
\lim_{\epsilon \to 0}s(\gamma_t |\omega^\e_t ) \le
C \delta^{-1} M T_0 \; e^{\delta^{-1} M T_0- c M^2}.
$$
We can now let $M \to \infty$ and conclude the proof of
Theorem \ref{thm-1}. We emphasize that we need the error term stemming from the
high-momentum cutoff to be smaller than $e^{- CM}$ for any $C> 0$ in order
to have our results hold for $t \le C T_0$ for arbitrary $T_0$. This is
guaranteed by the Maxwellian bound in the cutoff assumption II.1, expressed
by \eqref{maxwellian}.

\section{Thermodynamics and large deviation}

We now prepare the way for the proof of {Lemma \ref{ldlg}}. Our approach to large deviations for
quantum Gibbs states and local Gibbs states is quite different from the
explicit analysis in \cite{LLS} for the ideal gases. We first introduce
the following local Gibbs state with independent subcubes.

\subsection{Local Gibbs state with independent subcubes}

Divide the torus  $\Lambda= \Lambda_{\e^{-1}}$ into unions of non-overlapping
cubes of size $\ell$. To fix the grid, we assume that the origin is the
center of one small cube. Denote a typical cube by $\alpha$.
Recall the configuration  space ${\cal S}(\alpha^+)$
and define the configuration  space
$$
{\cal S}(\Lambda^{(+)}) = \otimes_{\alpha \in \ell \ZZ^3\cap
\Lambda}  {\cal S}(\alpha^+)
$$
An element in this configuration space
can be denoted by
$$
{\vec x}^\sharp = (\cdots, ({\alpha_j}, x_j), \cdots), \quad
x_j\in \alpha^+_j
$$
The (Fock) function space
$\Gamma(\Lambda^{(+)})$ is the $L^2$ space of
antisymmetric functions
on ${\cal S}  (\Lambda^{(+)})$. Notice that
${\cal S}(\Lambda^{(+)}) \not = {\cal S}(\Lambda^{+})$ and
$\Gamma(\Lambda^{(+)}) \not = \Gamma(\Lambda^{+})$.

Recall $\Lambda= \Lambda_{\e^{-1}}$  is a torus.
Define  ${ \cI^{\Lambda}_\ell}$ from $\Gamma
(\Lambda)$ to $\Gamma (\Lambda^{(+)})$ (cf \cite{CLY}) by
$$
(  \cI^{\Lambda}_\ell \psi ) ({\vec x}^\sharp) = \bigg [  \;
\prod_j \chi_{\alpha_j} (x_j) \; \bigg ]   \psi (\vec x) \; .
$$
The crucial fact is that  $\cI$ is an isometry.

\begin{lemma}\label{Iproperty}
${ \cI }$ is  an  isometric embedding, i.e.,
$$
\|\phi \|  =  \|{ \cI } \phi\|
$$

%

\end{lemma}

\begin{proof}
Recall from the construction of $\chi$ the relation
\eqref{=11} implies that
\begin{equation}
\sum_{\alpha \in \ell \ZZ^3} \chi_\alpha^2(x)=1
\label{=1}
\end{equation}
We can prove the isometry by the following identity: For any two
wave functions $f$ and $g$, we have
$$
({ \cI } f,  \; { \cI } g) = \sum_{\vec \alpha} \;   \prod_j
\left [ \int_{\alpha_j} dx_j  | \chi_{\alpha_j} (x_j)|^2 \right ]
\; \bar f(\vec x) g(\vec x) = (f,  \;  g)
$$
\end{proof}

For isometric embeddings, we have the following useful bound.

\begin{lemma}\label{Imono}
Suppose ${ \cI }: {\cal H}_1 \to {\cal H}_2 $ is an isometric
embedding. Then
$$
\Tr_{{\cal H}_1} \; e^{{ \cI }^* A \cI } \le \Tr_{{\cal H}_2} \;
e^{A}
$$
\end{lemma}

\begin{proof}
We can assume that ${\cal H}_1 $ is just a subspace of ${\cal
H}_2$ and $\cI$ is the natural embedding.  Let $\phi_j$
be the othornormal eigenvectors of $ A$
in ${\cal H}_1 $. The the claim follows from the following
Peierls' inequality: Suppose $\phi_j$ are othornormal. Then
$$
\sum_j e^{ (\phi_j, A \phi_j) } \le \Tr e^{A}
$$
\end{proof}

The following Lemma shows that $I^\ast X I  = \cI^\ast X \cI $
for a suitable class of observables.

\begin{proposition} \label{prop:I*XI}
Suppose $X$ is the observable
$$
X=\int_{\alpha^+}dx_1\cdots dx_k dy_1\cdots dy_k\,
f(x_1,\ldots,x_k;y_1,\ldots,y_k)a^+_{x_1,\alpha}\cdots
a^+_{x_k,\alpha}a_{y_k,\alpha}\cdots a_{y_1,\alpha}\quad ,
$$
Then we have $I^\ast X I  = \cI^\ast X \cI $.

\end{proposition}

\begin{proof}
The following identity is a direct consequence of the definition
of $ \cI $: For $n\geq k$, \beyn a_{y_k,\alpha}\cdots
a_{y_1,\alpha} \cI  \psi((z_1,\alpha_1),\ldots,(z_n,\alpha_n))
&=&a_{y_k,\alpha}\cdots a_{y_1,\alpha}
\chi_{\alpha_1}(z_1)\cdots\chi_{\alpha_n}(z_n)\psi(z_1,\ldots,z_n)\\
&=& \delta_{\alpha_1,\alpha}\cdots\delta_{\alpha_k,\alpha}
\chi_\alpha(y_1)\cdots\chi_\alpha(y_k)\chi_{\alpha_{k+1}}(z_{k+1})
\cdots\chi_{\alpha_n}(z_n) \\
&&\quad \times \psi(y_1,\ldots,y_k,z_{k+1},\ldots, z_n)\quad .
\eeyn It follows that, for any $\phi,\psi\in L^2(\Lambda^{\times
n})$, $n\geq k$, \beyn (\phi, \cI ^* X \cI \psi) &=&
\int_{\alpha^+}dx_1\cdots dx_k dy_1\cdots dy_k\,
f(x_1,\ldots,x_k;y_1,\ldots,y_k)\\
&&\quad\times (a_{x_k,\alpha}\cdots a_{x_1,\alpha} \cI
\phi,a_{y_k,\alpha}\cdots a_{y_1,\alpha} \cI
\psi)\\
&=&\sum_{\alpha_1,\ldots,\alpha_n} \int dz_{k+1}\cdots dz_n
dx_1\cdots dx_k dy_1 \cdots dy_k
\delta_{\alpha_1,\alpha}\cdots\delta_{\alpha_k,\alpha}\\
&&\quad\times
\chi_{\alpha_{k+1}}(z_{k+1})^2\cdots\chi_{\alpha_n}(z_n)^2\chi_\alpha(x_1)
\cdots\chi_\alpha(x_k)\chi_\alpha(y_1)\cdots\chi_\alpha(y_k)\\
&&\quad\times  f(x_1,\ldots,x_k;y_1,\ldots,y_k)
\overline{\phi(x_1,\ldots,x_k,z_{k+1},\ldots, z_n)}
\psi(y_1,\ldots,y_k,z_{k+1},\ldots, z_n) \eeyn The sum over
$\alpha_1,\ldots,\alpha_n$ can be carried out using the Kronecker
delta's and \eqref{=1}. As $n\geq k$, $\phi$, and $\psi$ are
arbitrary, we have $I^\ast X I  = \cI^\ast X \cI $ by the
formula of $I$ in \eqref{Idef}.
\end{proof}

We now construct a ``special local Gibbs state".
Recall that $\bua$ is defined in \eqref{4.8} and its component commute.
For a smooth $ \vec \lambda$, let $\tilde \omega^{\e, \ell}_{\vec
\lambda}$ be the  state
$$
\tilde \omega^{\e, \ell}_{\vec \lambda} = \Tr  \frac {1}{ \tilde
c_{\e, \ell} (\vec \lambda)}\exp \left [\, \e^{-3} \Av_\alpha
{\vec \lambda} (\e \alpha) \cdot \, { I }^*   \bua  \, { I }
\, \right ]
$$
where the average of $\vec \alpha$ is over
$\alpha \in \ell \ZZ^3\cap
\Lambda$ and
$\tilde c_{\e, \ell} (\vec \lambda)$ is the partition
function defined by
$$
\tilde c_{\e, \ell} (\vec \lambda) = \Tr  \exp \left [\, \e^{-3}
\Av_\alpha {\vec \lambda} ( \e \alpha) \cdot \, I^* \bua \,
I \, \right ]
$$
Here the trace is over $\Gamma(\Lambda_{\e^{-1}}^{(+)})$.


Assume for the moment we can
drop the $I $ and the small cubes are independent.
Then $\tilde c_{\e, \ell} (\vec \lambda)$ can be computed easily.
The following Lemma asserts that this is essentially correct.
Recall the partition function defined in \eqref{part}.

\begin{lemma}\label{localthermo}
\be
  \lim_{\ell \to \infty}
   \lim_{\e \to 0}  \Big ( \e^3   \log  { c_\e (\vec \lambda)}
 - \Av_\alpha  \ell^{-3}\log   { Z_\ell (\vec \lambda (\e \alpha))} \Big )
= \lim_{\ell \to \infty} \lim_{\e \to 0}    \Big ( \e^3   \log  {
\tilde c_{\e, \ell} (\vec \lambda)}
 - \Av_\alpha  \ell^{-3}\log   { Z_\ell (\vec \lambda (\e \alpha))} \Big ) = 0
\label{7.1} \ee
\end{lemma}

\begin{proof}
{\it Upper bound:}
We first state an upper to  $\e^3 \log \tilde c_{\e, \ell} $.
Notice that
it is an inequality with no limits or other constants.
\begin{equation}\label{7.10}
\e^3 \log \tilde c_{\e, \ell} (\vec \lambda)
\le \Av_\alpha  \ell^{-3}\log   { Z_\ell (\vec \lambda (\e \alpha))}
\end{equation}

From  Lemma \ref{prop:I*XI}, we have
$$
\log \tilde c_{\e, \ell} (\vec \lambda) = \log \Tr  \exp \left [\,
\e^{-3} \cI^\ast  \Av_\alpha {\vec \lambda} ( \e \alpha) \cdot \,
\bua \,  \cI \, \right ]
$$
If we can neglect $\cI^\ast$ and $\cI$, then  \eqref{7.10}
follows from the fact that
different cubes are considered independent.
Lemma \ref{Imono} shows that we can remove $\cI^\ast$ and $\cI$ to have
an upper bound. This concludes the proof of \eqref{7.10}.

{\it Lower bound:}
Since $\vec \lambda $ is fixed, we shall drop it in the subscript.
Consider the entropy
$$
0\le s\big ( \, \omega^{\e} \, | \,  \tilde \omega^{\e, \ell } \,
\big ) = R_2 + \e^3   \log  { \tilde c_{\e, \ell} }-\e^3   \log  {
c_\e}
$$
where
$$
R_2 = \omega_\e \big (\,  \e^3 \int d x \vec \lambda(\e x) \cdot
u_x -\Av_\alpha {\vec \lambda} (\e \alpha) \cdot \,
I^* \bua \,  I \big )
$$
From Lemma \ref{Ilemma}, we have
$$
\lim_{\ell \to \infty}\lim_{\e \to 0} R_2 = 0
$$
We have thus proved that \be\label{cc} \lim_{\ell \to \infty}
\lim_{\e \to 0} \big [ \, \e^3   \log  { \tilde c_{\e, \ell} } -
\e^3   \log  { c_\e}
 \, \big ] \ge 0
\ee
\donothing{ Similarly, we consider
$$
0\le s\big ( \, \tilde \omega^{\e, \ell }  \, | \,   \omega^{\e}
\, \big ) =R_3 +\e^3   \log  { c_\e}-  \e^3   \log  { \tilde
c_{\e, \ell} }
$$
where
$$
R_3 = \tilde \omega_{\e, \ell} \big (\,  \e^3 \int d x \vec
\lambda(\e x) \cdot u_x -\Av_\alpha {\vec \lambda} (\e t, \e
\alpha) \cdot \, { I }^*   \bu_\alpha  \,  { I } \big )
$$
From
$$
\lim_{\ell \to \infty}\lim_{\e \to 0} R_3 = 0
$$
we have the reversed inequality. Thus $\e^3   \log  { c_\e}=  \e^3
\log  { \tilde c_{\e, \ell} }$ in the limit. We now prove the
formula with $Z_\alpha$. }

To conclude Lemma \ref{localthermo}, we
now obtain a lower bound on $c_{\e}$. This is the standard
procedure on the thermodynamics and we shall give only sketch. We
first divide the cube of size $\e^{-1}$ into cubes of size $\ell
(1- \sqrt \ell )$ with corridors of size $ 2  \sqrt \ell $. Now we
impose the Dirichlet boundary conditions on the boundary to obtain
au upper bound on the kinetic energy. The partition function is
bounded below by restricting the configurations so that there is
no particle on the corridors. Now there is no interactions between
different cubes and we obtain a lower bound of $c_{\e}$ in terms
of average over Dirichlet boundary conditioned partition functions
in cubes of size $\ell (1- \sqrt \ell )$. Since we can take $\eta
\to 0$ after $\ell \to \infty$ and partition functions is
independent of boundary conditions, we have thus proved that
$$
\lim_{\ell \to \infty} \lim_{\e \to 0}   \e^3   \log  {  c_{\e} }
\ge \Av_\alpha  \ell^{-3}\log   { Z_\ell (\vec \lambda (\e \alpha))}
$$
This concludes the Lemma.
\end{proof}

\subsection{Large deviation for commuting variables}

Recall that $\bua$ is defined in \eqref{4.8} and its component commute.
We shall take $x=0$ and denote $\bua$ by $\bu _{\ell}$.
The following Lemma is a standard application of large deviation
theory (or thermodynamics) to commuting variables.

\begin{lemma}\label{ld}
Suppose  $\vec \lambda$ is a fixed  constant so the Gibbs state
with chemical potential $\vec \lambda$ is in  the one phase
region. For any bound smooth function $G$ satisfies that
$$
|G ( {\vec \lambda}, \,  \bq )| \le C (e+ \rho)
$$
where $e$ is the energy and $n$ is the density. Let $\omega_{\vec
\lambda, \ell}$ be the finite volume Gibbs state defined by
$$
\omega_{\vec \lambda, \ell}(X) =   \frac 1 { Z_\ell(\vec \lambda)}
\Tr \, \exp  \left [\,
 \lan {\vec \lambda},
 \,    \bu _{\ell} \ran_{\Lambda_{\ell}} \right]X
$$
with periodic boundary condition. Since the components of
$\bu_\ell$ are commuting, we have
$$
 \frac 1 { Z_\ell(\vec \lambda)} \Tr \, \exp  \left [\,
 \lan {\vec \lambda},
 \,    \bu _{\ell} \ran_{\Lambda_{\ell}} + \delta \ell^3
G ( {\vec \lambda}, \,  \bu _{\ell} ) \right] = \omega_{\vec
\lambda, \ell} \, \Big ( \exp  \left [\,
  \delta \ell^3
G ( {\vec \lambda}, \,  \bu _{\ell} ) \right] \Big )
$$
Then there is a $\delta_0> 0$ depending only on $C$ and a convex
functional $\tilde I$ such that for all $0\le \delta \le \delta_0$
$$
\tilde I (\bq^\prime, \lambda) = I (\bq^\prime, \lambda)
$$
in a small neighborhood of $\bq  =  \partial \psi (\vec
\lambda)/\partial \vec \lambda$ and
$$
|{\Lambda_\ell}|^{-1} \log \omega_{\vec \lambda, \ell} \, \Big (
\exp  \left [\,
  \delta \ell^3
G ( {\vec \lambda}, \,  \bu _{\ell} ) \right] \Big ) \le
\sup_{{\bq^\prime }}\big[ \delta  G ({\vec \lambda} , {\bq^\prime
}\, ) \, - \tilde I(\bq^\prime  ,{\vec \lambda}\, ) \, \big]
$$
Here the sup is over all constants $\bq^\prime$.
\end{lemma}

We first sketch the idea of the proof for Lemma \ref{ld}:
The rate function $I$ can be
understood  in the following way: The probability to find the
$\bu   _\ell$ with a given value ${\bq^\prime }$ is given by
$\exp [ -|\Lambda_\ell| s(\bq^\prime )] $ with the entropy given
by
$$
s(\bq^\prime )= \sup_{\vec \lambda} [ {\vec \lambda} \cdot
{\bq^\prime } - \psi({\vec \lambda})]
$$
We now write $ \Tr \frac 1 {  Z_\ell(\vec \lambda)} \exp  \left
[\,
 \lan {\vec \lambda},
 \,    \bu    \ran_{\Lambda_{\ell}} \right] $ as
$$
\int d {\bq^\prime }  \exp  \left [\, |\Lambda_\ell| \left \{
 {\vec \lambda} \cdot \,    {\bq^\prime }  -
\psi({\vec \lambda})- s({\bq^\prime }) \right \}  \right]
$$
This gives the last variational formula.

\begin{proofof}{\bf Lemma \ref{ld}}:
We shall drop the constant parameter $\lambda$ in $G$
in this proof. Since the components of $\bu$ commute, we
can define the joint distribution $\nu_\ell (d \bu)$ of $\bu$
w.r.t. the state $\omega_{\vec \lambda, \ell}$. Thus
$$
\omega_{\vec \lambda, \ell} \, \Big ( \exp  \left [\,
  \delta \ell^3
G ( {\vec \lambda}, \,  \bu _{\ell} ) \right] \Big ) = \int d
\mu_\ell  (\bu)  \, \Big ( \exp  \left [\,
  \delta \ell^3
G ( {\vec \lambda}, \,  \bu _{\ell} ) \right] \Big )
$$
We now approximate the integral by the summation so that
$$
 \int d \mu_\ell  (\bu)  \,  \exp  \left [\,
  \delta \ell^3
G ( {\vec \lambda}, \,  \bu  ) \right] \le   \sum_{ \bm  \in
\ZZ^d}  \; P_{\vec \lambda, \ell}[ |\bu - \e \bm| \le \e ] \,
\exp  \left [\,
  \delta \ell^3 G ( \e\bm)\right]
$$
where
$$
G_\e ( \by ) = \sup_{|\bx-\by|\le \e}G(\bx)
$$
and $P_{\vec \lambda, \ell}$ denotes the probability
of the event described in its argument, with respect to the state
$\omega_{\vec \lambda, \ell}$.
We can bound the summation by
$$
\e^{-5} \int d \bx  \; P_{\vec \lambda, \ell}[ |\bu -  \bx| \le \e
] \,  \exp  \left [\,
  \delta \ell^3 G_\e ( \bx)\right]
$$

We have
$$
P_{\vec \lambda, \ell}\big ( \,  |\bu -  \bx| \le \e \, \big ) \le
P_{\vec \lambda, \ell}\big (\,  \vec \xi \cdot \bu  \ge  \vec \xi
\cdot \bx - |\vec \xi| \e  \, \big )
$$
for all $\vec \xi$. Notice that from the Chebeshev inequality we

have
$$
P_{\vec \lambda, \ell}\big (\,  \vec \xi \cdot \bu  \ge  \vec \xi
\cdot \bx - |\vec \xi| \e  \, \big ) \le e^{-\ell^3 \vec \xi \cdot
\bx + \ell^3 |\xi| \e  } \int d \mu_\ell  (\bu) \; e^{ \ell^3 \vec
\xi \cdot \bu }
$$
Let $\psi_\ell(\vec \lambda)$ be the pressure defined by
$$
\psi_\ell(\vec \lambda) =  \ell^{-3} \log \Tr \, \big ( \, e^{
\ell^3 \vec \lambda \cdot \bu }\, \big )
$$
so that
$$
\int d \mu_\ell  (\bu) \; e^{ \ell^3 \vec \xi \cdot \bu } =
\exp\Big \{ \ell^3 \big [\,  \psi_\ell(\vec \xi + \vec \lambda)
-\psi_\ell(\vec \lambda) \,\big ] \, \Big \}
$$
Thus
$$
P_{\vec \lambda, \ell}\big (\,  \vec \xi \cdot \bu  \ge  \vec \xi
\cdot \bx - |\vec \xi| \e  \, \big )
 \le
\exp \Big \{   -\ell^3 \big [  \, \vec \xi \cdot \bx- \,
\psi_\ell(\vec \xi + \vec \lambda) - \e |\vec \xi| +
\psi_\ell(\vec \lambda)  \, \big ] \Big \}
$$
for all
$$
\lambda_4+ \xi_4 > 0 \; .
$$
In particular, we have
$$
P_{\vec \lambda, \ell}\big ( \,  |\bu -  \bx| \le \e \, \big )
 \le
\exp \Big \{   -\ell^3 \!\!\!\! \sup_{
{-\eta^{-1} \le \xi_{j}+ \lambda_j
\le \eta^{-1}, \, j = 0 \cdots 3}\atop
{ \eta \le \xi_4+ \lambda_4
\le \eta^{-1}} } \!\!\!\big( \, \vec \xi \cdot \bx- \,
\psi_\ell(\vec \xi + \vec \lambda) - \e |\vec \xi| +
\psi_\ell(\vec \lambda)  \, \big ) \Big \}
$$

The existence of thermodynamics states that
$$
\lim_{\ell \to \infty} \big | \psi_\ell(\vec \lambda) - \psi(\vec
\lambda)| = 0
$$
uniformly in compact interval away from $\lambda_4 =0$. Fix a
small constant $\eta > 0$. Define
$$
\tilde s_\eta(\bx)= \sup_{ -\eta^{-1} \le \xi_{j} \le \eta^{-1},
\, j = 0 \cdots 3 \atop \eta \le \xi_4 \le \eta^{-1} } \; \; \big(
\,  \vec \xi \cdot \bx-
 \,  \psi (\vec \xi)   \, \big )
$$
We have
$$
 \sup_{
-\eta^{-1} \le \xi_{j}+ \lambda_j  \le \eta^{-1}, \, j = 0 \cdots
3 \atop \eta \le \xi_4+ \lambda_4 \le \eta^{-1} } \; \;\big( \,
\vec \xi \cdot \bx- \,  \psi_\ell(\vec \xi + \vec \lambda) - \e
|\vec \xi|
  \, \big )
$$
$$
\le \tilde  s_\eta (\bx)-\vec \lambda \cdot \bx  - C(\eta, \vec
\lambda)  \e + C_\ell
$$
where
$$
\lim_{\ell \to \infty}  C_\ell = 0
$$
Define
$$
\tilde  I_\eta (\bx) = \tilde  s_\eta (\bx)-\vec \lambda \cdot \bx
+\psi (\vec \lambda)
$$
Thus we have
$$
P_{\vec \lambda, \ell}\big ( \,  |\bu -  \bx| \le \e \, \big ) \le
\exp \Big \{   -\ell^3 \big( \, \tilde  I_\eta (\bx) - C_\ell -
C(\eta, \vec \lambda) \e \, \big ) \Big \}
$$
We now have the estimate
$$
\ell^{-3} \log \int d \mu_\ell  (\bu)  \,  \exp  \left [\,
  \delta \ell^3
G ( {\vec \lambda}, \,  \bu  ) \right] \le \ell^{-3} \log \Big
\{ \e^{-5} \int d \bx  \; P_{\vec \lambda, \ell}\big ( \,  |\bu -
\bx| \le \e \, \big ) \,  \exp  \left [\,
  \delta \ell^3 G_\e (\bx) \right] \Big \}
$$
$$
\le \ell^{-3} \log \int d \bx \; \exp  \left [\, -\ell^3 \big( \,
\tilde  I_\eta (\bx) \, - \, \delta  G_\e (\bx) \,\big ) \,
\right] +C_\ell + C(\eta, \vec \lambda) \e + \ell^{-3} |\log \e |
$$
The error vanishes in the limit $\lim_{\e \to 0} \lim_{\ell \to
\infty}$. The integration can be calculated using the Laplace
method to give
$$
\lim_{\ell \to \infty} \ell^{-3} \log \int d \bx \; \exp  \left
[\, -\ell^3 \big( \, \tilde  I_\eta (\bx) \, - \, \delta  G_\e
(\bx) \,\big ) \, \right] \le \sup_{{\bq^\prime }}\big[ \delta
G_\e ({\vec \lambda} , {\bq^\prime }\, ) \, - \tilde  I_\eta
(\bq^\prime  ,{\vec \lambda}\, ) \, \big]
$$
Clearly, we have
$$
\lim_{\e \to 0} \sup_{{\bq^\prime }}\big[ \delta  G_\e ({\vec
\lambda} , {\bq^\prime }\, ) \, - \tilde  I_\eta (\bq^\prime
,{\vec \lambda}\, ) \, \big] = \sup_{{\bq^\prime }}\big[ \delta  G
({\vec \lambda} , {\bq^\prime }\, ) \, - \tilde  I_\eta
(\bq^\prime  ,{\vec \lambda}\, ) \, \big]
$$

We now collect some properties for $\tilde s_\eta$ and $\tilde
I_\eta $ . Notice that,  by definitions, $\tilde s_\eta$ and
$\tilde  I_\eta $ are still convex. Furthermore, we can check that
if $\eta$ is small then
$$
\tilde  I_\eta  (\bx) = I (\bx)
$$
in a small neighborhood of $\bq$. This proves the Lemma.
\end{proofof}

\subsection{Proof of Lemma \ref{ldlg}} \label{sec:ldlg}

Recall $\alpha$ indices disjoint subcubes of width $\ell$.
By the entropy inequality \eqref{entineq}, we have
$$
-\Av_{0\le t\le\e^{-1}T}  \gamma_\e \Av_\alpha \Big (  I ^*
\, G (\vec{\lambda}(\e \alpha),  \bua ) I  \, \Big )\le
 R \, + \, \delta^{-1} M \, s( \gamma_\e  \mid \tilde \omega^{\e, \ell}_t )
$$
where, for any $\delta >0$, \be R = \e^d \delta^{-1}   \log  \Tr
\frac {1}{ \tilde c_{\e, \ell} (\vec \lambda)}\exp \left [\,
\e^{-3} \Av_\alpha I^\ast \bigg \{ {\vec \lambda} ( \e \alpha)
\cdot \, \bua \;  -  \; \delta    G ({\vec \lambda}( \e
\alpha), \bua )  \, \bigg \}    { I }  \,  \right]
\label{RR} \ee
We first control the last term   $s( \gamma_\e  \mid \tilde
\omega^{\e, \ell}_t )$ by writing it as
$$
s( \gamma_\e  \mid \tilde \omega^{\e, \ell}_t ) =s( \gamma_\e
\mid  \omega^{\e}_t ) + \e^3   \log  {  \tilde c_{\e} (\vec
\lambda)} -\e^3   \log  { \tilde c_{\e, \ell} (\vec \lambda)} +
R_4
$$
where
$$
R_4 = \gamma_\e \big (\,  \e^3 \int d x \vec \lambda(\e x) \cdot
u_x -\Av_\alpha {\vec \lambda} (\e \alpha) \cdot \,
I^* \bua \,  I \big )
$$
From the Lemma \ref{Ilemma}, we have
$$
\lim_{\ell \to \infty}\lim_{\e \to 0} R_4 = 0
$$

We now estimate $R$. Using  the argument
in the proof of Lemma \ref{7.10}, we
can drop the operator ${ I }$ in \eqref{RR} to have an upper
bound. Since the cubes indexed by  $\alpha$ are independent, we
have
$$
R \le  \delta^{-1}  \Av_\alpha  Q_\alpha - \Big ( \e^3   \log  {
\tilde c_{\e, \ell} (\vec \lambda)}
 - \Av_\alpha  \ell^{-3}\log   { Z_\ell (\vec \lambda (\e \alpha))} \Big )
$$
where
$$
Q_\alpha
 = \ell^{-3} \log  \Tr  Z_\ell (\vec \lambda (\e \alpha))^{-1} \exp  \left [\,
\ell^{3}  \bigg \{ {\vec \lambda} (\e \alpha) \cdot \,
\bua \;  -  \; \delta G ({\vec \lambda}( \e \alpha),
 \bua )  \, \bigg \}
\right]
$$
The last term $\Big ( \e^3   \log  { \tilde c_{\e, \ell} (\vec
\lambda)}
 - \Av_\alpha  \ell^{-3}\log   { Z_\ell (\vec \lambda (\e \alpha))} \Big )$
vanishes by Lemma \ref{localthermo}. Notice that the components of
$\bua$ commute. Thus the trace is over functional of
commuting operators and we are essentially the same as in the
classical theory.  Thus we can apply Lemma \ref{ld}
to estimate $Q_\alpha$.  Summarizing, we have
$$
  \lim_{\ell \to \infty}\lim_{\e \to 0}  \gamma
\Av_\alpha \Big (  I ^*   \, G(\vec{\lambda}(\e \alpha),
 \bua ) I  \, \Big )
$$
$$
\le \int d X  \sup_{{\bq^\prime(X) }}\big[   G ({\vec \lambda}(X))
, {\bq^\prime }(X)\, ) \, - \delta^{-1}\tilde I(\bq^\prime(X)
,{\vec \lambda}(X)\, ) \, \big] +\, \delta^{-1} \, \lim_{\e \to 0}
s(\gamma \mid \omega^{\e}_{\vec \lambda} )+ R_4
$$
Notice that the right  side of the inequality is independent of the
location of the grid. If we average the grid over the cube of size $\ell$,
we can replace the left side of the inequality
from averaging  over $\alpha$ to
averaging  over all points $x$ on the torus. This proves
Lemma \ref{ldlg}.

\bigskip

We now state a corollary to Lemma \ref{ldlg}.

\begin{corollary}\label{lln}
Suppose  $\vec \lambda$ is a bounded smooth function so the Gibbs
state with chemical potential $\vec \lambda (X)$ is in  the one
phase region for all $X$. Suppose $\gamma_\e$ is a sequence of
states such that the specific entropy $s(\gamma_\e \mid
\omega^{\e}_{\vec \lambda} )$ satisfies
$$
\lim_{\e \to 0} s(\gamma_\e \mid \omega^{\e}_{\vec \lambda} )=0
$$
For any bound smooth function $\vec J$ on the unit torus, we have
$$
\lim_{\e \to 0}  \gamma_\e \e^3 \int dx  \; \vec J(\e x) \cdot \bu
( \e x) =\int d X \, \vec J(X) \cdot \bq ( X)
$$

\end{corollary}

\begin{proof} Since $\vec J$ is bounded smooth,
from Lemma \ref{Ilemma} we have
$$
  \lim_{\ell \to \infty}\lim_{\e \to 0}  \gamma\big ( \,
\e^3 \int dx  \; \vec J(\e x) \cdot \bu ( \e x)- \Av_x \vec
J(\e x) \cdot \bu_{x, \ell}^+ \, \big ) = 0
$$
We now apply Lemma \ref{ldlg} to have
$$
    \lim_{\ell \to \infty}\lim_{\e \to 0}
\gamma_\e \, \Av_x \vec J(\e x) \cdot \big ( \, \bu_{x, \ell}^+-
\bq ( \e x) \, \big )
$$
$$
\le  \int d X \;  \sup_{\bq^\prime (X)} \Big [   \vec J(X) \cdot
\big ( \, \bq^\prime(X)- \bq ( X) \, \big ) - \delta^{-1} \tilde I
( \bq^\prime(X), \vec \lambda(X)) \, \Big ] + \delta^{-1} \lim_{\e
\to 0} s(\gamma_\e |\omega_{\vec \lambda})
$$
Since $\tilde I \ge 0$ and $\tilde I ( \bq^\prime (X), \vec
\lambda(X))= 0$ only when $\bq^\prime (X)= \bq (X)$,  the sup is
bounded by $ C \delta$. To see this, consider the model problem
$$
\sup_x x - \delta^{-1}  x^2 \le \delta\quad .
$$
Recall the assumption $\lim_{\e \to 0} s(\gamma_\e |\omega_{\vec
\lambda})=0$. Since we can choose $\delta$ arbitrarily small, we
prove the corollary.
\end{proof}

%
%

\section{Calculation of the currents}\label{sec:calculations}

The current density operators $w^\mu_{k,x}$ are implicitly defined by
\eqref{re3}, i.e., for any test functions $\bJ=(J^\mu)$, $\mu=0,\ldots, 4$,
they should satisfy
$$
i\delta_H(\int dx \lan\bJ ,\bu_x\ran)
-\sum_{k=1}^3  \lan\nabla_k\bJ ,\bw_{k,x}\ran)
=\parbox{6cm}{terms containing second and higher derivatives of the $J^\mu$
integrated with densities of bounded expectation.}
$$
If we apply the same sign convention for dot products with $\bw$ as the
convention adopted in \eqref{dot-convention} for $\bu$, this means we are
looking for the definition of $w^\mu_{k,x}$, such that, for any test function
$J$, the following formal identity holds:
\be
i\int dx J(x) [H,u^\mu_x]=\sum_{k=1}^3  \int dx \nabla_k J(x) w^\mu_{k,x}
+\parbox{6.5cm}{integrals with higher derivatives of $J$},
\label{w-formal}\ee
where $H$ is the formal Hamiltonian $H=\int dx h_x$, with $h_x$ as defined in
\eqref{u-def}.

In order to compute the commutators we use the canonical anticommutation
relations \eqref{car} and integration by parts. The commutation relations
involving derivatives such as $\nabla_k a_x$ etc., are most easily derived by
taking derivatives of the appropriate commutation relations without derivatives.
E.g., the identity
$$
    \Big[ a_u^+ a_v, a_{x}^+  a_{y}\Big]
    =    \delta(x-v)a_u^+  a_y
       - \delta(y-u)a_{x}^+  a_v
$$
follows directly form \eqref{car} and, by taking derivatives with respect to
$u$, also leads to
$$
    \Big[ \nabla_k a_u^+ a_v, a_{x}^+  a_{y}\Big]
    =  \delta(x-v)\nabla_k a_u^+   a_{y}
    +\delta_k(y-u) a_{x}^+  a_v
$$
where $\delta_k$ is the derivative of the delta distribution with respect to the
$k$th component. It is straightforward to derive all other necessary relations
in the same way. E.g.,
\be
[\nabla_k a^+_y\nabla_k a_y, a^+_x a_x] = -\delta_k(x-y)\nabla_k a^+_y
a_x+\delta_k(x-y)\nabla_k a_y a^+_x\quad .
\label{[kin,mom]}\ee

There are essentially three cases to consider: i) $\mu=0$, ii)
$\mu=1,2,3$, and iii) $\mu=4$.

{\bf i) $\mu=0$: calculation of $w^0_{k,x}$:}\newline
As $n_x$ commutes with the potential part of the Hamiltonian we only have to
consider the kinetic energy term, which can be computed using \eqref{[kin,mom]}.
After integrating by parts, we get
$$
i\int\! dxdy\,J(x)[\frac{1}{2} \nabla a^+_y\nabla a_y , a^+_x a_x]
=\int\! dx\,\nabla J(x)\frac{1}{2}i[\nabla_k a^+_x a_x-a^+_x\nabla a_x]
$$
By comparing this result and \eqref{w-formal} we find agreement with the
definition of $w^0_{k,x}$ as given in \eqref{theta0}. Note that, in this case,
no higher order derivatives of $J$ appear.

{\bf ii) $\mu=j=1,2,3$: calculation of $w^j_{k,x}$.}\newline
Now, both the kinetic energy term and potential energy term both yield
non-trivial contributions. First, we compute the kinetic energy term.
$$
i\int dxdy J(x) [\frac{1}{2}\nabla a^+_y \nabla a_y , p^j_x]
=\frac{1}{4}\int dxdy J(x) [\nabla_j a^+_x a_x, \nabla a^+_y \nabla a_y]
+\mbox{h.c.}
$$
where here and in the following h.c. stands for the adjoint of the preceding
term(s). By using the commutation relations and integration by parts we find
the following expression for this quantity:
\beann
\lefteqn{i\int dxdy J(x) [p^j_x,
\frac{1}{2}\sum_{k=1}^3\nabla_k a^+_y \nabla_k a_y]}\\
&=&
-\int dx J(x)\left[\nabla_j a^+_x\Delta a_x+ \Delta a_x\nabla_j a^+_x\right]
-\sum_{k=1}^3\nabla_k J (x)\Delta a^+_x a_x +\mbox{h.c.}\\
&=&
\sum_{k=1}^3\int dx J(x) \left[\nabla_k\nabla_j a^+_x\nabla_k a_x
+\nabla_k a^+_x \nabla_k\nabla_j a_x\right]\\
&& + 2\sum_{k=1}^3\int dx  \nabla_k  J(x)  \nabla_j  a_x^+  \nabla_k   a_x
-\int dx \nabla_j J(x) \Delta a^+_x a_x +\mbox{h.c.}
\eeann
After further integration by parts and reorganization the result can be written
as
\beann
\lefteqn{i\int dxdy J(x) [\frac{1}{2}\sum_{k=1}^3\nabla_k a^+_y
\nabla_k a_y , p^j_x]}\\
&=&\sum_{k=1}^3\int dx \nabla_k J(x)\frac{1}{2}\left[\nabla_j a^+_x\nabla_k a_x
+ \nabla_k a^+_x \nabla_j a_x\right]
+ \frac{1}{4}\nabla_k\nabla_j J (x)\left[\nabla_k a^+_x a_x +
a^+_x\nabla_k a_x\right]
\eeann
The first term of the RHS in this expression determines the first term
\eqref{thetaj}. Note that this time higher derivative terms appear that are
not included in the definition of $w^j_{k,x}$, but they contribute to the
error terms.

To calculate the contribution from the potential energy term in the Hamiltonian,
we start from the identity
\beann
\Big[\nabla_j a_u^+ a_u, a_{x}^+ a_{y}^+ a_{y} a_{x}\Big]
    &=&  \delta(x-u)\nabla_j a_u^+  a_{y}^+ a_{y} a_{x}
    +   \delta(y-u)a_{x}^+ \nabla_j a_u^+ a_{y} a_{x}\\
    &&+\nabla_j\delta(y-u) a_{x}^+ a_{y}^+ a_u a_{x}
    +\nabla_j\delta(x-u)    a_{x}^+ a_{y}^+ a_{y} a_u
\eeann
which leads to
\bea
\lefteqn{\int dx dy W(x-y)\Big[\int du J(u)
\frac{1}{2}[\nabla_j a_u^+ a_u - a^+_u \nabla_j a_u],
a_{x}^+ a_{y}^+ a_{y} a_{x}\Big]}\label{forces}\\
&=&
\int dx dy W(x-y)\Big[J(x)[\nabla_j a^+_x a^+_y a_y a_x + \mbox{h.c.}]
+ J(y)[a^+_x \nabla_j a^+_y a_y a_x + \mbox{h.c.}]\nonumber\\
&&+ [\nabla_j J(x) +\nabla_j J(y)]a^+_x a^+_y a_y a_x\Big]\nonumber\\
&=&-\int dx dy \Big [J(x) \nabla_{j,x} W(x-y) +  J(y) \nabla_{j,y} W(x-y) \Big ]
a_x^+  a_{y}^+ a_{y} a_{x}\nonumber
\eea
Due to the spherical symmetry of the potential we have
\be
\nabla_j W(x-y) = W'(x-y)\frac{(x-y)_j}{\vert x-y\vert}
\label{W_spherical}\ee
Using this identity we can write \eqref{forces} in the form
$$
-\int \Big [(J(x)-J(y))   W'(x-y) \frac{(x-y)_j} {|x-y|}\Big ]
a_x^+  a_{y}^+ a_{y} a_{x}\; .
$$
As the range of $W$ is finite by assumption, we can Taylor expand $J(x)-J(y)$
to rewrite this quantity in the following form:
\be
-\sum_{k=1}^3\int dxdy \nabla_k J(x)   W'(x-y) \frac{ (x-y)_k(x-y)_j} {|x-y|}
a_x^+  a_{y}^+ a_{y} a_{x} + \mbox{higher order derivatives of $J$}\; .
\label{[pot,mom]}\ee
Recall that, by definition, only the coefficients of the first order derivatives
of $J$ are included in the $\bw$ tensor. Therefore, combining \eqref{[kin,mom]}
and \eqref{[pot,mom]} and also including the appropriate factors 1/2 and -
signs, we find the expression for $w^j_{k,x}$ claimed in  \eqref{thetaj}.

{\bf iii) $\mu=4$: calculation of $w^4_{k,x}$.}\newline
The calculation of the energy current proceeds in the same way as the previous
cases, but there are more terms and terms with higher derivatives. The
contribution from the kinetic energy in the Hamiltonian to the kinetic energy
current is, up to a trivial constant, given by
\beann
i\int du dx  J(u) [\nabla a_u^+ \nabla a_u, \nabla a_{x}^+
\nabla a_{x}]
&=&i \sum_{k,l=1}^3\int du dx  J(u) \delta_{k,l}(u-x)
\Big[\nabla_k a_u^+ \nabla_l a_x
- \nabla_l a_{x}^+ \nabla_k  a_{u}\Big]\\
&=&i\sum_{k=1}^3\int dx \nabla_k J(x) \Big[ \nabla_k a_x^+  \Delta a_x
 - \Delta a_x ^+ \nabla_k a_x  \Big]
\eeann
where $\delta_{k,l}$ is shorthand for $\nabla_k\nabla_l\delta$.
This yields the first term of the energy current.

The potential energy term in the Hamiltonian does not contribute to the
potential energy portion of the energy current due to the fact that the
following commutators vanish:
$$
\Big[a_u^+ a_v^+ a_v  a_u, a_{x}^+ a_{y}^+ a_{y} a_{x} \Big]=0
$$
To calculate its contribution to the kinetic energy current we start form
\beann
\lefteqn{\int   J(u) \Big[\nabla a_u^+ \nabla a_u, a_{x}^+ a_{y}^+ a_{y}
a_{x} \Big]}\\
&=&  \sum_{k=1}^3\int du  J(u)\Big[ \delta_k (u-x)\nabla_k a_u^+
a_{y}^+ a_{y} a_{x}
+\delta_k(u-y) a_x^+  \nabla_k  a_{u}^+ a_{y} a_{x} \\
&& -\delta_k(u-y)a_{x}^+ a_{y}^+ \nabla_k a_{u}  a_x
-\delta_k(u-x)) a_{x}^+ a_{y}^+  a_y  \nabla_k a_{u} \Big ]\\
&=&   -\nabla_x  (J(x) \nabla  a_x^+)  a_{y}^+ a_{y} a_{x}
    - a_x^+  \nabla_y  (J(y) \nabla  a_{y}^+ ) a_{y} a_{x}\\
&& + a_{x}^+ a_{y}^+  \nabla_y ( J(y) \nabla a_y ) a_{x}
    + a_{x}^+ a_{y}^+ a_{y} \nabla_x ( J(x) \nabla a_x )
\eeann
By multiplying this expression by \ $W(x-y)$, and integrating over $x$ and $y$,
and integrating by parts, we find
\beann
\lefteqn{\int dx dy du W(x-y) J(u) \Big[\nabla a_u^+ \nabla a_u,
a_{x}^+ a_{y}^+ a_{y} a_{x} \Big]}\\
&=&\sum_{k=1}^3\int J(x) \nabla_{k,x}  W(x-y)\Big [ \nabla_k  a_x^+
a_{y}^+ a_{y} a_{x}
- a_{x}^+ a_{y}^+ a_{y}  \nabla_k  a_x  \Big ]\\
&&+ \sum_{k=1}^3\int J(y) \nabla_{k,y}  W(x-y)
 \Big [a_x^+ \nabla_k a_{y}^+  a_{y} a_{x}
-  a_{x}^+ a_{y}^+   \nabla_k a_y  a_{x} \Big ]
\eeann
The contribution of the kinetic energy to the potential energy current is
obtained in a similar way. The result is
\beann
\lefteqn{\int dx dy du W(x-y)J(x)
\Big[ a_{x}^+ a_{y}^+ a_{y}a_{x},\nabla a_u^+\nabla a_u \Big]}\\
&=&-\sum_{k=1}^3 \int  \nabla_{k,x} ( J(x) W(x-y) )
\Big [ \nabla_k  a_x^+  a_{y}^+ a_{y} a_{x}
- a_{x}^+ a_{y}^+ a_{y}  \nabla_k  a_x  \Big ]\\
&&- \sum_{k=1}^3\int  J(x) \nabla_{k,y}  W(x-y)
\Big [a_x^+  \nabla_k   a_{y}^+  a_{y} a_{x}
    -  a_{x}^+ a_{y}^+   \nabla_k a_y  a_{x} \Big ]
\eeann
The last four terms become the last two terms of the energy current:
\beann
\lefteqn{i \int dx J(x) [ h_x, H]}\\
&=& \frac i 4 \sum_{k=1}^3\int    \nabla_k J(x)  \Big [  \nabla_k a_x^+  \Delta
a_x - \Delta a_x ^+     \nabla_k a_x  \Big ]\\
&&
 - \frac i 4 \sum_{k=1}^3 \int  (\nabla_k J)(x) W(x-y)
 \Big [ \nabla_k  a_x^+  a_{y}^+ a_{y} a_{x}
    - a_{x}^+ a_{y}^+ a_{y}  \nabla_k  a_x  \Big ]\\
&&+ \frac i 4 \sum_{k=1}^3 \int (J(y)-J(x) ) \nabla_{k,y}  W(x-y)
 \Big [a_x^+    \nabla_k  a_{y}^+  a_{y} a_{x}
 -  a_{x}^+ a_{y}^+   \nabla_k a_y  a_{x} \Big ]
\eeann
By the same argument as for \eqref{[pot,mom]}, the last term can be rewritten
in the form
$$
+ i\sum_{k=1}^3 \int \nabla_k  J(x)  \Big [
 W'(x-y) \frac{ (x-y) \otimes (x-y)} {|x-y|}\Big ]
 \Big [a_x^+    \nabla   a_{y}^+  a_{y} a_{x}
 -  a_{x}^+ a_{y}^+   \nabla a_y  a_{x} \Big ] + O(J'')\quad .
$$

\section{The virial Theorem}\label{sec:virial}

The purpose of this section is to relate the expectation values of the RHS of
the dynamical equations \eqref{re3} in a Gibbs state with specified values of
the densities of the conserved quantities (local equilibrium), to these
quantities themselves in order to obtain a closed set of equations. To achieve
this we will make use of canonical transformations relating Gibbs states with
respect to reference frames with different velocities. This will allow us to use
reflection symmetry of Gibbs states at zero total momentum. A second element we
will need is the Virial Theorem to relate the so-called virial to the
thermodynamic pressure. We start with the latter. For the convenience of the
reader we first recall the main definitions.

Consider a system of particles in a finite volume $\Lambda\subset\Rl^d$,
interacting via a pair potential $W$. The {\em pressure\/} at inverse
temperature $\beta$ and chemical potential $\mu$, $P(\beta,\mu)$,  is defined by
\be
P(\beta,\mu)=\lim_{\Lambda\to\Rl^d}\frac{1}{\beta\vert\Lambda\vert} \log \Tr
e^{-\beta(H_{0,\Lambda}+V_\Lambda-\mu N_\Lambda)}
\label{def-pressure}\ee
where
\beann
H_{0,\Lambda}&=&\frac{1}{2}\int_\Lambda\!dx\, \nabla a^+_x\nabla a_x\\
V_\Lambda&=&\frac{1}{2}\int_\Lambda\!\int_\Lambda\!dxdy\,
W(x-y)a^+_x a^+_y a_y a_x\\
N_\Lambda&=&\int_\Lambda\!dx\,a^+_x a_x
\eeann
The trace is taken over the Fermion Fock space with one-particle space
$L^2(\Lambda)$. For our purposes, we can simply consider $\Lambda$ to be a cube
of side $L$, and define the operators with periodic boundary conditions. We will
write $V_\Lambda(W)$ when we wish to indicate the pair potential function
explicitly. By our general assumptions, the limit \eqref{def-pressure} exists
and we will restrict ourselves to the one-phase region of the  phase diagram. In
particular we assume that the pressure is continuously differentiable.

Gibbs states at non-vanishing total momentum are defined by introducing an
additional Lagrange multiplier for the momentum as follows.
\be
\omega_\lambda(X)=\lim_{\Lambda\to\Rl^d} \frac{1}{Z(\lambda)}\Tr
Xe^{-\beta(H_{0,\Lambda}+ V_\Lambda-\balpha\cdot P_\Lambda-\mu N_\Lambda)}
\end{equation}
where $\lambda=(\beta,\balpha,\mu)$, $\balpha=(\alpha_1,\alpha_2,\alpha_3)$,
are constants, and $P_\Lambda$ is the total momentum operator in the volume
$\Lambda$ defined by
$$
P_\Lambda=\frac{i}{2}\int_\Lambda\!dx\, \nabla a^+_x a_x - a^+_x\nabla a_x\ .
$$
and
$$
Z(\lambda)= \Tr e^{-\beta(H_{0,\Lambda}+V_\Lambda
-\balpha P_\Lambda-\mu N_\Lambda)}
$$
is the partition function.

The kinetic energy density is defined by
$$
e_{\rm kin}(\beta,\balpha,\mu)=\lim_{\Lambda\to\Rl^d}
\frac{1}{\vert\Lambda\vert}\omega_{\beta,\pi,\mu}(H_{0,\Lambda})
$$
The limits $\Lambda\to\Rl^3$ exist and are independent of the boundary
conditions under general stability assumptions \cite{Rue}. We will use the
abreviations $\omega_{\beta,\mu}=\omega_{\beta,0,\mu}$ and $e_{\rm
kin}(\beta,\mu)=e_{\rm kin}(\beta,0,\mu)$.

The {\em virial\/} of the potential $W$ in the volume $\Lambda$ is denoted by
$\cV_\Lambda(W)$ and is defined by
\be
\cV_\Lambda(W)=\frac{1}{2}\int_\Lambda\!\int_\Lambda\!dxdy\,
\nabla W(x-y)\cdot(x-y)a^+_x a^+_y a_y a_x \label{virial-def}
\ee
and the density of the local density virial is given by
\be
\nu_x=\frac{1}{2}\int_{\Rl^3}\!dy\, \nabla W(x-y)\cdot
(x-y)a^+_x a^+_y a_y a_x
\label{virialdensity-def}\ee
As we have assumed that $W$
has compact support, $\nu_x$ is well-defined.

Due to Galileo invariance, the Gibbs states for different values of $\balpha$
are related by a canonical transformation, which is  why in the statistical
mechanics of global equilibrium situations the total momentum is usually assumed
to vanish.

The canonical transformations relating $\omega_{\beta,\mu}$ and the states
$\omega_{\beta,\balpha,\mu}$, are defined as follows. Let $\bs\in\Rl^d$, and
consider the unitary $U_\bs$ on $L^2(\Rl^d,dx)$ defined by $(U_\bs\psi(\bx))=
e^{i\bs \cdot \bx}\psi(\bx)$. The second quantization of $U_\bs$ implements an
automorphism $\gamma_\bs$ on the Fermion algebra given by
$$
\gamma_\bs(a(f))=a(U_\bs f), \quad \gamma_\bs(a^+(f))=a^+(U_\bs f)\ .
$$
One can easily verify that the action of $\gamma_\bs$ on the operator-valued
distributions $a_\bx, \nabla a_\bx$, and their adjoints, is given by: \beann
\gamma_\bs(a_\bx)=e^{-i\bs \cdot \bx}a_\bx,&&\quad
\gamma_\bs(a^+_x)=e^{i\bs \cdot \bx}a^+_\bx\\
\gamma_\bs(\nabla a_\bx) =e^{-i\bs\cdot \bx}\nabla a_\bx-i\bs e^{-i\bs\cdot
\bx}a_\bx,&&\quad \gamma_\bs(\nabla a^+_\bx) =e^{i\bs\cdot \bx}\nabla a^+_x+i\bs
e^{i\bs\cdot \bx}a^+_\bx \eeann Clearly, $\gamma_\bs^{-1}=\gamma_{-\bs}$.

With these relations it is easy to check that
$$
\gamma_\bs(\nabla a^+_\bx \nabla a_\bx) =\nabla a^+_\bx \nabla a_\bx
+\vert\bs\vert^2 a^+_\bx a_\bx +i
\bs\cdot(a^+_\bx\nabla a_\bx-\nabla^+_\bx a_\bx)\ .
$$
Hence, the kinetic energy transforms as follows:
\be
\gamma_\bs(H_{\Lambda,0})=
H_{\Lambda,0}+\frac{1}{2}\vert \bs\vert^2 N_\Lambda -\bs\cdot P_\Lambda
\label{gamma_kinetic}\end{equation}
In the same way we see that \bea
\gamma_\bs(N_\Lambda)&=&N_\Lambda\label{gaugeN}\\
\gamma_\bs(P_\Lambda)&=&P_\Lambda -\bs N_\Lambda\label{gaugeP}\\
\gamma_\bs(V_\Lambda)&=&V_\Lambda \\
\gamma_\bs(\cV_\Lambda(W))&=&\cV_\Lambda(W))
\eea
It follows that
$$
\gamma_\bs (e^{-\beta(H_\Lambda  -\mu N_\Lambda)}) =e^{-\beta(H_\Lambda
-\bs \cdot P_\Lambda- (\mu -\frac{1}{2}\vert\bs\vert^2 N_\Lambda)}
$$
By putting $\bs =\balpha$, replacing $\mu$ by
$\mu-\frac{1}{2}\vert\balpha\vert^2$, we obtain
$$
e^{-\beta(H_\Lambda - \balpha \cdot P_\Lambda -\mu N_\Lambda)}
=\gamma_\balpha(e^{-\beta(H_\Lambda
-(\mu+\frac{1}{2}\vert\balpha\vert^2) N_\Lambda)})
$$
As the trace is invariant under $\gamma_\balpha$, this implies
\be
Z(\lambda)=\Tr
e^{-\beta(H_\Lambda - \balpha \cdot P_\Lambda - \mu N_\Lambda)} = \Tr
e^{-\beta(H_\Lambda -(\mu+\frac{1}{2}\vert\balpha\vert^2) N_\Lambda)} =
Z(\tilde\lambda) \label{Zlambdatilde}
\ee
where, for $\lambda=(\beta,\balpha,\mu)$,
we define $\tilde\lambda=(\beta,0,\mu +\frac{1}{2}\vert\balpha\vert^2)$.
Using this relation between partition functions and the invariance of the
trace under canonical transformations, we immediately get
\bea
\omega_\lambda(\gamma_\balpha(X)) &=&\frac{1}{Z(\lambda)} \Tr
\gamma_\balpha(X)e^{-\beta(H_\Lambda - \balpha \cdot P_\Lambda -
\mu N_\Lambda)} \nonumber\\
&=&\frac{1}{Z(\tilde\lambda)} \Tr X e^{-\beta(H_\Lambda -
(\mu+\frac{1}{2}\vert\balpha\vert^2) N_\Lambda)} =\omega_{\tilde\lambda}(X)
\label{omega_gamma} \eea
In combination with \eqref{gaugeN} this implies
$$
\omega_\lambda(N_\Lambda)=\omega_{\tilde\lambda}(N_\Lambda),
$$
and hence $\rho(\lambda)=\rho(\tilde\lambda)$, and we will simply
write $\rho$. Also
$$
\omega_{\tilde\lambda}(P_\Lambda)=\balpha\omega_\lambda(N_\Lambda)\; .
$$
If we apply this to $X=H_{\Lambda, 0}$ and combine this
with \eqref{gamma_kinetic}, to relate the kinetic energy densities of
$\omega_\lambda$ and $\omega_{\tilde\lambda}$, we find
\be e_{\rm
kin}(\beta,\balpha,\mu)=e_{\rm kin}(\beta,\mu+\frac{1}{2}\vert\balpha\vert^2)
+\frac{1}{2}\vert\balpha\vert^2\rho\; , \label{e_kin_lambdatilde}
\ee
where we have also used $\omega_{\tilde\lambda}(P_\Lambda)=0$.
The relation
\eqref{Zlambdatilde} between partition functions immediately implies the
following property of the pressure:
\be
P(\beta,\balpha,\mu)=P(\beta,0,\mu+\frac{1}{2}\vert\balpha\vert^2)
\label{pressure_lambdatilde}\ee
One interpretation of this relation is that the chemical potentials at different
values of $\balpha$, when regarded as a function of the particle density $\rho$,
satisfy
$$
\mu_\balpha(\rho)=\mu_0(\rho)-\frac{1}{2}\vert\balpha\vert^2\quad.
$$

We can now prove the virial theorem in the form we need.

\begin{theorem}[Virial Theorem]\label{thm:virial}
For a three-dimensional translation innvariant system with a continuously
differentiable pressure function, one has
$$
2\left[e_{\rm kin}(\beta,\balpha,\mu)
-\frac{1}{2}\vert\balpha\vert^2\rho\right] -
\lim_{\Lambda\to\Rl^d}\frac{1}{\vert\Lambda\vert}
\omega_{\beta,\balpha,\mu}(\cV_\Lambda(W))=dP(\beta,\balpha,\mu)
$$
\end{theorem}
The quantity between square brackets can be considered as the gauge invariant
kinetic energy.

\begin{proof}
Suppose that the theorem holds for $\balpha=0$. We can then use a canonical
transformation to obtain the result for arbitrary $\balpha$ as a consequence of
\eqref{omega_gamma}:
$$
\omega_\lambda(\cV(W))=\omega_\lambda(\gamma_{\balpha}^{-1}(\cV(W)))
=\omega_{\tilde\lambda}(\cV(W))= 2e_{\rm kin}(\beta,\mu+\frac{1}{2}\balpha^2) -
dP(\beta,\mu+\frac{1}{2}\vert\balpha\vert^2)
$$
By using \eqref{e_kin_lambdatilde} and \eqref{pressure_lambdatilde}, this is
equivalent to the statement of the theorem.

We now prove the theorem for $\balpha=0$. As the pressure is independent of the
boundary conditions, we can use periodic boundary conditions to compute it,
i.e., we choose $\Lambda$ to be a $d$-dimensional torus. For $t>0$, let
$t\Lambda$ be the torus rescaled by $t$. Then $\vert
t\Lambda\vert=t^d\vert\Lambda\vert$, and
$$
U_t: L^2(t\Lambda, dx)\to  L^2(\Lambda, dx): (U_t\psi)(x)=t^{d/2}\psi(tx)
$$
is unitary. The Laplacians on $\Lambda$ and $t\Lambda$are related as follows:
$$
U_t\Delta_{t\Lambda} U_t^*=t^{-2}\Delta_\Lambda\ .
$$
This relation carries over to the kinetic energy in second quantization:
$$
U_t H_{0,t\Lambda} U_t^*=t^{-2}H_{0,\Lambda}\ .
$$
where we have used the same notation for the corresponding unitary on the Fock
space $\cF(L^2(\Lambda))$ with one-particle space $L^2(\Lambda)$. Similarly, one
easily finds that the scaling behavior of the potential energy terms in the
Hamiltonians is as follows:
$$
U_t V_{t\Lambda}(W)U_t^*=V_\Lambda((W(t\cdot))\ .
$$
and for the particle number we have
$$
U_t N_{t\Lambda}U_t^*=N_\Lambda\ .
$$
By using these unitary equivalences we obtain \beann
P(\beta,\mu)&=&\lim_{\Lambda\to\Rl^d}\frac{1}{\vert t\Lambda\vert} \log
\Tr_{\cF(L^2(t\Lambda))} e^{-\beta(H_{0,t\Lambda}+V_{t\Lambda}(W)
-\mu N_{t\Lambda})}\\
&=&\lim_{\Lambda\to\Rl^d}\frac{1}{t^d\vert\Lambda\vert} \log
\Tr_{\cF(L^2(\Lambda))} e^{-\beta(t^{-2}H_{0,\Lambda}
+V_{\Lambda}(W(t\cdot)) -\mu
N_{\Lambda})}
\eeann
This shows that the last expression is independent of $t$.
Setting its derivative in $t=1$ equal to zero yields the following equation
$$
2e_{\rm kin}(\beta,\mu) -\lim_{\Lambda\to\Rl^d}\frac{1}{\vert\Lambda\vert}
\omega_{\beta,\mu}(\cV_\Lambda(W)) -dP(\beta,\mu)=0\ .
$$
\end{proof}

In order to close the dynamical equations, we need to express the expectation
values of the currents $w^j_k$, given in (\ref{theta0}-\ref{theta4}), in the
states $\omega_\lambda$ in terms of the  expectations of the conserved
quantities $u^j$ of \eqref{u-def}.

\begin{proposition}\label{prop:expectations}
The expectations of the local currents $w^j_k$ in a Gibbs state $\omega_\lambda$
are given by
\beann
\omega_\lambda(w^0_{k,x}) &=& \omega_\lambda( u  ^k_x)\\
\omega_\lambda(w^j_{k,x}) &=& \alpha_j\alpha_k\omega_\lambda( u  ^0_x)
+\delta_{k,j}P(\lambda)\\
\omega_\lambda(w^4_{k,x}) &=& \alpha_k \omega_\lambda( u  ^4_x) +\alpha_k
P(\lambda)
\eeann
where $P(\lambda)$ is the pressure defined in \eqref{def-pressure} and $u^j_x$
are the local densities of the five conserved quantities defined in
\eqref{u-def}. With the definitions of \eqref{A-def} and \eqref{Xhat}, this is
equivalent to $A=\hat{\bw}$. Explicitly: $q^0=\rho$, and
\beann
\omega_\lambda(w^0_{k,x}) &=& \alpha_k\rho=q^k\\
\omega_\lambda(w^j_{k,x}) &=& \alpha_j\alpha_k\rho+\delta{j,k}P
=q^jq^k/q^0+\delta{j,k}P\\
\omega_\lambda(w^4_{k,x}) &=& \alpha_k (q^4+P)= q^k (q^4+P)/q^0
\eeann
\end{proposition}

\begin{proof}
The first equation, $j=0$, follows directly from \eqref{theta0},
\eqref{gaugeP}, and \eqref{omega_gamma}. The expressions for
$\bw^j_{k,x}, j=1,2,3$, contain the virial of the potential $W$, which we can
relate to the thermodynamic pressure by using the virial theorem, Theorem
\ref{thm:virial}:
$$
w^j_{k,x} =  \nabla_j  a_x^+  \nabla_k   a_x - \frac 1 2 \Big [ W'(x-y) \frac{
(x-y)_j  (x-y)_k} {|x-y|}\Big ]
  a_x^+  a_{y}^+ a_{y} a_{x}
$$
For $j\neq k$, the expectation of the second term vanishes as it changes sign
under rotation over $\pi$ about the $j$th axis, which is a symmetry of the
potential and the Gibbs states. Due to the rotation invariance of the potential,
we also have $W^\prime(x)x/\vert x \vert=\nabla W(x)$. Therefore, the
expectation of the second term in a Gibbs state $\omega_\lambda$ is given by
$$
-\frac{1}{3}\delta_{j,k}\omega_\lambda(\nu_x)
$$
To treat the first term of $w^j_{k,x}$, as well as the first two terms of
$w^4_{k,x}$, we will transform these these terms to a frame where the Gibbs
state has zero total moment, so that we can more easily use invarance under
reflections in space. E.g., from \eqref{omega_gamma} we get
\beann
\omega_\lambda(\nabla_j  a_x^+ \nabla_k   a_x)
&=&\omega_{\tilde\lambda}(\gamma_\alpha(\nabla_j a_x^+ \nabla_k a_x))\\
&=&\omega_{\tilde\lambda}(\nabla_j a_x^+ \nabla_k a_x)
+\alpha_j\alpha_k\omega_{\tilde\lambda}(a_x^+ a_x)
+i\omega_{\tilde\lambda}(\alpha_ja_x^+ \nabla_k a_x
- \alpha_k \nabla_j a_x^+ a_x)
\eeann
As the total momentum has zero expectation in $\omega_{\tilde\lambda}$, the last
term vanishes for all $j,k=1,2,3$. By reflection symmetry and the defintion of
the kinetic energy we have
$$
\omega_{\tilde\lambda}(\nabla_j a_x^+ \nabla_k a_x) =\frac{2}{3}e_{\rm
kin}(\beta,\mu+\frac{1}{2}\vert\balpha\vert^2)
$$
By combining the above relations we obtain
\beann
\omega_\lambda(w^j_{k,x}) &=&
\alpha_j\alpha_k\omega_\lambda(u^0_x) + \frac{1}{3}\delta_{j,k}\left[ 2e_{\rm
kin}(\beta,\mu+\frac{1}{2}\vert\balpha\vert^2)
- \omega_\lambda(\nu_x)\right] \\
&=&\alpha_j\alpha_k\omega_\lambda(u^0_x) + \frac{1}{\beta}P(\beta,\balpha,\mu)
\eeann
where, for the last equality, we have used the virial theorem and
\eqref{pressure_lambdatilde}.

To compute the energy current, $\omega_\lambda(\bw^4_{k,x})$, we need to
consider the following expectations:
\beann
&&i\omega_\lambda(\nabla_k a_x^+a_{y}^+a_{y}a_{x}
-a_x^+ a_{y}^+ a_{y} \nabla_k a_{x})\\
&&i\omega_\lambda(a_x^+ \nabla_j a_{y}^+ a_{y} a_{x}
-a_x^+ a_{y}^+ \nabla_j a_{y} a_{x})\\
&&i\omega_\lambda(\nabla_k a_x^+  \Delta a_x - \Delta a_x ^+ \nabla_k a_x)\\
\eeann
Again, we use \eqref{omega_gamma} to relate these expectation to expectations
in $\omega_{\tilde\lambda}$. The first expectation becomes:
\be
i\omega_{\tilde\lambda}(\nabla_k a_x^+ a_{y}^+ a_{y} a_{x} -a_x^+ a_{y}^+ a_{y}
\nabla_k a_{x}) + 2\alpha_k\omega_{\tilde\lambda}(a_x^+ a_{y}^+ a_{y} a_{x})
\label{energycurrent1}\ee
The first term of this expression vanishes by symmetry.
In the same way we find \be i\omega_\lambda(a_x^+ \nabla_j a_{y}^+ a_{y} a_{x}
-a_x^+ a_{y}^+ \nabla_j a_{y} a_{x}) = 2\alpha_j\omega_{\tilde\lambda}(a_x^+
a_{y}^+ a_{y} a_{x})\quad . \label{energycurrent2}\ee

We treat the third expression with similar arguments:
\beann
i\omega_\lambda(\nabla_k a_x^+  \Delta a_x - \Delta a_x ^+ \nabla_k a_x)
&=&i\omega_{\tilde\lambda}
(\gamma_\balpha((\nabla_k a_x^+  \Delta a_x - \Delta a_x ^+ \nabla_k a_x))\\
&=&i\omega_{\tilde\lambda} (\nabla_k a_x^+  \Delta a_x + 2i\nabla_k
a_x^+(\balpha\cdot\nabla)a_x -\vert\balpha\vert^2\nabla_k a_x^+a_x\\
&& - i\alpha_ka^+_x\Delta a_x +2\alpha_k a^+_x (\balpha\cdot\nabla)a_x
+ i\alpha_k\vert\balpha\vert^2 a^+_x a_x)\\
&& + \mbox{ complex conjugate}\\
&=&- \alpha_k\omega_{\tilde\lambda}( 2\vert\balpha\vert^2 a^+_x a_x + 4 \nabla_k
a^+_x \nabla_k a_x -2 a^+_x\Delta a_x)
\eeann
Then, by using integration by parts and reflection symmetry we get the
following expression:
\be
i\omega_\lambda(\nabla_k a_x^+  \Delta a_x - \Delta a_x ^+ \nabla_k a_x)
= - 4\alpha_k\left[\frac{1}{2}\vert\balpha\vert^2\omega_{\tilde\lambda}
(u^0_x)+ \frac{5}{3}e_{\rm kin}(\tilde\lambda)\right]
\label{energycurrent3}\ee
Recall the expression for the energy current:
$$
w^4_{k,x} (t) = - \frac i 4   \Big [  \nabla_k a_x^+  \Delta a_x
 -  \Delta a_x ^+     \nabla_k a_x  \Big ]
 + \frac i 4 \int dy  W(x-y)
       \Big [ \nabla_k  a_x^+  a_{y}^+ a_{y} a_{x}
    - a_{x}^+ a_{y}^+ a_{y}  \nabla_k  a_x  \Big ]
$$
$$
- \frac i 4  \int   \Big [ W'(x-y) \frac{ (x-y)_k  (x-y)_j} {|x-y|}\Big ]
     \Big [a_x^+    \nabla_j   a_{y}^+  a_{y} a_{x}
    -  a_{x}^+ a_{y}^+   \nabla_j a_y  a_{x} \Big ]
$$
Using \eqref{energycurrent3}, we see that the expectation
of the first term in $\omega_\lambda$ equals
$$
\alpha_k \left[ \frac{5}{3} e_{\rm kin}(\tilde\lambda) +
\frac{1}{2}\vert\balpha\vert^2 \omega_\lambda(u^0_x)\right]
$$
For middle term we use \eqref{energycurrent2} and find
$$
\alpha_k\omega_\lambda(\frac{1}{2} \int d y  W(x-y) a_x^+ a_{y}^+ a_{y} a_{x})
$$
Similarly,  for the last term we get
\beann
-\frac{1}{2} \alpha_k\omega_\lambda(\int d y
\left[W^\prime(x-y)\frac{ (x-y)_k  (x-y)_j}
{|x-y|}\right] a_x^+ a_{y}^+ a_{y} a_{x})
&=&-\frac{1}{3}\alpha_k\omega_\lambda(\nu_x)\\
&=&\alpha_k\left[P(\lambda)-\frac{2}{3}e_{\rm kin}(\tilde\lambda)\right]
\eeann
where we have used the definition of $\nu_x$ \eqref{virialdensity-def}
and the virial theorem (Theorem \ref{thm:virial}).

By combining the three terms and applying the relation \eqref{e_kin_lambdatilde}
one obtains the expression for $\omega_\lambda(w^4_{k,x})$ given in the
statement of this proposition.
\end{proof}

\section{Appendix. The entropy inequality}

Our arguments rely in a crucial way on the following entropy inequality
\eqref{entineq}: For any pair of density matrices $\gamma$ and
$\omega$, and for all self-adjoint $h$, and any $\delta > 0$, one has
\begin{equation}
\gamma(h) \le \delta^{-1} \log\Tr \; e^{ \delta h+ \log \omega }
+\delta^{-1} S(\gamma|\omega) \label{entineq2}\end{equation} The
inequality holds in the more general context of normal faithful
states on a von Neumann algebra \cite{Pet}. Here we give a proof
for density matrices that emphasizes the  connection with
the variational principle of statistical mechanics.

\begin{proof}
Let $h$ be self-adjoint, and $\beta>0$. The variational principle
of statistical mechanics \cite{Rue} states that
$$
-\frac{1}{\beta}\Tr e^{-\beta H} =\inf_\gamma \left[\Tr\gamma
H-\beta^{-1}S(\gamma)\right]
$$
where the infimum is taken over density matrices $\gamma$, and
$S(\gamma) :=-\Tr \gamma\log\gamma$, is the von Neumann entropy of
$\gamma$. For any non-singular density matrix $\omega$, define
$H=-(\beta^{-1}(h+\log \omega)$, and take $\beta=\delta$, use
$$
S(\gamma\mid\omega)=\Tr \gamma(\log \gamma-\log \omega)
=-\Tr\gamma \log(\omega)-S(\gamma)\quad ,
$$
and rearrange the resulting inequality to obtain \eqref{entineq2}.
\end{proof}

Equality in \eqref{entineq2} holds if and only if
$$
\gamma=\frac{e^{h+\log\omega}}{\Tr e^{h+\log\omega}}\quad .
$$
The inequality \eqref{entineq2} can also be turned around:
\begin{equation}
S(\gamma\mid\omega)\leq \gamma(h)-\log\Tr e^{h+\log\omega}\quad ,
\end{equation}
and one can then take the sup over $h$ to obtain a
characterization of the relative entropy (as was done \cite{Pet}):
\begin{equation}
S(\gamma\mid\omega) =\sup_h\left[\gamma(h)-\log\Tr
e^{h+\log\omega}\right]
\end{equation}
with equality iff $\omega=e^{-h}/\Tr e^{-h}$, i.e., iff $h=\log
D_\omega + constant\times \idty$.

In contrast to the classical case, if $\log \omega$ and $h$ do not
commute, we generally have
$$
\log\Tr e^{h+\log\omega}\neq\log\Tr\omega e^h \quad.
$$
However, due to the Golden-Thompson inequality, i.e., for any pair
of self-adjoint $A$ and $B$,
$$
Tr e^{A+B} \leq Tr e^A e^B,
$$
we still have
$$
Tr\gamma h-\log\Tr \omega e^h\leq S(\gamma\mid\omega)\quad .
$$
Whenever $\omega$ and $\gamma$ do not commute, the equality will
be strict for all $h$.

\end{document}